\documentclass[final,nopreprintline,3p,times,twocolumn]{elsarticle}

\usepackage{amssymb}
\usepackage{amsmath}

\usepackage{graphicx}
\usepackage{algorithm}
\usepackage{algpseudocode}
\usepackage[colorlinks=true, allcolors=blue]{hyperref}
\usepackage{listings}
\usepackage[sets]{cryptocode}
\usepackage{booktabs}
\usepackage{multirow}
\usepackage{colortbl}
\usepackage{xspace}
\usepackage{tabularx}
\newtheorem{remark}{Remark}
\usepackage{fontawesome5}
\usepackage{xurl} 
\usepackage{paralist}

\newif\ifshowblock
\showblockfalse   
\newcommand{\vecdim}{$\ell$}
\newcommand{\numslots}{\texttt{numSlots}\xspace}
\newcommand{\numvectors}{$K$\xspace}

\newcommand{\numbaby}{\ensuremath{\texttt{n}_1\xspace}}
\newcommand{\matrixgroups}{$G$\xspace}
\newcommand{\groupmat}{\ensuremath{\mathbf{M}}\xspace}
\newcommand{\groupmatk}{\ensuremath{\mathbf{M}_k}\xspace}

\newcommand{\code}[1]{\text{\texttt{#1}}} 
\newcommand{\geo}[1]{\textcolor{orange}{{\sf (GP:} {\sl{#1})}}}
\newcommand{\gabrielle}[1]{\textcolor{blue}{{\sf (GD:} {\sl{#1})}}}
\newcommand{\syed}[1]{\textcolor{purple}{{\sf (SMH:} {\sl{#1})}}}

\newcommand{\ApproachA}{HyDia-RTX-DG\xspace}
\newcommand{\ApproachB}{BSGS-RTX-TBS\xspace}
\newcommand{\ApproachBprime}{BSGS-RTX-TBE\xspace}
\newcommand{\cpubsgs}{BSGS-CPU-DG}

\newcommand{\compdepth}{\kappa}

\newcommand{\mat}[1]{\textbf{#1}}
\renewcommand{\vec}[1]{\textbf{#1}}
\newcommand{\pp}[1]{\left( #1 \right)}
\newcommand{\dotp}[1]{\left\langle #1 \right\rangle}
\newcommand{\size}[1]{\left| #1 \right|}
\newcommand{\thales}[1]{}

\journal{Computer and Security}

\begin{document}

\begin{frontmatter}


 \cortext[cor1]{Equal contribution from these authors.}
\author{Gabrielle De Micheli\fnref{label1}\corref{cor1}}
\ead{gabrielle.demicheli@lge.com}
\author{Syed Mahbub Hafiz\fnref{label1}\corref{cor1}}

\author{Geovandro Pereira\fnref{label1}\corref{cor1}}

\author{Eduardo L. Cominetti\fnref{label1}\corref{cor1}}

\author{Thales B. Paiva\fnref{label1,label2}\corref{cor1}}

\author{Jina Choi\fnref{label3}}

\author{Marcos A. Simplicio Jr\fnref{label1,label2}}

\author{Bahattin Yildiz\fnref{label1}}

\title{Lightweight, Practical Encrypted Face Recognition with GPU Support} 

 \affiliation[label1]{organization={Advanced Security Team, LG Electronics, USA},
             }

 \affiliation[label2]{organization={
Universidade de São Paulo, São Paulo, Brazil},
             }

\affiliation[label3]{organization={
Next-Generation Computing Research Lab, CTO, LG Electronics, South Korea}
             }

\begin{abstract}
Face recognition typically operates in a client-server setting, where the client extracts a compact face embedding and the server performs similarity search over a template database.
Since facial data is highly sensitive, this raises significant privacy concerns.
Fully homomorphic encryption (FHE) addresses these concerns by enabling end-to-end encrypted similarity search.
However, existing FHE-based protocols are computationally costly and, especially, impose high memory overhead due to large rotation-key sets and bandwidth-bound homomorphic operations.
Building on prior work, \textit{HyDia} (PoPETS 2025), we introduce algorithmic and system-level improvements targeting real-world deployment with resource-constrained (edge) clients.
First, we propose BSGS-Diagonal, a fast and memory-efficient similarity computation algorithm that applies a Baby-Step/Giant-Step strategy with precomputed rotations reused across consecutive matrix--vector products.
This yields a $91\%$ reduction in rotation keys (${\sim}14$\,GB less client memory) and cuts peak server-side CPU RAM usage from over $33$\,GB to $11$\,GB for databases up to 1M entries, with runtime improvements of up to $1.57\times$ for membership verification and $1.43\times$ for identification.
Second, we introduce GPU-optimized similarity computation kernels, including an efficient homomorphic Chebyshev evaluator built upon FIDESlib (ISPASS 2025), a CKKS-level GPU library based on OpenFHE.
Rather than offloading individual CKKS primitives, our integrated kernels fuse operations to avoid repeated CPU--GPU ciphertext movement and costly FIDESlib/OpenFHE data-structure conversions.
Our HyDia and BSGS GPU results achieve up to $9\times$ and $21\times$ speedups on single GPU (and up to $287\times$ and $211\times$ using multi-GPUs), respectively, enabling sub-second encrypted face recognition for databases up to $2^{16}$ entries (or $2^{19}$ entries in a multi-GPU setting), while further reducing host memory usage.

\end{abstract}



\end{frontmatter}



\section{Introduction}

Face recognition automatically verifies or identifies a person from an image of their face, which is useful across application domains such as healthcare, law enforcement, and banking~\cite{survey-face-recon:2024}. In a typical client-server architecture, the client extracts a real-valued compact face embedding from facial images using deep neural networks, and the server compares those embeddings against templates stored in a database, using a similarity metric (typically, cosine similarity). The goal may be either membership verification (``is this person in the database?'') or identification (``which database entries match?'').

Transmitting facial data to a remote server raises privacy concerns, since facial images are personally identifiable information (PII) subject to strict regulations. This motivates privacy-preserving face recognition approaches for limiting identity leakage and protect against reconstruction attacks~\cite{mai2018reconstruction,dong2023reconstruct,shahreza2023comprehensive,yan2025black}. Transform-based techniques, such as perturbation methods~\cite{wang2023privacy,liu2025patronus,liu2026adversarial} or frequency-domain transforms~\cite{ji2022privacy,mi2022duetface}, apply an irreversible transformation before data leaves the device, but provide only empirical guarantees and can still leak identity information under strong or adaptive attacks.

Cryptographic schemes that enable matching operations directly on encrypted data are a more robust alternative. Using fully homomorphic encryption (FHE), garbled circuits, or multi-party computation (MPC), biometric data remains encrypted throughout the pipeline, with privacy derived from cryptographic security rather than obfuscation. FHE is particularly attractive for this purpose, as it requires only a single round of communication between the client and server, and recent advances in software and hardware acceleration have significantly improved its practicality~\cite{fhe-acceleration-survey:2024,agullo2025fideslib}.

Despite these advances, encrypted similarity search under FHE remains computationally demanding. In CPU-based implementations of state-of-the-art schemes like CKKS~\cite{cheon2017homomorphic}, the primary bottlenecks are ciphertext rotations (in diagonal linear transforms), ciphertext--ciphertext multiplications, and relinearizations. These operations are largely memory-bandwidth bound, making them promising candidates for GPU acceleration. However, naively offloading individual homomorphic operations can degrade performance, as costly CPU--GPU ciphertext transfers may dominate latency in a carelessly designed pipeline.

In this work, we propose an FHE-based face recognition protocol supporting both membership testing and identification, with emphasis on edge-device clients constrained in computation, memory, and energy. We build upon HyDia~\cite{martin2025hydia} and address its main limitation, shared by many FHE-based solutions: high RAM usage on both clients and servers, largely from rotation keys that must be generated, transmitted, and stored. Our contributions are both algorithmic and system-wide.

First, we incorporate a Baby-Step/Giant-Step (BSGS) strategy into the encrypted matrix--vector multiplication underlying the scheme's similarity computation. Our variant, BSGS-Diagonal, precomputes and caches structured rotations across a sequence of similarity matrices, reducing the rotation-key set by $91\%$, client-side memory by approximately $14$\,GB, and server-side peak RAM by up to ${\approx}4.5\times$ for databases from $2^{10}$ to $2^{20}$ entries, while also delivering up to $1.57\times$ speedups.

Second, we design an end-to-end GPU-resident architecture for encrypted similarity search. The principle is simple: upload the encrypted query once, execute similarity, threshold comparison, and aggregation predominantly on the GPU, and minimize host--device transfers. 
We integrate FIDESlib~\cite{agullo2025fideslib}
with our OpenFHE-based framework, implement a custom GPU-native Chebyshev evaluator 
to avoid CPU round-trips, and use a deferred (lazy) relinearization in similarity accumulation. Moreover, the BSGS path uses a different strategy than the CPU-based BSGS-Diagonal, based on pre-rotated diagonals with on-demand giant-step rotations.
The result is up to $9\times$ and $21\times$ speedups for HyDia and BSGS, respectively, enabling sub-second end-to-end runtimes for databases with up to $2^{15}$ entries while further reducing host RAM usage. We further improve our results by accelerating the single-GPU solution with a multi-GPU approach, achieving $287\times$ and $211\times$ speedups respectively over HyDia's and our BSGS single-GPU results, using a cluster of up to eight GPUs. This setup enables sub-second runtimes for databases with up to $2^{19}$ entries.

Our main contributions are:
\begin{enumerate}
\item Significant client- and server-side memory reduction for FHE-based face recognition, using an improved BSGS-style variant tailored for multi-similarity matrix computation.
\item To the best of our knowledge, the first GPU-accelerated implementation of an end-to-end encrypted face recognition protocol, bringing membership and identification scenarios to sub-second execution for databases having up to $2^{15}$ entries. The acceleration comes from new dedicated kernels with a custom comparison approach and a ciphertext multiplication kernel avoiding fused relinearization.
\item An open-source implementation based on OpenFHE~\cite{OpenFHE} and FIDESlib~\cite{agullo2025fideslib} tested with databases of varying sizes up to 1M entries for CPU and $2^{16}$ entries for GPU. Our implementation code (CPU and GPU) can be found here: \url{https://github.com/FastHE-Search/FastHE-Search/releases/tag/V1.0.0}\footnote{Regarding reproducibility: for the BSGS-Diagonal numbers, use the \emph{benchmark.sh} script with approach 8 (named ``BSGS-Precom-Opt'' in the code); for the GPU numbers, use approach 81 (``BSGS-GPU'') and 812 (``BSGS-GPU-PreRot'').}.
\item A demonstration of our solution's scalability using multiple GPUs and sharding techniques, achieving sub-second execution for databases having up to $2^{19}$ entries. Release available at \url{https://github.com/FastHE-Search/FastHE-Search/releases/tag/v2.2.0}.
\end{enumerate}

\paragraph{Paper organization} Section~\ref{sec:related_work} reviews related works. Section~\ref{sec:background} describes the CKKS scheme, the diagonal and baby-step/giant-step methods, the HyDia protocol~\cite{martin2025hydia}, and our security and privacy assumptions. Section~\ref{sec:bsgs} presents our BSGS-Diagonal optimization. Section~\ref{sec:gpu-bsgs} details our GPU implementation. Section~\ref{sec:experiments} discusses experimental results.
Section~\ref{sec:conclusion} concludes the work.
\section{Related works}\label{sec:related_work}
\paragraph{Secure K-NN}  

A core building block in similarity search is the $K$-nearest neighbors ($K$-NN) problem, sometimes also called top-$K$ selection.
Given a query vector, the $K$-NN problem identifies the $K$ closest points in the dataset under a chosen metric (e.g., Euclidean distance, cosine similarity, or inner product). 
%
%
Many studies in the literature focus on secure $K$-NN schemes, using a variety of privacy preserving techniques to provide client's query and/or database confidentiality.
Early solutions~\cite{10.1145/1559845.1559862,5767862,6816690,10.1145/2808425.2808430,6930802} often combine HE with techniques such as MPC, oblivious RAM or garbled circuits. 
Most of these proposals suffer from significant practical limitations, such as long execution times, large communication complexity, access pattern leakage or multiple rounds of interaction with the client.
Many subsequent works have focused on overcoming these limitations.
One example is the HE-based \emph{approximate} classifier proposed by Shaul et al.\cite{shaul2018secure}, which provides an efficient solution to the \emph{$k$-ish} NN problem.
However, their implementation based on BGV \cite{bgv:2014} still takes several hours to run for medium-sized datasets.
Later on, Chen et al.~\cite{chen2020sanns} introduced SANNS, two secure $K$-NN classifiers relying on homomorphic encryption, MPC, oblivious RAM and garbled circuits that keep both the client’s query and the search result confidential. 
As drawbacks, their solutions require asymptotically linear communication and are interactive, making them less suitable for outsourced computations.
%
\emph{Non-interactive} fully homomorphic $K$-NN algorithms also exist, such as theschemes by Zuber et al.~\cite{zuber2021efficient} and Ameur et al.~\cite{ameur2022secure}, both of which use TFHE~\cite{tfhe:2026} and have a quadratic cost in the size of the database. 
Without relying on HE, Servan-Schreiber et al.~\cite{servan2022private} presented PRECO, a first lightweight protocol for private nearest neighbor search where the server-side computation is efficient and the communication overhead is reduced to sublinear. 
PRECO relies on two non-colluding MPC servers running private NN searches, and achieves obliviousness by using distributed point functions, partial batch retrieval and oblivious masking.
%
Finally, Cong et al.~\cite{cong2024revisiting} focused on oblivious top-$K$ selection, describing a TFHE-based non-interactive and secure $K$-NN classification with significant speedups compared to~\cite{zuber2021efficient}.

\paragraph{Private database search}
While secure $K$-NN protocols address the problem of privately retrieving nearest neighbors, some works have explored private search in broader contexts. 
In these private search solutions, the goal is often to protect the client's query itself, while the database may remain in plaintext in some scenarios.
Ahmad et al.~\cite{ahmad2021coeus} presented Coeus, a private Wikipedia search system that uses BFV scheme \cite{bfv-eprint:2012} for the underlying homomorphic operations.
The protocol relies on secure matrix-vector product using private information retrieval (PIR) and the Term Frequency-Inverse Document Frequency (TF-IDF) method \cite{if-book-schutze:2008} to protect the privacy of the users' queries.
Henzinger et al.~\cite{henzinger2023private} introduced Tiptoe, a private web search engine where clients can efficiently search over millions of public documents while protecting their queries. 
Tiptoe relies on semantic embeddings, private NN search with fast linearly homomorphic encryption (using BFV), and clustering to reduce communication.
The server-side computational cost remains linear per query.
More recently, Asi et al.~\cite{asi2024scalable} introduced Wally, another private search system which is efficient for very large databases and many clients. 
Wally relies both on Somewhat Homomorphic Encryption (SHE) and differential privacy to guarantee the client's query privacy.
Concurrently to Wally, Zhou et al.~\cite{zhou2024pacmann} presented Pacmann, where clients run a graph-based approximate NN search combined with recent sublinear PIR schemes \cite{10646686} to achieve better performance and accuracy than \cite{henzinger2023private}.
Focusing on a specific application, Wang et al.~\cite{wang2025pathe} presented PATHE, a privacy-preserving pattern search framework that integrates FHE (CKKS and TFHE) with hyperdimensional computing (HDC) to achieve high-performance secure mass spectrometry database search.

\paragraph{Graph-based similarity search} 
Other recent contributions have explored fully homomorphic encryption for encrypted search more broadly. 
Notably, the Graph‑based Secure Similarity Search (GraSS) proposal~\cite{kim2024grass} is a secure and scalable FHE‑based framework for graph‑structured similarity queries with end‑to‑end privacy. 
It defines an FHE‑friendly graph representation and index encoding that reduces the complexity of neighborhood retrieval, including novel homomorphic algorithms tailored to graph operations. 
%
%
More recently, Zhu et al.~\cite{zhu2025compass} introduced Compass, a search system over encrypted embedding data.
It uses a search index that includes the traversal of a graph-based ANN algorithm on top of Oblivious RAM for improved performance.

\paragraph{Privacy-preserving facial recognition} 
Bodetti~\cite{boddeti2018secure} is one of the first studies on the practicality of using FHE for 1:1 private face recognition systems.
The suggested algorithm relies on BFV and uses batching techniques for the homomorphic computation of inner products between vectors, so multiple elements are encrypted into a single encrypted block. 
In addition to these standard batching techniques, Bodetti also utilizes dimensionality reduction algorithms (in particular, the classical Principal Component Analysis) to further tackle the high computational cost of face matching over encrypted data.

Subsequently, Engelsma et al.~\cite{engelsma2022hers} introduced a novel data encoding scheme that is tailored for efficient ${1:\texttt{many}}$ representation matching in the encrypted domain, also using BFV. 
Compared to~\cite{boddeti2018secure}, their proposed HERS algorithm scales to larger databases due to a slower increase in computational complexity (namely, number of homomorphic additions and multiplications) and the absence of required rotations.

Later, Ibarrondo et al.~\cite{ibarrondo2023grote} proposed replacing element-wise testing by the notion of group testing to significantly reduce the number of non-linear operations in the encrypted domain. 
More specifically, rather than evaluating all scores individually, their proposed GROTE algorithm derives representative maxima for each group and performs threshold checks only on these values for better efficiency. 
Their algorithm relies on the CKKS homomorphic scheme.

Aiming to improve the efficiency of FHE-based face recognition, Choi et al.~\cite{choi2024blind} proposed Blind-Match, a CKKS-based scheme where the feature vector is divided into smaller parts and each part is then processed individually. 
Their custom cosine similarity computation method allows them to reduce the number of additions and rotations compared to prior work~\cite{boddeti2018secure} and outperform GROTE for varying database sizes. 
%

More recently, Martin et al.~\cite{martin2025hydia} presented HyDia, another CKKS-based protocol for $1 : \texttt{many}$ face recognition that exploits diagonal packing for efficient matrix multiplication and scalability. 
The proposed algorithm uses a hybrid approximation method to compute homomorphic thresholding with better accuracy. 
Their study also provides a comprehensive comparison with the aforementioned works, including detailed plots and tables across multiple metrics such as server overhead, network bandwidth, circuit depth, and ciphertext size. 
%
Since HyDia is one of the latest works in private face recognition, and delivers state-of-the-art performance for large-scale $1 : \texttt{many}$ matching, we primarily compare our improvements to it, besides building upon their open-source implementation.

Another recent related work is CryptoFace~\cite{ao2025cryptoface}, which targets an end-to-end setting where feature extraction itself is performed under FHE. 
The client encrypts the probe face image and the server evaluates an FHE-friendly convolutional neural network (CNN) to produce encrypted features for thresholded matching. 
Even though CryptoFace offers stronger protection by moving more of the pipeline into the encrypted domain, it is primarily evaluated for 1:1 verification, with identification demonstrated only via a small 1:128 closed-set retrieval experiment. 
In addition, its online latency is dominated by the homomorphic CNN feature-extraction stage (rather than by database-scale $1 : \texttt{many}$ matching), whereas HyDia and our work target and optimize the large-scale encrypted similarity search stage.

\paragraph{GPU-based similarity search}

To the best of our knowledge, no public research articles report results specifically for GPU-accelerated encrypted face recognition. 
Nevertheless, public information on the topic can be found for CryptoLab’s Encrypted Facial Recognition (EFR) solution, described primarily through product and press material~\cite{cryptolab_prnewswire_efr_rsac2025}.
In those documents, the authors claim that GPU acceleration enables matching against tens of millions of encrypted face-template vectors in milliseconds, with horizontal throughput scaling as additional GPUs are provisioned.

There are also many studies involving GPU acceleration for the FHE backend, rather than face-recognition applications directly. 
For example, the FIDESlib~\cite{agullo2025fideslib} library adopted in this work is an open-source, server-side CKKS-on-GPU library designed to interoperate with client-side OpenFHE workflows, including optimized GPU-oriented primitives and multi-GPU support. 
%
More broadly, proprietary and emerging libraries such as HEAAN2~\cite{cryptolab_heaan2} and Desilo~\cite{desilo_fhe_docs} explicitly emphasize CUDA-based GPU acceleration for standard CKKS operations (e.g., rescaling, rotations/key-switching, and bootstrapping).
Finally, the Cheddar GPU library~\cite{choi2026cheddar} was very recently open-sourced. 
However, based on our initial investigation, it does not appear to provide support for encrypted matrix--vector multiplication.
Therefore, a thorough analysis of its performance and eventual comparison with our FIDESlib-based implementation is left as a topic for future work.

\section{Preliminaries}\label{sec:background}

%

\subsection{CKKS}\label{sec:ckks}
CKKS~\cite{cheon2017homomorphic, ckks2} is a single instruction multiple data (SIMD)-style homomorphic encryption scheme that operates over a vector of complex or real values. 
It relies on RingLWE ciphertexts in ${\mathcal{R}_Q^2}$ for a given ring $\mathcal{R}=\mathbb{Z}[X]/(X^N+1)$, where $N = 2^k$ (for $k > 0$) is the ring dimension and $\mathcal{R}_Q=\mathcal{R}/Q\mathcal{R}$ denotes $\mathcal{R}$ reduced modulo an integer $Q = \prod_{i=0}^{L}q_{i}$. 
The integer $L$ is known as the \emph{multiplicative depth} and represents the maximum number of rescaling levels available before decryption fails.
CKKS supports element-wise operations, such as addition and multiplication, as well as cyclic rotation.
We now describe some core operations supported by CKKS that are most relevant to this work.

\noindent \textbf{Encoding.} Given a real (or complex) vector $\textbf{x} \in \mathbb{C}^{N/2}$, it is encoded into elements of $\mathcal{R}_Q$ using an approximate inverse of a scaled complex canonical embedding.
More precisely, one applies an inverse Fast Fourier Transform on the elements of $\textbf{x}$ and scales each output by a factor~$\Delta$. 
Each element is then rounded to the nearest integer, since encryption is performed over integers modulo~$Q$.
This encoding step outputs a plaintext polynomial $\textbf{m}(X)$ that 
packs~$N/2$ complex values, named the available \emph{slots}.

\noindent\textbf{Encryption.} A plaintext polynomial $\textbf{m}(X)$ is encrypted into a ciphertext $(\textbf{a}, \textbf{b}) \in \mathcal{R}^2_Q$ with a given public key and the addition of random noise.


\noindent\textbf{Addition.} CKKS supports element-wise plaintext-ciphertext and ciphertext-ciphertext additions. The output ciphertext represents the SIMD addition of the underlying complex vectors.

\noindent\textbf{Multiplication.} CKKS also supports plaintext-ciphertext and ciphertext-ciphertext multiplication. 
To avoid an exponential growth in the scaling factor over repeated multiplications, each multiplication is followed by a rescaling step. 
The output polynomial has coefficients in $\mathbb{Z}_{Q_{\ell -1}}$ instead of $\mathbb{Z}_{Q_{\ell}}$ where $Q_{\ell} = \prod_{i=0}^{\ell}q_i$ for $0 < \ell \leq L$. 
%
This corresponds to the consumption of one level from the available depth $L$.

\noindent \textbf{Rotation.} Cyclic rotations shift the elements of the input vector $\textbf{x}$ by $0 < k < N/2$ slots. 
The output ciphertext represents the same underlying vector with shifted elements by $k$ slots.



\subsection{Diagonal method}\label{sec:diag}
The diagonal encoding method is a well-known technique for efficiently computing homomorphic matrix-vector products in a SIMD-friendly manner. 
Originally introduced by Halevi and Shoup~\cite[Section 4.3]{10.1007/978-3-662-44371-2_31}, the method starts by extracting the diagonals from a given matrix $\groupmat\in\mathbb{R}^{\ell\times\ell}$ before computing the product homomorphically by summing rotated, element-wise products. 
More precisely, if $\text{diag}_i = (\groupmat_{0, i}, \groupmat_{1, i+1}, \ldots, \groupmat_{\ell-1, i-1})$ denotes the $i$th generalized diagonal of \groupmat, 
then the product $\groupmat \cdot\vec{v}$ is computed as
\begin{equation}\label{equ:diag}
\groupmat \cdot\vec{v} = \sum_{i=0}^{\ell-1} \text{Rot}_i(\vec{v}) \odot \text{diag}_i,
\end{equation}
where $\text{Rot}_i$ denotes a cyclic rotation by $i$ slots.
In a standard plaintext-matrix setting, the vector $\vec{v}$ is encrypted as a single ciphertext and each diagonal $\text{diag}_i$ is a packed plaintext. 
The homomorphic evaluation involves, for each $i$, one ciphertext rotation, one plaintext--ciphertext multiplication, and one addition. 
Therefore, the method uses $\ell-1$ rotations, $\ell$ multiplications, and $\ell-1$ additions in total, i.e., the complexity is $\mathcal{O}(\ell)$.
Efficient implementations may take advantage of commonplace optimizations such as hoisting (see Section \ref{sec:prelim:bsgs:hoisting}), but those only affect constants underlying such asymptotic costs. 
%
The diagonal packing method also requires $\ell-1$ (i.e., $\mathcal{O}(\ell)$ memory) rotation keys, which typically dominate the RAM footprint on both client and server sides.

We emphasize that Equation~\ref{equ:diag} describes the classical plaintext-matrix variant of the diagonal method, where the diagonals are known in advance and encoded as plaintext constants. 
In our setting, however, the database matrix is encrypted.
This results in additional challenges, as further discussed in Section \ref{sec:prelim:bsgs:plainVSencM}.

\subsubsection{Diagonal method for rectangular matrices}
\label{diag-ct-packing}

In real-world applications, such as $1 : \texttt{many}$ facial matching, the matrix $\mathbf{M}$ is not necessarily square. 
Indeed, standard face feature extractors~\cite{martin2025hydia,deng2019arcface} often set the embedding vectors dimension~$\ell$ to a small power of two (e.g., FRGC2 dataset employs $\ell=2^9$).
Meanwhile, facial template databases are often large, reaching a size $K$ that may surpass $2^{20}$ entries.
This unbalance between $\ell$ and $K$ implies that Equation~\ref{equ:diag} cannot be directly applied to produce the desired $K$ inner products.

We briefly recall how HyDia~\cite{martin2025hydia} handles this problem via a packing strategy (see \cite[Section 5.1]{martin2025hydia} for details):
%
\begin{enumerate}
    \item Consider a database DB containing $K$ vectors of dimension $\ell$, say $\text{DB} = \{v_0, ..., v_{K-1}\}$.
    First, partition DB into groups of $\ell$ vectors, each getting $\ell \times \ell$ square matrices $S_j$, for $0 \leq j < \lceil{K/\ell}\rceil$. 
    The rows of matrix $S_j$ will contain the vectors $v_{j\ell}, \cdots, v_{(j+1)\ell-1}$.
    
    \item {\sloppy For each square matrix $\textbf{S}$, build a diagonal matrix $\textbf{D}$ where the $i$-th diagonal entry is denoted $\text{diag}_i(\textbf{S}) = (\textbf{S}_{0, i}, \textbf{S}_{1, i+1}, \cdots, \textbf{S}_{\ell-1, i-1})$, for all $i$. \par}
    
    \item \label{step:matrix-packing}Since each CKKS ciphertext encodes $\numslots$ coefficients, concatenate $\numslots / \ell$ diagonal matrices horizontally.
    The resulting rectangular matrix is hereby denoted as \groupmatk, for $0 \leq k < G=\lceil K/\numslots \rceil$. Each row of $\groupmatk$ is then encrypted into a separate ciphertext, yielding $\ell$ ciphertexts. 
    
    \item Repeat this process for all remaining diagonal matrices. 
\end{enumerate}

\begin{figure*}[!t]
  \centering
  \includegraphics[trim=38pt 50pt 8.5cm 50pt, clip=true, width=\textwidth]{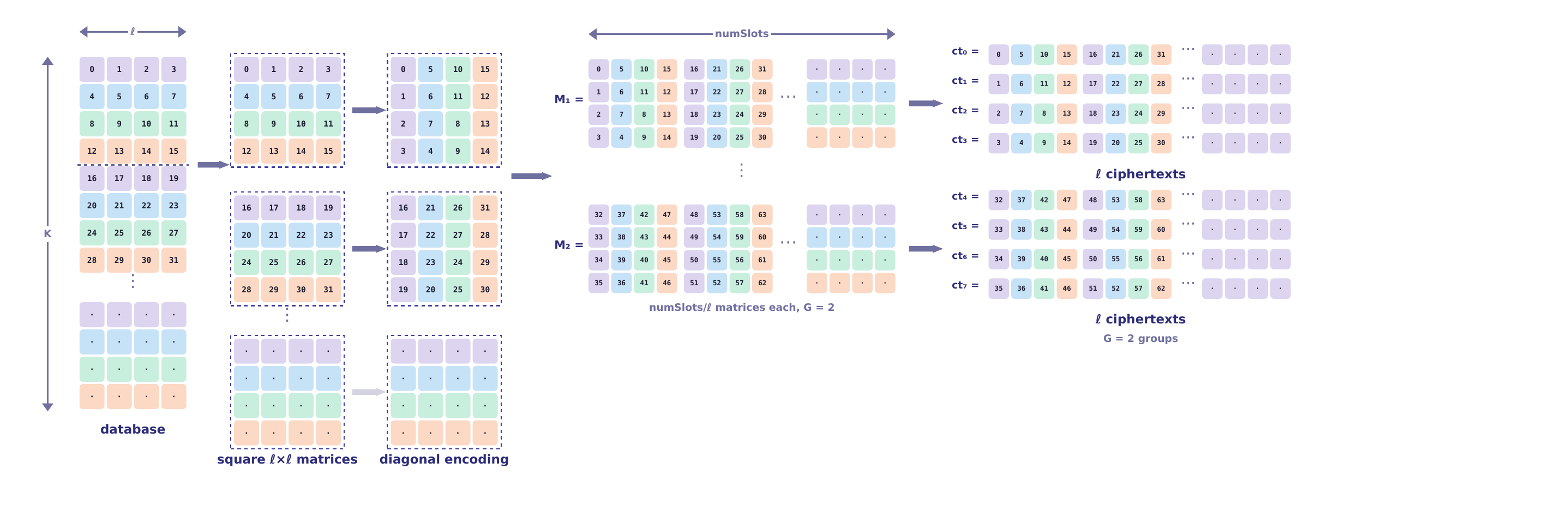}
  \vspace{-12px}
  \caption{Pipeline showing the diagonal packing from HyDia generalized to multiple groups. When $K > \numslots$, more than one \groupmatk matrix is produced (in this example, we have $\groupmat_1$ and $\groupmat_2$) to fit all the diagonal matrices. This results in $G \cdot \ell$ ciphertexts (in this figure, $\ell = 4$ and $G =2$).}
\label{fig:packing}
\end{figure*}

Figure~\ref{fig:packing} illustrates this process.


\subsubsection{Packed matrices and batched multiplications}
In the original diagonal method, the product $\groupmat \cdot \vec{v}$ is defined for a single square matrix $\groupmat \in \mathbb{R}^{\ell \times \ell}$ and a vector $\vec{v} \in \mathbb{R}^{\ell}$.
In contrast, following HyDia, we leverage the SIMD semantics of CKKS to batch multiple matrix--vector products. 
As explained in Section \ref{diag-ct-packing}, we pack $\numslots / \ell$ square matrices side-by-side into one enlarged representation.
The query ciphertext follows a similar structure, carrying $\numslots / \ell$ replicated copies of the same $\ell$-dimensional query vector, aligned with the packed matrix blocks.
The computation follows the diagonal method using component-wise (Hadamard) multiplication with rotations and slot-wise accumulations. 
Because all $\numslots / \ell$ matrix blocks and replicated query vectors are packed via SIMD, the evaluation performs $\numslots / \ell$ independent matrix--vector products in parallel. 
Hence, $\numslots / \ell$ matrices are processed at essentially the cost of one matrix--vector multiplication.

\subsection{Textbook Baby-step Giant-Step}\label{prelim:bsgs}

The Baby-Step Giant-Step (BSGS) optimization reduces the number of ciphertext rotations needed in a homomorphic matrix-vector product by grouping the diagonals $\text{diag}_i$ of \groupmat
into $n_2$ groups of size $n_1$, such that $\ell = n_1 \cdot n_2$. 
By writing a diagonal index as $\ j n_1 + i$, with $0 \leq j < n_2$ and $0 \leq i < n_1$, the matrix-vector product given in Equation~\ref{equ:diag} can be re-indexed as
 \begin{equation}
 \label{equ:diagbsgs}
 \groupmat \cdot \vec{v} = \sum_{j=0}^{n_2-1}\sum_{i=0}^{n_1-1} \text{Rot}_{jn_1 + i}(\vec{v}) \odot \text{diag}_{jn_1+i}(\groupmat)
 \end{equation}

Note that Equation~\ref{equ:diagbsgs} still represents the fully unrolled diagonal method and, if evaluated directly, requires $\ell-1$ ciphertext rotations. 
The actual BSGS gain appears after grouping the terms and using the following rotation identities: $\text{Rot}_{k + w}(\vec{a}) = \text{Rot}_{k}(\text{Rot}_{w}(\vec{a}))$, $\text{Rot}_k(\vec{a} \odot \vec{b}) = \text{Rot}_k(\vec{a}) \odot \text{Rot}_k(\vec{b})$, and $\text{Rot}_k(\vec{a}) \odot \vec{b} = \text{Rot}_k(\vec{a} \odot \text{Rot}_{-k}(\vec{b}))$.
Factoring out $\text{Rot}_{jn_1}(\cdot)$ from Equation \ref{equ:diagbsgs} gives
\begin{equation}
\label{equ:diagbsgs-grouped}
\begin{aligned}
\groupmat \cdot \vec{v} &= \sum_{j=0}^{n_2-1} \text{Rot}_{jn_1} \left( \sum_{i=0}^{n_1 -1} \text{Rot}_{i}(\vec{v}) \odot \text{Rot}_{-jn_1} \left( \text{diag}_{jn_1+i}(\groupmat) \right) \right) \\
&= \sum_{j=0}^{n_2-1} \text{Rot}_{jn_1} \left( P_j \right) 
\end{aligned}
\end{equation}
where the inner sum of products is defined as:
\[
P_j := \sum_{i=0}^{n_1 -1} \text{Rot}_{i}(\vec{v}) \odot \text{Rot}_{-jn_1} \left( \text{diag}_{jn_1+i}(\groupmat) \right).
\]

Equation \ref{equ:diagbsgs} can thus be rewritten as the following \textit{grouped} matrix-vector multiplication:
\begin{equation}\label{eq:bsgs2}
\groupmat \cdot \vec{v} = \sum_{j=0}^{n_2-1}\text{Rot}_{jn_1}(P_j).
\end{equation}
Equation \ref{eq:bsgs2} matches what was originally introduced in~\cite[Section 6.3]{halevi2018faster}.

A key observation is that the baby-step rotations
$\mathrm{Rot}_{i}(\vec{v})$, for $0 \leq i < n_1$, are shared across all groups
$P_j$ and therefore can be computed once and reused. 
As a result, the number of distinct ciphertext rotations drops from about $\ell-1$ to $(n_1-1) + (n_2-1)$, namely the baby steps $1,\dots,n_1-1$ and the giant steps
$n_1,2n_1,\dots,(n_2-1)n_1$. 
This quantity is minimized when $n_1 \approx n_2 \approx \sqrt{\ell}$, yielding approximately $2\sqrt{\ell}-2$ distinct rotations. 
Thus, the main advantages of BSGS is that it lowers the rotation cost from $\mathcal{O}(\ell)$ to $\mathcal{O}\left(\sqrt{\ell} \right)$, together with a corresponding reduction in the number of rotation keys and the memory needed to store them.
In the standard literature, this grouped form is especially useful when the
matrix is in \textit{plaintext} form.
In that case, the rotated diagonals $\mathrm{Rot}_{-jn_1}(\mathrm{diag}_{jn_1+i}(\groupmat))$ can be preprocessed offline before the online homomorphic evaluation. 
Equivalently, one may view BSGS as using a table of baby-step rotations
\[
\mathrm{TB}[i] \leftarrow \mathrm{Rot}_{i}(\vec{v}), \qquad 0 \leq i < n_1,
\]
which is computed once and reused across all giant-step groups. 
In the case of rectangular matrices, where Equation~\ref{eq:bsgs2} is evaluated for many
matrices \groupmatk, the same baby-step table can be reused across all of them.

The number of homomorphic multiplications remains unchanged from the diagonal method, since each of the $\ell$ diagonals still contributes one product. 
Similarly, the number of homomorphic additions remains $\ell-1$ because one must still combine all diagonal contributions. 
Thus, BSGS does not improve the asymptotic multiplication complexity, which remains $\mathcal{O}(\ell)$.
Nevertheless, it does reduce the number of expensive ciphertext rotations and the associated key-storage cost.

\subsubsection{Plaintext vs encrypted matrices}
\label{sec:prelim:bsgs:plainVSencM}
Most of the standard BSGS literature, including the grouped formulation
given above, considers the setting where the matrix is in plaintext and only the input vector is encrypted. 
In that setup, the matrix diagonals are known in advance and can be pre-rotated, pre-encoded, and reused across many evaluations. 
The online computation therefore consists of ciphertext rotations applied to the encrypted query vector, followed by plaintext--ciphertext Hadamard products.

Conversely, the setting considered in this paper is different as the matrix itself is encrypted. 
Consequently, the usual plaintext-side preprocessing of diagonals is no longer available, and the online products are no longer plaintext--ciphertext but instead ciphertext--ciphertext.
Therefore, the textbook grouped BSGS derivation above should be viewed as the reference BSGS construction from the standard literature, rather than as something that transfers directly to our setting. 
We define our own variant of grouped BSGS in Section~\ref{sec:bsgs-precomp}, which we refer to as BSGS-Diagonal.

\subsubsection{Hoisting and double hoisting for BSGS}
\label{sec:prelim:bsgs:hoisting}
Recall that, in Ring Learning With Errors (RLWE)-based schemes such as CKKS, a slot rotation by index $k$ is implemented using a Galois automorphism $\sigma_k$ on the ciphertext components, followed by a key-switching operation that maps the secret key from $\sigma_k(\mathbf{sk})$ back to $\mathbf{sk}$.

A standard key-switch can be decomposed into two stages. 
The first stage is a ciphertext-dependent gadget decomposition $G_B(\mathbf{c})$ of the ciphertext $\mathbf{c}$ into $\lceil \log_B(q)\rceil$ digits, where $B$ is the gadget base and $q$ is the current ciphertext modulus.
The second stage is a ciphertext-independent accumulation that multiplies those digits by the corresponding rotation key and sums the result.
%
Let $\mathcal{C}_1$ denote the cost of the first stage and $\mathcal{C}_2$ the cost of the second, where $\mathcal{C}_1$ is usually significantly higher than $\mathcal{C}_2$ in practice.

Hoisting applies whenever the same ciphertext is rotated by several indices $j \in \mathcal{J}$.
In this case, instead of recomputing the gadget decomposition for every rotation, one computes $G_B(\mathbf{c})$ once and reuses it, reducing the cost
from
$|\mathcal{J}|(\mathcal{C}_1+\mathcal{C}_2)$
to $\mathcal{C}_1 + |\mathcal{J}|\mathcal{C}_2.$
In grouped BSGS, this applies directly to the baby-step rotations
$\{\mathrm{Rot}_i(\mathbf{c})\}_{i=0}^{n_1-1}$, since they all rotate the same
input ciphertext $\mathbf{c}$. 
Hence, textbook BSGS is naturally \emph{single-hoistable}.

Double hoisting, introduced by Bossuat et al.~\cite{10.1007/978-3-030-77870-5_21}, is a further optimization on top of textbook BSGS. 
In the standard plaintext-matrix setting, it extends hoisting beyond the baby-step reuse and reduces the cost associated with the giant-step side of the computation.


In an encrypted-matrix setting, single hoisting still applies whenever the same ciphertext is rotated several times, since this only depends on reusing the ciphertext-dependent decomposition across multiple rotations.
However, the standard double-hoisting optimization from the plaintext-matrix literature does not necessarily carry over.
The reason is that the grouped products now involve encrypted matrix data rather than preprocessed plaintext diagonals, so the usual giant-step evaluation schedule and reuse pattern no longer apply. 
Any additional giant-step reuse beyond this would need to be specific to the chosen packing and evaluation order, rather than being directly taken from the textbook double-hoisted BSGS method for plaintext matrices.
In our work, we propose a BSGS-style algorithm that supports double-hoisting in Section~\ref{sec:bsgs-precomp}.

\subsection{HyDia~\cite{martin2025hydia}}\label{sec:hydia-intro}
As a starting point to our work, we consider the HyDia protocol proposed in~\cite{martin2025hydia}, as it already demonstrated improved performance over prior works.
We refer to the actual paper for its technical details and hereby simply recall the main ideas behind HyDia, as that work serves as our main comparison point.

HyDia is a three-party protocol for one-to-many ($1:K$) facial matching where both the query face embedding and the database embeddings are encrypted end-to-end using CKKS.
The scheme supports both membership verification and identification use cases. 
The overall system model can be described as follows. 
An enroller performs a one-time setup: it collects face images, runs a feature extractor to produce database template vectors (face embeddings), preprocesses them, encrypts them under a public key, and uploads the resulting template ciphertexts to the server's database.
The server stores the encrypted data and homomorphically evaluates client queries over it. 
The client captures a face image, derives a query template vector, preprocesses and encrypts it to form a query ciphertext, and asks whether the query matches any enrolled subject.
The overall description of the protocol and the entities involved are illustrated in Figure~\ref{fig:hydia}.
\begin{figure}[h]
\centering
\resizebox{\columnwidth}{!}{%
\begin{tikzpicture}[
  font=\small,
  box/.style={draw, rounded corners, thick, fill=gray!10, align=center,
              minimum width=28mm, minimum height=10mm},
  arr/.style={->, >=stealth, thick} 
  ]

\node[box] (C) at (-3.2,  1.8) {\textbf{Client}:\\
takes a picture\\
extracts features from facial image \\ \faLock\ encrypts facial query embedding vector\\ decrypts result};
\node[box] (E) at ( 3.2,  1.8) {\textbf{Enroller}: \\
holds many facial images (database DB) \\
preprocesses facial embedding vectors \\
\faLock\ encrypts DB facial embedding vectors};
\node[box] (S) at ( 0.0, -1.4) {\textbf{Server}:\\
\faLock\ computes homomorphic matrix-vector multiplication \\
\faLock\ computes homomorphic comparison with threshold};

\draw[arr] (C) to[bend left=10]
  node[midway, right, align=center] {\faLock\ query} (S);

\draw[arr] (S) to[bend left=10]
  node[midway, below left, align=center] {\faLock\ result} (C);

  \draw[arr] (E) to[bend left=10]
  node[midway, left, align=center] {\faLock\ DB} (S);

\end{tikzpicture}%
}
\caption{Illustration of the HyDia protocol (and ours).}
\label{fig:hydia}
\end{figure}


One important characteristic of HyDia is that its design avoids revealing per-entry similarity scores.
This prevents a semi-honest client from learning additional information about the database from their queries.
This is accomplished by returning only a thresholded match decision or matching indices.
In addition, HyDia outperforms methods proposed in prior works~\cite{ibarrondo2023grote,choi2024blind,engelsma2022hers} by adapting diagonalized matrix multiplication to the ${1:K}$ scenario, thus reducing expensive HE overhead and enabling higher parallelism.
Moreover, the protocol supports single ciphertext queries (instead of sending hundreds of ciphertexts per query as in prior work), which significantly reduces communication and end-to-end latency. 
Finally, it uses a hybrid polynomial thresholding approximation that combines a lower depth Chebyshev approximation with a fixed polynomial degree (from Cheon et al.~\cite{cheon2020efficient}), improving accuracy near the decision boundary without increasing the multiplicative depth budget. 
\subsection{Security and privacy assumptions}

\paragraph{Entities and threat model}

We closely follow the threat model of~\cite{martin2025hydia}. The system involves three roles: a \textbf{Client} (\textbf{C}) who submits encrypted queries and holds decryption keys; a \textbf{Server} (\textbf{S}) who stores the encrypted database and performs homomorphic computations; and an \textbf{Enroller} (\textbf{E}) who collects plaintext samples, encrypts them, and uploads them to the server ensuring only ciphertexts are stored, in compliance with regulations such as GDPR and CCPA. Both \textbf{C} and \textbf{S} are considered semi-honest: they follow the protocol but may attempt to learn additional information. \textbf{E} is fully trusted and assumed not to collude with \textbf{S} or \textbf{C}~\cite{martin2025hydia}.
\paragraph{Privacy goals}
Under this semi-honest setting, the system must preserve: \emph{query privacy} (embeddings and biometric templates remain secret), \emph{database privacy} (stored templates are not leaked), \emph{response privacy} (similarity scores or indices reveal no more than the match result), and \emph{key confidentiality} (only the client holds decryption keys).
\paragraph{Security}
The security of our protocol relies on the CKKS scheme, whose hardness is based on the Ring Learning with Errors problem, yielding post-quantum security with appropriate parameter choices.

\paragraph{Model limitations}
Since homomorphic multiplication requires operands encrypted under a common key, database encryption uses the client's public key. Database confidentiality towards the client is therefore enforced at the protocol level, and would not survive a client--server collusion.
A multi-client deployment would also require either a shared decryption key or a per-client encrypted copy of the database. Separately, a malicious client could obtain thresholded predicates on individual embedding coordinates submitting an unnormalized query and, via repeated queries, recover information about database entries. Query confidentiality holds even against a malicious server, which never obtains decryption capability. Result integrity, however, does not: clients cannot detect incorrect evaluations if a malicious server returns an arbitrary ciphertext.

\paragraph{Possible extensions}
Even though a malicious server setting is out of scope for this work, integrity in this scenario can be addressed by verifiable computations over FHE.
Those could employ homomorphic MACs and authenticators or SNARK-based proofs of correct CKKS evaluation, although current constructions impose significant overheads.
Furthermore, the key-ownership limitation mentioned above may be addressed by threshold or multi-key FHE, under which database and query are encrypted to a joint key and decryption requires cooperation among parties, removing the client's unilateral ability to decrypt database ciphertexts. 
We leave this prominent multi-client direction for future work.

\subsection{Summary of parameters involved}

We summarize the main parameters used in our algorithms and pseudo-codes in Table~\ref{tab:variables-enroller-sender}.

\begin{table}[t]
\centering
\scriptsize
\setlength{\tabcolsep}{3pt}
\renewcommand{\arraystretch}{1.1}

\begin{tabularx}{\columnwidth}{@{}p{0.26\columnwidth}X@{}}
\hline
\multicolumn{2}{@{}l}{\textbf{Layout parameters}}\\
\hline
\numvectors & Total number of database vectors. \\
\vecdim     & Vector dimension of each database/query vector. \\
\numslots   & CKKS batch size (number of SIMD slots). \\
$\matrixgroups=\left\lceil \frac{\code{\numvectors}}{\numslots} \right\rceil$ &
Number of groups (each containing up to $\lceil \numslots / \ell \rceil$ square diagonal matrices). \\
\hline
\multicolumn{2}{@{}l}{\textbf{BSGS parameters}}\\
\hline
\numbaby & Baby-step size (fixed by $\ell$; independent of Server GPU hardware) \\
$\texttt{n}_2$ & Giant-step size ($\texttt{n}_2 = \lceil \ell/\texttt{n}_1 \rceil$). \\
\hline
\multicolumn{2}{@{}l}{\textbf{Similarity computation parameters}}\\
\hline
\groupmatk & Concatenation of $\numslots / \ell$ square diagonal matrices in encrypted form  \\
$\vec{q}$ & Encrypted query vector (single ciphertext). \\
 $\mathsf{diag}_{i}(\groupmatk)$ & $i^{th}$ diagonal ciphertext of  matrix \groupmatk \\

$\compdepth$ & Comparative depth budget\\
\hline
\multicolumn{2}{@{}l}{\textbf{GPU parameters}}\\
\hline
$\mathsf{diag}'[k]$ & Pre-rotated diagonal ciphertext, written as $\mathsf{diag}'[k] \;=\; \textsf{Rot}_{{-}\lfloor k/n_1 \rfloor \cdot n_1}\!\bigl(\mathsf{diag}[\groupmatk]\bigr).$ \\
$n$ & Chebyshev poly degree (e.g.,\ $n{=}13$ for $\kappa{=}8$). \\
$S$ & CUDA stream pool size (Equation~\ref{eq:stream-pool}). \\
$\text{SMs}$ & GPU streaming multiprocessor count. \\
$F$ & Free GPU memory (bytes). \\
$C$ & Estimated per-ciphertext size on GPU. \\
\hline
\end{tabularx}

\caption{Parameters and variables used in the diagonal BSGS enrollment and query processing.
}
\label{tab:variables-enroller-sender}
\end{table}
\section{Improved HyDia with BSGS Key Reduction}\label{sec:bsgs}
For improving HyDia’s diagonal encoding method~\cite{martin2025hydia}, we integrate a BSGS algorithm to reduce the number of rotation keys generated and transmitted by the client and processed by the server.
This greatly lowers RAM requirements while improving performance. 
This improvement specifically targets the homomorphic similarity computation on the server side for both membership and identification scenarios.



\subsection{BSGS with precomputation across groups}
\label{sec:bsgs-precomp}

We now present an optimized BSGS-style algorithm for our encrypted matrix setting, called BSGS-Diagonal. 
Unlike the textbook BSGS method, 
our scenario depends on encrypted matrix groups \groupmatk. 
The main idea of BSGS-Diagonal is therefore different: instead of relying on the usual plaintext-side giant-step reuse, we precompute the required rotations of the encrypted query once and reuse them across all matrix groups.
When the number of database vectors $K$ exceeds the packing capacity, the database is partitioned into multiple encrypted matrix groups \groupmatk, for $0 \le k < G$ (Section~\ref{diag-ct-packing}).
The server must then evaluate $G$ matrix-vector products $\groupmatk \cdot \vec{q}$, where $\vec{q}$ is the query ciphertext produced by the client. It follows the packing from HyDia (Section 5.2), where $\numslots/\ell$ copies of the plaintext query vector are concatenated and encrypted. 
In the standard grouped BSGS formulation, each group would be evaluated using grouped baby-step and giant-step rotations.
However, since the same query ciphertext $\vec{q}$ is used for all groups, the full set of query rotations needed across all \groupmatk can instead be generated
once and reused throughout the entire computation.

More precisely, rather than evaluating each group through the grouped form of Equation~\ref{eq:bsgs2}, we use the unrolled diagonal expression
\begin{equation}
	\label{equ:diagbsgsgroupsprecomp}
	\groupmatk \cdot \vec{q}
	=
	\sum_{i=0}^{\ell-1}
	\mathsf{rotQ}[i]
	\odot
	\mathrm{diag}_{i}(\groupmatk),
\end{equation}
for $0 \le k < G$, where
$\mathsf{rotQ}[i]
\leftarrow
\mathrm{Rot}_{r}(\vec{q}).$
Thus, $\mathsf{rotQ}$ is a table containing all $\ell$ rotated versions of the query ciphertext required by the diagonal decomposition. 
This table can be computed a single time and then reused across all $G$ encrypted matrix groups.

This changes the role of BSGS in our setting. 
In the plaintext-matrix literature, BSGS reduces the number of distinct online rotation offsets by
grouping terms into baby steps and giant steps.
Conversely, in our encrypted-matrix case the main benefit comes from amortizing the cost of query rotations across many groups. 
Once the table $\mathsf{rotQ}$ has been generated, no further query rotations are needed while processing the remaining matrix groups. 
Hence, the rotation cost associated with the query no longer scales linearly with the number of groups $G$, but is paid once and then reused.

The precomputation of $\mathsf{rotQ}$ can itself exploit hoisting, since all required rotations are applied to the same input ciphertext $\vec{q}$.
Therefore, the ciphertext-dependent decomposition of $\vec{q}$ can be shared across the rotations used to build the table. 
We emphasize, however, that this is different from the textbook double-hoisted BSGS method for plaintext matrices.
Specifically, since the matrices \groupmatk are encrypted, the usual plaintext-diagonal preprocessing and giant-step reuse pattern from the standard double-hoisting literature does not apply. 
We can thus view BSGS-Diagonal as a BSGS-inspired precomputation method for encrypted matrices, rather than as a direct instance of the standard double-hoisted BSGS algorithm.

The tradeoff is that BSGS-Diagonal stores more rotated query ciphertexts
simultaneously in memory, since the table $\mathsf{rotQ}$ contains $\ell$ ciphertexts. 
As a compensation, it removes the need to recompute or re-rotate the query for every group, which yields a substantial reduction in total runtime when the number of groups $G$ is large.
In addition, the relinearizations of the accumulated sums of ciphertexts can be postponed to the very end when evaluating Equation \ref{equ:diagbsgsgroupsprecomp}, as no intermediate rotation is applied in the process.
Hence, a single relinearization is applied per matrix group. 
In comparison to a textbook BSGS approach, $n_2$ relinearizations have to be performed per group (Equation \ref{eq:bsgs2}), since, before each giant rotation by $jn_1$, it is recommended to relinearize the input ciphertext to avoid memory explosion issues\footnote{OpenFHE, for example, does not implement the rotation operation on quadratic or higher degree ciphertext due to this memory issue. \url{https://openfhe.discourse.group/t/which-operations-can-be-followed-multiplication-w-o-relinearization/1148}}. 
Moreover, to mitigate the additional memory overhead, we also introduce a memory-aware multi-threaded processing strategy that limits the number of temporary ciphertexts kept live during evaluation.

\textbf{Memory-Efficient Accumulation.}
When generating the partial products in Equation \ref{equ:diagbsgsgroupsprecomp} in a multi-threaded implementation, we employ an \emph{incremental accumulation} method for reducing RAM memory usage. 
More precisely, for each matrix group we compute all $\ell$ ciphertext products and store them before summing:
\begin{equation*}
    \texttt{products}[i] \leftarrow \mathsf{rotQ}[i] \odot \text{diag}_i(\groupmatk). 
\end{equation*}
This requires concurrent storage of \(O(\ell)\) ciphertexts; for $\ell = 512$ and $\approx 8\,$MB ciphertexts (in OpenFHE), this amounts to \(\approx 4\,\)GB per matrix group.

On a multi-core processor, BSGS-Diagonal can be adapted to employ per-thread running accumulators:
\begin{equation*}
    \texttt{acc}_t \mathrel{+}= \mathsf{rotQ}[i] \odot \text{diag}_i(\groupmatk),
\end{equation*}
where products are added to thread-local accumulators immediately after computation and then freed.
With $T$ threads, memory consumption drops from $O(\ell)$ to $O(T)$ ciphertexts, reducing peak RAM from $\approx 4\,$GB to $\approx T \times 8\,$MB per matrix group. 
The full version of BSGS-Diagonal with multi-threading is illustrated by Algorithm \ref{alg:bsgs-precomp}.

\begin{algorithm}[h]
\caption{BSGS-Diagonal: BSGS variant with Incremental Accumulation}
\label{alg:bsgs-precomp}
\begin{algorithmic}[1]
\footnotesize
\Require Query ciphertext $\vec{q}$, database diagonal ciphertexts
\Ensure Similarity scores for all database vectors
\Statex

\State $\triangleright$ \textbf{Step 1: Baby-step rotations with hoisting}
\State $\mathsf{queryPre} \gets \textsf{rotPrecompute}(\vec{q})$
\For{$b = 0$ \textbf{to} $n_1-1$ \textbf{in parallel}}
    \State $\mathsf{rotQ}[b] \gets \textsf{hoistedRot}(\vec{q}, b, \mathsf{queryPre})$
\EndFor
\Statex

\State $\triangleright$ \textbf{Step 2: Double hoisting - precompute for giant steps}
\For{$b = 0$ \textbf{to} $n_1-1$ \textbf{in parallel}}
    \State $\mathsf{babyPre}[b] \gets \texttt{rotPrecompute}(\mathsf{rotQ}[b])$
\EndFor
\Statex

\State $\triangleright$ \text{Step 3: Compute all rotations upfront}
\For{$i = n_1$ \textbf{to} $\ell-1$ \textbf{in parallel}}
    \State $(b_1, b_2) \gets (i / n_1, i \mod n_1)$
    \State $\mathsf{rotQ}[i] \gets \textsf{hoistedRot}(\mathsf{rotQ}[b_2], b_1 \cdot n_1, \mathsf{babyPre}[b_2])$
\EndFor
\Statex

\State $\triangleright$ \textbf{Step 4: Process groups with incremental accumulation}
\For{each matrix group $\groupmatk$}
    \State Initialize per-thread accumulators $\mathsf{acc}_t \gets \mathbf{0}$
    \For{$i = 0$ \textbf{to} $\ell-1$ \textbf{in parallel}}
        \State $\mathsf{prod} \gets \mathsf{rotQ}[i] \odot \text{diag}_{i}$(\groupmatk) \Comment{No relinearization}
        \State $\mathsf{acc}_{t} \gets \mathsf{acc}_{t} + \mathsf{prod}$ \Comment{Free product immediately}
    \EndFor
    \State $\mathsf{result}[k] \gets \textsf{TreeReduce}(\mathsf{acc}_0, \ldots, \mathsf{acc}_{T-1})$
    \State $\mathsf{result}[k] \gets \textsf{Relinearize}(\mathsf{result}[k])$
\EndFor
\Statex

\State \Return $\mathsf{result}$
\end{algorithmic}
\end{algorithm}

As our optimizations focus on the computation of the homomorphic inner product itself, the proposed Algorithm \ref{alg:bsgs-precomp} can easily substitute the diagonal method used by the server in Hydia. 
The membership and index scenarios remain the same as in the original HyDia -- for better self-containing, we recall their pseudo-code in \ref{appendix:pseudo_code}, Algorithms~\ref{alg:membership} and~\ref{alg:index}.

\subsection{Complexity analysis and memory consumption}

\subsubsection{Complexity analysis}

We summarize the algorithmic complexity of HyDia and BSGS-Diagonal for server-side encrypted similarity computation, referring to~\cite[Appendix E]{martin2025hydia} for details on HyDia. 
Let $G$ be the number of groups, $K$ the number of database vectors, and $\ell$ the embedding dimension. Both algorithms perform $G \cdot (\ell-1)$ ciphertext--ciphertext additions, $G \cdot \ell$ ciphertext--ciphertext multiplications, $G$ relinearizations and rescalings, and $\ell-1$ hoisted rotations. 
Nevertheless, BSGS-Diagonal reduces the number of rotations by a factor $G$ compared to textbook BSGS (Equation~\ref{equ:diagbsgs}) and saves significant memory compared to HyDia's diagonal method.

There is a tradeoff between BSGS-Diagonal and a single-hoisted grouped BSGS strategy, depending on parameter $G$. The total nontrivial query rotations are $(n_1-1) + G(n_2-1)$ for single-hoisted grouped BSGS, versus $n_1 n_2 - 1$ for BSGS-Diagonal; single-hoisted grouped BSGS is preferable for small $G$ (fewer rotations and live ciphertexts), while BSGS-Diagonal wins for large $G$ by amortizing the query-rotation cost paid once upfront. The memory cost reflects this: single-hoisted grouped BSGS stores about $n_1$ rotated query ciphertexts, whereas BSGS-Diagonal stores the full table of $\ell = n_1 n_2$ (approximately $\ell - n_1$ additional ciphertexts). In our experiments, single-hoisted grouped BSGS is preferable for database sizes up to $2^{18}$, while BSGS-Diagonal is more time-efficient at $2^{19}$ and above, reducing the query-rotation count by roughly $3\times$ for $\approx 10^6$ entries. Table~\ref{tab:bsgs-diagonal-comparison} in~\ref{appendix:pseudo_code} summarizes this tradeoff for our CKKS parameters.

\subsubsection{Lowering RAM consumption}
\paragraph{Client-side} HyDia requires $\ell-1$ rotation keys $\{1, 2, \ldots, \ell-1\}$ for the within-block diagonal matrix--vector multiplication. BSGS-Diagonal instead requires only keys for the baby and giant steps, $S_{\mathrm{baby}} = \{1, 2, \ldots, n_1-1\}$ and $S_{\mathrm{giant}} = \{jn_1 : 1 \le j < n_2\}$. Section~\ref{sec:performance} shows that, under the same parameter settings, this significantly reduces the number of client-generated rotation keys.

\subsection{Error growth and scaling behavior}\label{sec:noise}
 
Recall that the database is partitioned into $\matrixgroups=\left\lceil \code{\numvectors} / \numslots \right\rceil$ groups, each evaluated through an identical fixed-depth circuit. 
Within a group, each output slot is the inner product of two $l_2$-normalized vectors of dimension $\ell$, obtained by accumulating $\ell$ ciphertext-ciphertext products (see Equation~\ref{equ:diagbsgsgroupsprecomp}). 
Let the decrypted slot values of the rotated query and of the database diagonals be $q_i + \eta_{q,i}$ and $v_i + \eta_{v,i}$, respectively.
The first-order error of the accumulated sum is
$$\sum_i ( v_i \cdot \eta_{q,i} + q_i \cdot \eta_{v,i} ),$$
where $q_i$ and $v_i$ are the $i$-th components of the plaintext normalized query and database embeddings, and $\eta_{q,i}$ and $\eta_{v,i}$ are the CKKS errors carried by the slots encoding them, with standard deviations $\sigma_q$ and $\sigma_v$ respectively. 
The variance of this error is then $\sigma_q^2 \cdot ||v||^2 + \sigma_v^2 \cdot ||q||^2 = \sigma_q^2 + \sigma_v^2,$ since $||q|| = ||v|| = 1$. 
This already captures the amplification of the input encryption errors through the $\ell$ multiply-accumulate operations: because the embeddings are $\ell_2$-normalized, the $\sqrt{\ell}$ growth from accumulating $\ell$ terms is exactly offset by the $O(1/\sqrt{\ell})$ magnitude of the components, leaving $err_{sim}$ independent of $\ell$. 
Including the single deferred relinearization and the rescaling applied once per matrix group (see Algorithm~\ref{alg:bsgs-precomp}, lines 22-23), the similarity error is therefore
$$err_{sim} \approx \sqrt{\sigma_q^2 + \sigma_v^2} + err_{relin} + err_{rs}$$
In this approximation: $\sigma_q$ is the per-slot error magnitude of the rotated query ciphertexts rotQ[i], \emph{i.e.,} the fresh encryption error of the query and the key-switching error of the hoisted rotation that produces rotQ[i];
$\sigma_v$ is the per-slot error magnitude of the encrypted database diagonals $\text{diag}_i(\groupmatk)$, \emph{i.e.,} the fresh encryption error introduced by the enroller;
$err_{relin}$ is the error introduced by the single relinearization applied to the accumulated products of each group; and $err_{rs}$ is the error introduced by the subsequent rescaling.
Various conclusions can be made. First, in the identification scenario, the noise growth
is independent of $\ell$ or $K$, \emph{i.e.,} the per-slot precision is not affected by the database size. 
The database size affects the identification pipeline only through the number of groups $G$, as each group is evaluated by the same circuit on distinct ciphertexts and no operation aggregates across groups before the decision.
Second, the comparison stage preserves this property: the threshold comparison amplifies $err_{sim}$ by a fixed factor determined by the comparison depth $\kappa$, independent of $K$. The error at each output slot is therefore a function of the CKKS parameters only, while enrolling additional identities increases the number of groups without changing it, so the precision of a score at $K = 2^{20}$ and $K = 2^{10}$ are identical. 
The situation is different in the membership scenario. Indeed, in this case the $G$ comparison outputs are aggregated before the result is returned, so $K$ matters for the error. 
More precisely, this scenario sums the $G$ comparison ciphertexts and their errors are independent, so the per-slot variance after this stage is $G \cdot err_c^2$, where $err_c$ is the per-slot error of a single comparison output, including the CKKS noise of $err_{sim}$ amplified through the depth-$\kappa$ Chebyshev evaluation and the approximation residual $\delta_i = p(s_i) - \mathbf{1}[s_i > \tau]$ of the comparison polynomial $p$. 
The membership scenario then reduces all slots to the first ($s_0$) via $\log_2(\numslots)$ rotate-and-add steps, each doubling the variance already accumulated, and injects one key-switching term $err_{ks}^2$, where $err_{ks}$ is the error introduced by the key-switching of a single rotation, independent of $K$ and of the message. 
Hence, we get the relation $\mathsf{var}_t = 2 \cdot \mathsf{var}_{t-1} + err_{ks}^2$, $\mathsf{var}_0 = G \cdot err_c^2$,
unrolled as $\mathsf{var} = 2^{\log_2(\numslots)} \cdot \mathsf{var}_0 + (2^{\log_2(\numslots)} - 1) \cdot err_{ks}^2$. The output error is, thus,
$$err_{out} = \sqrt{G \cdot \numslots \cdot err_c^2 + (\numslots - 1) \cdot err_{ks}^2}.$$
The first term grows with the database size through $G$. The second is a key-switching term proportional to $\numslots$ instead of $G$ or $K$, because the same $\log_2(\numslots)$ rotations are performed regardless of database size, while the summation over groups involves $G-1$ additions and no rotations. 
%
Hence, only the first term grows with the database, and it does so as $\sqrt{G \cdot \numslots} \approx \sqrt{K}$. Indeed, the variance accumulates linearly in the number of slots aggregated: each of the $G \cdot \numslots$ slots contributes one independent $err_c^2$, so the variance is proportional to $G \cdot \numslots$ and the error magnitude, being its square root, grows as $\sqrt{G \cdot \numslots} \approx \sqrt{K}$.
We emphasize that in both scenarios the multiplicative depth is fixed and independent of $K$, meaning that database scaling consumes no additional modulus chain levels.
Its only effect is a growth in the aggregated error magnitude, which manifests as a loss of precision (in bits) rather than as depth exhaustion. 
This analysis characterizes the growth of the error variance. For the membership decision, however, the important metric is the error mean, which we now analyze.

In a membership query, the server outputs a ciphertext whose slot $s_0$
ideally corresponds to an approximation of $2m$, where $m$ is the number of matches.
The client then checks if $s_0 \geq 1$ to decide whether there is a match or not.
Let FP and FN be the events when the result of a membership query is falsely
positive (when $m = 0$) and negative (when $m > 0$).
Let $E_\text{out}$ be the random variable representing the total error
after aggregating all $G\cdot\texttt{numSlots}$ entries, 
so $\text{Var}\pp{E_\text{out}} = err_\text{out}^2$.
Then, the probabilities of observing FP and FN results in membership queries 
are given by
\begin{align*}
\begin{cases}
  \Pr\pp{\text{FP}} = \Pr\pp{E_\text{out} \geq 1},  &\text{if $m = 0$},\\
  \Pr\pp{\text{FN}} = \Pr\pp{2m + E_\text{out} < 1}, &\text{if $m > 0$}.
\end{cases}
\end{align*}

$\Pr\pp{\text{FN}}$ decreases as $m$ gets larger,
since $2m + E_\text{out}$ tends to get farther from~$1$.
We can then consider the upper bound $\Pr\pp{\text{FN}} \leq \Pr\pp{2 + E_\text{out} < 1}$
for the analysis.
Remember that $E_\text{out}$ is dominated by the sum of 
$G \cdot \texttt{numSlots}$ terms corresponding to the homomorphic comparison error.
However, for large values of $K$, we know that
$G \cdot \texttt{numSlots} \approx K$, so we can appply the
Central Limit Theorem (CLT) to obtain the approximation
\[
  E_\text{out} \sim
    \mathcal{N}\pp{K \cdot \mathbb{E}\pp{E_c},
                   K \cdot \text{Var}\pp{E_c}},
\]
where $E_c$ is the random variable representing the homomorphic comparison error.
Therefore, both $\Pr\pp{\text{FP}}$ and $\Pr\pp{\text{FN}}$ can be
estimated by considering tail bounds for this normal distribution.
The aggregated error in large databases is dominated by its mean, which scales with $\eta$, while the standard deviation scales with $\sqrt{\eta}$. The per-comparison error $E_c$ carries a small but nonzero bias
$\mathbb{E}[E_c]$ that accumulates, whereas its variance
contributes only negligibly. The client's decision may then fail not because of the noise in a statistical sense, but due to this deterministic drift once
$\eta \cdot |\mathbb{E}[E_c]|$ becomes comparable to the threshold value, as experimentally shown in Section~\ref{sec:acc_synth}.


\section{GPU-accelerated encrypted search}\label{sec:gpu-bsgs}

Two operations bottleneck the CPU-based similarity computation described in Section~\ref{sec:bsgs}:
(i)~ciphertext rotations during the diagonal linear transform; and (ii)~ciphertext--ciphertext multiplications, relinearization, and rescaling during similarity computation and Chebyshev comparison. Profiling reveals that the homomorphic matrix--vector product dominates per-query time, while individual HE operations are limited by memory bandwidth rather than arithmetic throughput, making them ideal for GPU acceleration.

Naively offloading individual HE operations to the GPU, however, can be \emph{less} efficient than CPU-only execution, since converting each ciphertext between the host (OpenFHE~\cite{OpenFHE}) and device (FIDESlib~\cite{agullo2025fideslib}) data structures can take tens to hundreds of milliseconds, rivaling the cost of several GPU rotations. Our GPU design is therefore guided by a single principle: \emph{minimize host--device transfers by keeping the entire query processing pipeline on the GPU}, uploading only the baby-step query ciphertexts and downloading only the final comparison result. Therefore, this GPU-oriented BSGS strategy adapts, rather than copies, the CPU-optimized approach described in Algorithm~\ref{alg:bsgs-precomp}.

We apply the aforementioned principle to port two CPU variants to the GPU: \ApproachA, following HyDia's original diagonal method, and \ApproachB, using textbook BSGS. These variants occupy different points in a \emph{memory--flexibility tradeoff}: \ApproachA is algorithmically simpler but requires ${\sim}27$\,GB of peak GPU memory (dominated by rotation keys), while \ApproachB needs only ${\sim}9$\,GB for a database instance of $8{,}192$ vectors via on-demand BSGS rotations and a reduced key set. \ApproachB is preferred over BSGS-Diagonal (Section~\ref{sec:bsgs}) on GPU because the latter requires all $\ell$ query rotations resident in GPU memory, whereas \ApproachB keeps only $n_1$ of them---an $n_2$-factor saving. A further variant, \ApproachBprime (Section~\ref{sec:gpu-prerot-enroller}), shifts the diagonal pre-rotation to the enroller, eliminating negative rotation keys entirely.

Both GPU variants share a three-layer architecture: FIDESlib~\cite{agullo2025fideslib} provides CUDA-accelerated CKKS primitives (\texttt{mult}, \texttt{add}, \texttt{rotate}, \texttt{rescale}, \texttt{square}, etc.) that map one-to-one onto OpenFHE, a middleware layer orchestrates diagonal and rotation caching, Chebyshev evaluation, and aggregation, and variant-specific sender modules implement the full query pipelines. Following the \emph{upload-once, compute-many, download-once} pattern, similarity computation, Chebyshev comparison, and aggregation all execute on the GPU, with at most $n_1$ ciphertexts uploaded per query and one (membership) or $G$ (identification) ciphertexts downloaded. \ApproachB uploads $n_1$ baby-step ciphertexts rather than computing them on the GPU because that would require uploading $n_1{-}1$ baby-step rotation keys (${\sim}814$\,MB for $n_1 = 23$), negating its memory advantage. A side benefit of this design is reduced \emph{host} RAM: offloading cached diagonals and intermediates to the GPU consumes $24{-}53\%$ less host RAM than the CPU variants.

\subsection{Common GPU infrastructure}\label{sec:gpu-common}

Both GPU variants share four infrastructure components, described below.

\subsubsection{Offline caching of encrypted database and keys}
The encrypted database and all required rotation keys are uploaded to GPU memory \emph{once} during system setup and remain resident across queries. During the offline phase: (i)~all encrypted diagonal ciphertexts $\{\mathsf{diag}_i(\groupmatk)\}$ are deserialized from disk using OpenMP parallelism and uploaded to the GPU; (ii)~all rotation keys required by the variant are pre-initialized on the GPU, reducing first-query latency; (iii)~the CUDA stream pool is initialized (Section~\ref{sec:stream-pool}).

GPU memory is dominated by the $G \times \ell$ diagonal ciphertexts represented in FIDESlib (${\sim}5$\,MB each) and rotation keys (${\sim}37$\,MB each, including key-switching buffers). Each variant estimates its total footprint before loading and may fall back to CPU-only computation if GPU capacity is exceeded. The stream pool reserves additional memory for concurrent per-matrix intermediates to prevent runtime out-of-memory (OOM) failures.

\subsubsection{Multi-stream matrix parallelism}\label{sec:stream-pool}
The $G = \lceil K/\numslots \rceil$ matrix groups can be processed independently. Both variants allocate a pool of $S$ CUDA streams and dispatch matrices in waves of up to $S$ concurrent tasks, with pool size
\begin{equation}\label{eq:stream-pool}
    S = \min\Bigl(\bigl\lfloor \tfrac{\text{SMs}}{4} \bigr\rfloor,\;\; \bigl\lfloor \tfrac{0.2 \cdot F}{40 \cdot C} \bigr\rfloor,\;\; 32\Bigr), \quad S \ge 2,
\end{equation}
where $\text{SMs}$ is the GPU's streaming multiprocessor count, $F$ the free GPU memory, and $C$ the estimated ciphertext size. The first term allocates ${\sim}4$ SMs per stream for good occupancy without contention; the second prevents OOM from concurrent intermediates (each in-flight matrix uses ${\sim}40$ ciphertext-sized buffers for BSGS products and Chebyshev intermediates). The cap of $32$ averts oversubscription, beyond which we observed negligible throughput gains. All streams use \texttt{cudaStreamNonBlocking} to avoid implicit synchronization with the default stream.

\subsubsection{GPU-resident rotation caching}\label{sec:gpu-rotation-cache}
\ApproachA computes and caches all $\ell{-}1$ rotated query ciphertexts directly on the GPU via FIDESlib's \texttt{copy} and \texttt{rotate} at ${\sim}1.8$\,ms per hoisted rotation---roughly $1.7\times$ faster than CPU double hoisting (${\sim}3.1$\,ms). A naive variant that downloaded rotated ciphertexts for subsequent use was slower than CPU-only due to transfer overhead.

\ApproachB instead computes its $n_1{-}1$ baby-step rotations on the CPU and uploads the resulting ciphertexts, avoiding the ${\sim}1$\,GiB cost of uploading baby-step rotation keys and preserving VRAM for diagonals. Total latency is comparable to the GPU alternative while using $8\times$ less peak GPU memory for the baby-step phase ($115$\,MB vs.\ $929$\,MB). In both variants, cached baby-step ciphertexts are reused across all $G$ matrix groups.

\subsubsection{GPU-native Chebyshev comparison}\label{sec:gpu-chebyshev}
We implement the full pipeline, including Chebyshev evaluation, natively on the GPU. A hybrid design offloading only similarity to the GPU yields minimal gains because per-matrix similarity dominates query time. The optimized CPU ``f4'' path from~\cite[Section 5.4]{martin2025hydia} is tightly coupled with CPU-side memory locality; on GPU, a single Paterson--Stockmeyer path gives better end-to-end behavior by keeping intermediates device-resident and mapping to batched kernels.

Besides performance, returning only binary match/non-match indicators (rather than raw encrypted similarity scores) prevents database-reconstruction attacks that triangulate embeddings from repeated queries under our semi-honest client assumption. Given an encrypted score $\mathsf{ct}_s$ and threshold $\delta$, we compute
\begin{equation}\label{eq:cheb-sign}
    f(x) \;=\; \tfrac{1}{2}\bigl(\mathrm{sign}(x - \delta) + 1\bigr) \;\approx\; \begin{cases} 1 & x \ge \delta,\\ 0 & x < \delta, \end{cases}
\end{equation}
slot-wise. The sign function is approximated by $\mathrm{sign}_\delta(x) = \sum_{i=0}^{n} c_i\, T_i(x)$, where $T_i$ is the $i$-th Chebyshev polynomial of the first kind ($T_0=1$, $T_1=x$, $T_{k+1} = 2x\,T_k - T_{k-1}$) and the coefficients $\{c_i\}$ are computed offline via DCT-based interpolation. The degree $n$ is selected from the comparison depth $\kappa$ via a PS-aware lookup (e.g., $\kappa{=}8 \mapsto n{=}13$). Both GPU variants use $\kappa = 8$.

\paragraph{Paterson--Stockmeyer evaluation}
Naive Horner evaluation of a degree-$n$ polynomial requires $O(n)$ multiplicative depth~\cite{low-mult-depth-he:2025}. Paterson--Stockmeyer (PS)~\cite{paterson1973number} restructures the evaluation as a baby-step/giant-step decomposition using only $O(\sqrt{n})$ depth. Denoting the PS baby-step degree $d_1$ and giant-step count $d_2$ (distinct from BSGS $n_1, n_2$), PS minimizes $d_1 + d_2$ subject to $d_1 \cdot 2^{d_2-1} \ge n$; for $n=13$, this gives $d_1{=}2$, $d_2{=}4$. Evaluation proceeds in three stages (Algorithm~\ref{alg:gpu-chebyshev}): (1)~baby-step powers $T_1,\ldots,T_{d_1}$ via a binary doubling tree; (2)~giant-step powers $T_{d_1 \cdot 2^j}$ by repeated squaring; (3)~chunk polynomials $Q_j$ combined via Horner's rule using the giant-step powers.

\begin{algorithm}[!t]
\caption{GPU Chebyshev sign approximation (Paterson--Stockmeyer)}\label{alg:gpu-chebyshev}
\begin{algorithmic}[1]
\footnotesize
\Require Encrypted similarity ciphertext $\mathsf{ct}_s$ on GPU,
Chebyshev coefficients $\{c_i\}_{i=0}^{n}$ (precomputed on CPU),
baby-step degree~$d_1$, giant-step count~$d_2$, threshold~$\delta$
\Ensure Encrypted comparison ciphertext on GPU (slot values $\approx 1$ for match, and $\approx 0$ otherwise)

\Statex
\Statex $\triangleright$ \textit{Step 1: Baby-step Chebyshev powers via binary tree}
\State $T[1] \gets \mathsf{ct}_s$ \Comment{$T_1(x) = x$}
\For{$i = 2$ to $d_1$}
    \If{$i$ is a power of $2$}
        \State $T[i] \gets 2 \cdot T[i/2]^2 - 1$ \Comment{doubling: $T_{2j}(x) = 2\,T_j(x)^2 - 1$}
    \Else \Comment{$i$ odd}
        \State $T[i] \gets 2 \cdot T[\lfloor i/2 \rfloor] \cdot T[\lceil i/2 \rceil] - T[1]$ \Comment{product: $T_{a+b} = 2\,T_a T_b - T_{|a-b|}$}
    \EndIf
    \State $\textsf{Rescale}_{\text{GPU}}(T[i])$
    \State $\textsf{MatchLevel}_{\text{GPU}}(T[1],\, T[i])$ \Comment{drop $T[1]$ to match if needed}
\EndFor

\Statex
\Statex $\triangleright$ \textit{Step 2: Giant-step powers by repeated doubling}
\State $T_2[0] \gets T[d_1]$ \Comment{$T_{d_1}(x)$}
\For{$j = 1$ to $d_2 - 1$}
    \State $T_2[j] \gets 2 \cdot T_2[j{-}1]^2 - 1$ \Comment{$T_{d_1 \cdot 2^j}(x)$}
    \State $\textsf{Rescale}_{\text{GPU}}(T_2[j])$
\EndFor

\Statex
\Statex $\triangleright$ \textit{Step 3: Chunk polynomials and Horner combination}
\For{$j = 0$ to $2^{{d_2}-1} - 1$} \Comment{compute chunk polynomials $Q_j$}
    \State $Q[j] \gets c_{j \cdot d_1}$ \Comment{constant term (scalar)}
    \For{$i = 1$ to $d_1 - 1$}
        \If{$c_{j \cdot d_1 + i} \ne 0$}
            \State $Q[j] \gets Q[j] + c_{j \cdot d_1 + i} \cdot T[i]$ \Comment{scalar--ciphertext multiply + add}
        \EndIf
    \EndFor
\EndFor

\Statex
\State $\mathsf{result} \gets Q[2^{d_2-1} - 1]$ \Comment{start from highest chunk}
\For{$j = d_2 - 1$ down to $1$} \Comment{Horner-like combination via giant steps}
    \State $\textsf{MatchLevel}_{\text{GPU}}(\mathsf{result},\, T_2[0])$
    \State $\mathsf{result} \gets \mathsf{result} \cdot T_2[0]$
    \State $\textsf{Rescale}_{\text{GPU}}(\mathsf{result})$
    \State $\textsf{MatchLevel}_{\text{GPU}}(\mathsf{result},\, Q[j{-}1])$
    \State $\mathsf{result} \gets \mathsf{result} + Q[j{-}1]$
\EndFor

\Statex
\State $\mathsf{result} \gets \mathsf{result} + 1.0$ \Comment{shift from $[-1,1]$ to $[0,2]$}
\State \Return $\mathsf{result}$
\end{algorithmic}
\end{algorithm}
\paragraph{Depth and level management}
The implementation uses CKKS \textsc{FixedManual} rescaling with two invariants: (i)~before every ciphertext--ciphertext multiplication, operands' noise level must be $1$ (a pending rescale is applied first); (ii)~before every addition, operands must share the same RNS level (achieved via \textsf{MatchLevel}). Scalar--ciphertext multiplications do not consume a level. FIDESlib's \texttt{square()} strictly enforces (i), requiring explicit rescale before every squaring, even where OpenFHE's CPU path would tolerate a pending rescale. The binary-tree construction uses $O(\log d_1)$ depth and Horner combination adds $O(d_2)$, totaling ${\sim}7$ multiplicative levels (well below the $13$ needed by naive Horner).

\subsection{HyDia on GPU}\label{sec:gpu-diag}

\ApproachA employs HyDia's original diagonal method (described in Section~\ref{sec:hydia-intro}) because its parallelizability favors GPU implementations. 
%
Specifically, although our BSGS optimization reduces the CPU rotation count from $\ell$ to $O \left( \sqrt{\ell} \right)$, in the naive diagonal method 
all $\ell$ multiplications per matrix are independent and can execute in parallel, while the $\ell$ query rotations are computed once and reused across all $M$ matrices.
Because hoisted GPU rotations are 
roughly $1.7\times$ 
faster than in CPU, 
the cost of computing all $\ell$ rotations on the GPU 
is lower than 
on the CPU.
However, the trade-off lies in GPU memory usage: storing $\ell-1$ rotation keys requires ${\sim}18.7$\,GB (including key-switching decomposition buffers), so the total peak GPU memory, including cached queries, diagonals, and runtime intermediates, reaches ${\sim}27$\,GB even at $G{=}1$.
This limits the database size on memory-constrained GPUs, an issue addressed by \ApproachB. 
We now describe the different steps of \ApproachA.

\paragraph{Offline phase}
Uploads the deserialized $G \cdot \ell$ standard diagonal ciphertexts and $\ell{-}1$ rotation keys to GPU memory.

\paragraph{Online phase}
\begin{enumerate}
    \item \textbf{Query rotations:} Computes $\mathsf{rotQ}[i] \gets \textsf{Rot}_i(\vec{q})$ for $0 \leq i < \ell$ on the GPU using FIDESlib's native rotation. Caches all $\ell$ rotated queries in the GPU.
    \item \textbf{Similarity:} For each matrix \groupmatk, computes on the GPU $\mathsf{sim}_{\groupmatk} \gets \sum_{i=0}^{\ell-1} \mathsf{rotQ}[i] \odot \mathsf{diag}[k \ell + i]$, where $\mathsf{diag}[k\ell + i]$ denotes the GPU-resident ciphertext encoding $\mathrm{diag}_i(\groupmatk)$ from Equation~\ref{equ:diag}. This is done via element-wise multiplication with lazy/deferred relinearization (accumulate products first, then relinearize once), followed by rescale.
    \item \textbf{Chebyshev and aggregation:} Each similarity ciphertext passes through the GPU Chebyshev comparison (Algorithm~\ref{alg:gpu-chebyshev}). Membership aggregation (tree-reduce addition and rotate-and-sum) also runs on the GPU.
\end{enumerate}

A full-pipeline function processes all matrices in CUDA waves using the stream pool, performing similarity, Chebyshev, and aggregation entirely on the GPU before downloading a single resulting ciphertext.

\subsection{BSGS with pre-rotated diagonals on GPU}\label{sec:gpu-simple-bsgs}

The primary motivation for the BSGS-based GPU variant \ApproachB is \emph{GPU memory efficiency}: \ApproachA requires $\ell-1=511$  rotation keys (${\sim}18.7$\,GB), limiting the database size to GPU memory capacity.
By reducing the online rotation-key set to a much smaller BSGS subset (giant-step and EvalSum keys only, as baby-step keys remain on the CPU), \ApproachB reduces the peak GPU memory footprint from ${\sim}27$\,GB to ${\sim}9$\,GB for $8{,}192$ vectors, enabling deployment on GPUs with as little as $12$\,GB of VRAM.
\ApproachB also has a significantly faster setup phase, since uploading this reduced key set is much faster than uploading all HyDia rotation keys.

The key algorithmic aspect of \ApproachB is that diagonals are {pre-rotated} during the offline phase, enabling \emph{on-demand} BSGS execution entirely on the GPU.
Without pre-rotation, one could still use BSGS to reduce the rotation key count: compute only $n_1$ baby-step rotations plus $G$ giant-step rotations via hoisted composition, then run the standard diagonal method with all $\ell$ pre-composed rotations. 
This is exactly how our CPU variant (Section~\ref{sec:bsgs}) operates. 
However, this approach requires storing all $\ell$ composed rotations in GPU memory before the matrix--vector product begins.
On-demand BSGS avoids this by computing each giant-step contribution \emph{during} the matrix--vector product, accumulating $n_1$ baby-step products and then applying a single giant-step rotation per group.
The difficulty is that ciphertext rotation distributes over products --- $\textsf{Rot}_r(A \cdot B) = \textsf{Rot}_r(A) \cdot \textsf{Rot}_r(B)$ --- so a giant-step rotation of the accumulated baby-step products would incorrectly rotate the diagonal components as well.
Pre-rotating the diagonals during the offline phase absorbs this offset in advance, restoring correctness (see Equation~\ref{eq:prerotation}).
We do not apply this pre-rotation on CPU because the CPU pipeline uses hoisted composition effectively and lacks GPU-resident rotation cache constraints, diminishing offline preprocessing benefits.

\subsubsection{Pre-rotated diagonals}

During enrollment, for each giant-step index $j \in \{0,\ldots,n_2{-}1\}$ and baby-step index $i \in \{0,\ldots,n_1{-}1\}$, instead of storing the diagonal ciphertext $\mathrm{diag}_{jn_1+i}$ (cf.\ Equation~\ref{equ:diag}) directly, we store
\begin{equation}\label{eq:prerotation}
    \mathrm{diag}'_{jn_1+i} \;=\; \textsf{Rot}_{{-}j n_1}\!\bigl(\mathrm{diag}_{jn_1+i}\bigr).
\end{equation}
This pre-rotation absorbs exactly the $\mathrm{Rot}_{-jn_1}(\cdot)$ factor that would otherwise be applied to the diagonal term during the online evaluation.
Substituting Equation~\ref{eq:prerotation} into the standard BSGS identity and noting that $\textsf{Rot}_{jn_1} \odot \textsf{Rot}_{-jn_1}$ is the identity, the on-demand computation
$\textsf{Rot}_{j n_1}\!\bigl(\sum_{i} \textsf{Rot}_i(\vec{q}) \odot \mathrm{diag}'_{j n_1 + i}\bigr)$
recovers the standard BSGS result.

Operationally, this means the server computes only $n_1$ baby-step rotations and, for each giant step $j$, applies one rotation $\textsf{Rot}_{j \cdot n_1}(\cdot)$ to the accumulated baby-step products.

\subsubsection{Online pipeline}\label{sec:gpu-online-pipeline}
\paragraph{Offline phase}
Standard diagonal ciphertexts from disk are loaded and \emph{pre-rotated on the GPU}: 
for each diagonal index $jn_1+i$, applies the rotation $\textsf{Rot}_{-j n_1}(\mathrm{diag}_{jn_1+i})$ using FIDESlib's GPU rotation before caching.
Negative giant-step rotation keys ($-n_1, -2n_1, \ldots$) are uploaded to the GPU for this pre-rotation and then \emph{freed} to reclaim GPU memory once all diagonals have been cached.
Giant-step keys ($n_1, 2n_1, \ldots$) and power-of-two rotation keys (for the membership scenario's slot-wise \textsf{RotateAndSum} aggregation) are uploaded and remain GPU-resident for the online phase. (The power-of-two keys are not needed for the identification scenario, which downloads per-matrix results without aggregation.)
Baby-step rotation keys are \emph{not} uploaded to the GPU; instead, baby-step rotations are computed on the CPU, and the resulting ciphertexts are uploaded (see Section~\ref{sec:gpu-rotation-cache}).

\paragraph{Online phase}
\begin{enumerate}
    \item \textbf{Baby-step rotations:} Compute $n_1$ baby-step rotations of the query ciphertext on the CPU and upload the resulting ciphertexts to GPU memory (see Section~\ref{sec:gpu-rotation-cache}).
    \item \textbf{On-demand BSGS:} For each matrix $m$ and each giant step $j$:
    \begin{itemize}
        \item Multiply each baby step $\textsf{Rot}_i(\vec{q})$ with the pre-rotated diagonal $\mathrm{diag}'_{jn_1 + i}$, accumulate with lazy/deferred relinearization, and rescale.
        \item Apply on-demand giant-step rotation $\textsf{Rot}_{j \cdot n_1}(\mathsf{sum}_j)$, which correctly aligns the baby-step component without affecting the diagonal (since it was pre-rotated).
    \end{itemize}
    \item \textbf{Chebyshev plus aggregation:} The full-pipeline function processes all matrices in CUDA waves using the stream pool.
\end{enumerate}

\paragraph{Rotation keys} 
To leave more GPU space for database ciphertexts, \ApproachB uses a reduced online key set: ${\sim}36$, including giant-step and powers-of-two EvalSum keys, listed in the \emph{Keys on GPU (online)} row in Table \ref{tab:gpu-bsgs-complexity}. 
This number is roughly one order of magnitude smaller than \ApproachA's $(\ell-1) + \lceil \log_2(N/\ell) \rceil$ key set (which translates to 511 diagonal rotation plus 5 power-of-two EvalSum keys), and the one relevant for estimating steady-state VRAM usage.
Table \ref{tab:gpu-bsgs-complexity} also shows $75$ as the \emph{Rotation keys (total)}, which additionally counts two groups of keys that are not resident on the GPU during queries: 
(i) $n_1-1$ baby-step rotation keys, which exist only in the CPU-side OpenFHE context and are used solely to compute baby-step rotations of the query on the CPU; and (ii) $n_2-1$ negative giant-step keys, uploaded to the GPU to pre-rotate the diagonals once during the offline phase and freed immediately afterward (never contributing to the online-phase footprint reported in the \emph{Peak GPU memory} row).


\subsubsection{Variant: enroller-side plaintext pre-rotation}
\label{sec:gpu-prerot-enroller}

\ApproachB applies the diagonal pre-rotation of Equation~\ref{eq:prerotation} \emph{homomorphically on the GPU} during the offline phase, requiring $n_2{-}1$ negative giant-step rotation keys.
An alternative variant, \ApproachBprime, shifts this pre-rotation to the \emph{enroller} by applying it in the plaintext domain \emph{before} encryption.
For each diagonal index $jn_1+i$, the enroller cyclically rotates the plaintext coefficient vector by $-j n_1$ positions and then encrypts the rotated plaintext.
Because the rotation is applied before encryption, it is exact and incurs only $O(\ell)$ data movement per diagonal, a negligible cost compared to CKKS encryption (${\sim}8\%$ enrollment overhead at $K{=}65{,}536$).

The main advantage of \ApproachBprime is that it eliminates all negative giant-step rotation keys: the server never performs homomorphic rotation during diagonal upload, reducing the total client key count from 75 to 53 (Table \ref{tab:gpu-bsgs-complexity}) and peak GPU memory from $\sim9$ GB to $\sim 7$ GB at K=8,192. The \emph{Keys on GPU (online)} count remains at 36 as in \ApproachB, since both variants use identical online key sets (giant-step and power-of-two EvalSum keys); this reduction comes entirely from removing the $n_2-1$ temporary negative giant-step keys, which BSGS-RTX-TBS uploads and frees during the offline phase but \ApproachBprime never generates.
The offline diagonal upload reduces to a bulk ciphertext transfer without any GPU rotation, yielding a faster setup phase.

The online phase of \ApproachBprime is \emph{identical} to \ApproachB. 

\paragraph{Number Theoretic Transform (NTT) representation tradeoff}
Despite eliminating offline homomorphic rotations, \ApproachBprime exhibits a small overhead in online query time compared to \ApproachB.
This is attributed to a FIDESlib implementation detail: in \ApproachB, the homomorphic rotation applied during diagonal upload leaves the ciphertext polynomials in NTT form, the native representation for subsequent multiplications.
In \ApproachBprime, diagonals are uploaded directly from disk in coefficient form, requiring an implicit NTT conversion when first used during multiplication.
With $\ell = 512$ diagonals per matrix group and each NTT costing ${\sim}40{-}50\,\mu$s, the accumulated overhead (${\sim}20{-}25$\,ms per matrix group) is consistent with the observed slowdown.
A warm-up NTT pass after uploading could move this cost offline, but would not eliminate it.

\paragraph{Tradeoff summary}
\ApproachBprime trades a modest online overhead (from NTT conversions) and an enrollment-time dependency on $n_1$ for fewer rotation keys, lower GPU memory, faster offline setup, and simpler key management.
It is best suited for deployments where $n_1$ is fixed, and re-enrollment is infrequent.
\ApproachB retains full flexibility, since the server can re-derive the pre-rotated diagonals for any $n_1$ without re-enrolling the database.

\paragraph{Hardware/algorithm coupling} 
We emphasize that $n_1$ is a purely algorithmic parameter of the BSGS decomposition, fixed by the embedding dimension $\ell$ (we use $n_1 = \lceil \sqrt{\ell} \rceil$, following Section \ref{prelim:bsgs}). Since it is independent of the Server's GPU model, VRAM capacity, or SM count, the Enroller requires no hardware knowledge in advance. Instead, it only needs to agree with the Server on $n_1$, just as with all other CKKS parameters. 
The Server's GPU affects only two independent online quantities, neither of which the Enroller requires:
the maximum database size fitting in VRAM, and the CUDA stream-pool size $S$ of Equation \ref{eq:stream-pool}, both computed during start-up considering the Server's free memory (e.g., via \emph{cudaMemGetInfo}).

\paragraph{Maintenance implications} 
A Server-side hardware upgrade (e.g., replacing an RTX 8000 with an H200) that preserves $\ell$ and the chosen $n_1$ requires no re-enrollment: the pre-rotated diagonals produced by BSGS-RTX-TBE remain valid, since $n_1$ is unchanged, and the new GPU simply permits a larger $S$ and a bigger resident database. Re-enrollment is only required if: (1) $n_1$ is modified deliberately, for instance, to trade a slightly different online/offline balance on a new deployment; or (2) $\ell$ changes, e.g., when adopting a different face embedding model. In such cases, because the pre-rotation of BSGS-RTX-TBE is applied to the plaintext diagonal before encryption, changing $n_1$ requires decrypting and re-encrypting the affected diagonals, i.e., a full re-enrollment at the original cost ($\sim 8\%$ overhead over plain CKKS encryption, according to Section \ref{prelim:bsgs}).
BSGS-RTX-TBS avoids this maintenance cost entirely: since its pre-rotation is applied homomorphically to already-encrypted diagonals at GPU cache-load time (offline phase) rather than at enrollment time, it can be re-parameterized with a new $n_1$ purely on the Server side. 
We recommend BSGS-RTX-TBS over BSGS-RTX-TBE whenever $n_1$ is expected to change across the database's lifetime (e.g., during active \mbox{hardware/parameter} experimentation), and BSGS-RTX-TBE once $n_1$ has been fixed for a long-lived deployment.

\subsection{Overall algorithm description}\label{sec:gpu-algorithm}

The complete GPU-accelerated query processing algorithm comprises an offline phase (executed once at system setup) and an online phase (executed for each query).
The online phase is presented in Algorithms ~\ref{alg:gpu-bsgs-membership} and~\ref{alg:gpu-bsgs-index} for the membership and identification scenarios, respectively.
The algorithm is parameterized by the rotation strategy: \ApproachA uses $n_1 = \ell$ baby steps and no giant steps; \ApproachB and \ApproachBprime use $n_1 = \lceil\sqrt{\ell}\rceil$ baby steps and $n_2 = \lceil \ell/n_1 \rceil$ on-demand giant steps with pre-rotated diagonals (the two variants differ only in \emph{where} and \emph{how} the pre-rotation is performed, following Section~\ref{sec:gpu-prerot-enroller}).
\begin{algorithm}[htbp]
\caption{GPU query processing --- Membership}\label{alg:gpu-bsgs-membership}
\begin{algorithmic}[1]
\scriptsize
\Require Encrypted query $\vec{q}$, diagonal ciphertexts $\{\mathsf{diag}[\groupmatk]\}$ cached on GPU (pre-rotated for \ApproachB/\ApproachBprime),
rotation parameter $n_1$,
comparison threshold~$\delta$, Chebyshev depth~$\kappa$,
stream pool size~$S$, number of matrices~$G$
\Ensure Encrypted membership result (single ciphertext)

\Statex

\Statex $\triangleright$ \textit{Step 1: Baby-step rotations}
\State Upload $\vec{q}$ to GPU
\For{$i = 0$ to $n_1 - 1$}
    \State $\mathsf{baby}[i] \gets \textsf{Rot}_{i}(\vec{q})$ \Comment{on GPU for \ApproachA ($n_1{=}\ell$); on CPU then uploaded for \ApproachB ($n_1{=}\lceil\sqrt{\ell}\rceil$)}
\EndFor

\Statex

\Statex $\triangleright$ \textit{Step 2: Per-matrix similarity + Chebyshev (multi-stream on GPU)}
\State $n_2 \gets \lceil \ell / n_1 \rceil$ \Comment{$n_2=1$ for \ApproachA; $n_2=23$ for \ApproachB}
\For{each wave $w = 0, S, 2S, \ldots$ up to $G$}
    \For{$m = w$ \textbf{to} $\min(w{+}S, G){-}1$ \textbf{in parallel} (stream $m \bmod S$)}
        \State $\mathsf{acc}_m \gets \mathbf{0}$
        \For{$j = 0$ to $n_2 - 1$}
            \State $\mathsf{sum}_j \gets \sum_{i=0}^{\min(n_1,\, \ell - j n_1) - 1} \mathsf{baby}[i] \odot \mathsf{diag}[m \cdot \ell + j n_1 + i]$
            \State $\textsf{Rescale}_\text{GPU}(\mathsf{sum}_j)$
            \If{$j > 0$ \textbf{and} $n_2 > 1$}
                \State $\mathsf{sum}_j \gets \textsf{Rot}_{\text{GPU}}(\mathsf{sum}_j,\; j \cdot n_1)$ \Comment{on-demand giant-step rotation (\ApproachB only)}
            \EndIf
            \State $\mathsf{acc}_m \gets \mathsf{acc}_m + \mathsf{sum}_j$
        \EndFor
        \State $\mathsf{cheb}_m \gets \textsf{ChebyshevCompare}_{\text{GPU}}(\mathsf{acc}_m,\, \delta,\, \kappa)$ \Comment{Algorithm~\ref{alg:gpu-chebyshev}}
    \EndFor
    \State \textsf{SynchronizeStreams}$(w,\ldots,w{+}S{-}1)$
\EndFor

\Statex

\Statex $\triangleright$ \textit{Step 3: Aggregation (on GPU)}
\State $\mathsf{result} \gets \sum_{m=0}^{G-1} \mathsf{cheb}_m$ \Comment{tree-reduce addition on GPU}
\State $\mathsf{result} \gets \textsf{RotateAndSum}_{\text{GPU}}(\mathsf{result},\, N)$ \Comment{slot-wise aggregation}

\Statex

\Statex $\triangleright$ \textit{Step 4: Download}
\State Download $\mathsf{result}$ from GPU to CPU
\Statex

\State \Return $\mathsf{result}$ \Comment{slot 0 encodes the total match count}
\end{algorithmic}
\end{algorithm}

\begin{algorithm}[htbp]
\caption{GPU query processing --- Identification}\label{alg:gpu-bsgs-index}
\begin{algorithmic}[1]
\scriptsize
\Require Same as Algorithm~\ref{alg:gpu-bsgs-membership}
\Ensure Vector of $G$ encrypted comparison ciphertexts

\Statex

\Statex $\triangleright$ \textit{Steps 1--2: identical to Algorithm~\ref{alg:gpu-bsgs-membership}, yielding $\{\mathsf{cheb}_m\}_{m=0}^{G-1}$ on GPU}

\Statex

\Statex $\triangleright$ \textit{Step 3: Bulk download}
\For{$m = 0$ to $G - 1$}
    \State Download $\mathsf{cheb}_m$ from GPU to CPU
\EndFor
\State \Return $\{\mathsf{cheb}_0,\, \mathsf{cheb}_1,\, \ldots,\, \mathsf{cheb}_{G-1}\}$ \Comment{non-zero slots identify matching indices}
\end{algorithmic}
\end{algorithm}

\subsection{Complexity analysis}\label{sec:gpu-complexity}

Table~\ref{tab:gpu-bsgs-complexity} summarizes the per-query costs of both GPU variants alongside the CPU baseline, where the per-query upload count refers to uploading the query ciphertext $\vec{q}$.
\begin{table*}[htbp]
\centering
\small
\caption{Per-query operation and transfer costs. $\ell{=}512$, $n_1{=}23$, $n_2{=}23$, $G{=}\lceil K/\numslots\rceil$ matrix groups. \ApproachA's $516$ keys comprise $\ell{-}1 = 511$ diagonal rotation keys plus $5$ power-of-two EvalSum keys. For BSGS-RTX-TBS and BSGS-RTX-TBE, "Rotation keys (total)" counts every key the client generates, including those that never reach the GPU; "Keys on GPU (online)" counts only the subset resident in GPU memory throughout the online query phase (giant-step and power-of-two EvalSum keys) and is the figure relevant to the "Peak GPU memory" row. The gap between the two rows refers to the $n_1-1$ baby-step keys, which live solely in the CPU-side OpenFHE context, plus (for BSGS-RTX-TBS only) the $n_2-1$ negative giant-step keys uploaded to the GPU for offline diagonal pre-rotation and then freed (see Section \ref{sec:gpu-online-pipeline}).}
\label{tab:gpu-bsgs-complexity}
\setlength{\tabcolsep}{7pt}
\begin{tabular}{lrrrrrr}
\toprule
\textbf{Resource} & \textbf{BSGS-CPU-DG} & \textbf{\ApproachA} & \textbf{\ApproachB} & \textbf{\ApproachBprime} \\
\midrule
Baby-step rotations & $\ell{-}1$ & $\ell{-}1$ (GPU) & $n_1{-}1$ (CPU) & $n_1{-}1$ (CPU) \\
Giant-step rotations & --- & --- & $(n_2{-}1){\cdot}G$ & $(n_2{-}1){\cdot}G$ \\
Total online rotations & $\ell{-}1$ & $\ell{-}1$ & $(n_1{-}1) {+} (n_2{-}1){\cdot}G$ & $(n_1{-}1) {+} (n_2{-}1){\cdot}G$ \\
Mult.\ per matrix & $\ell$ & $\ell$ & $\ell$ & $\ell$ \\
CPU $\to$ GPU upload & --- & $1$ ct & $n_1$ cts & $n_1$ cts \\
GPU $\to$ CPU download & --- & $1$/$G$ cts & $1$/$G$ cts & $1$/$G$ cts \\
Rotation keys (total) & $44$ & $516$ & ${\sim}75$ & ${\sim}53$ \\
Keys on GPU (online) & --- & $516$ & $36$ & $36$ \\
Peak GPU memory ($K{=}8192$) & --- & ${\sim}27$\,GB & ${\sim}9$\,GB & ${\sim}7$\,GB \\
Pre-rotated diags & No & No & Server (HE) & Enroller (PT) \\
Negative rot.\ keys & No & No & Yes (freed) & No \\
\bottomrule
\end{tabular}
\end{table*}

\paragraph{Transfer overhead}
Per-query host--device transfer is dominated by the baby-step ciphertext uploads: $1$ ciphertext (${\sim}5$\,MB) for \ApproachA versus $n_1$ ciphertexts (${\sim}115$\,MB) for \ApproachB/\ApproachBprime---in both cases orders of magnitude less than uploading all $\ell$ rotations (${\sim}2.5$\,GB). Detailed transfer cost measurements are reported in Section~\ref{sec:gpu}.

\paragraph{Rotation key tradeoff}
\ApproachB requires ${\sim}10\times$ fewer online rotation keys on the GPU than \ApproachA, freeing GPU memory for larger databases. The trade-off is that \ApproachB requires $n_2{-}1$ on-demand giant-step rotations per matrix,
whereas \ApproachA performs no giant-step rotations but pre-computes all $\ell{-}1$ baby-step rotations.

\paragraph{Pre-rotation tradeoff (\ApproachB)}
\ApproachB trades additional offline computation (pre-rotating all diagonals during GPU caching) for a smaller online rotation count. Quantitatively, pre-rotation reduces the online giant-step work from $n_1 \times (n_2{-}1)$ rotations per matrix to $n_2{-}1$, which is the key cost reduction for GPU execution.

\paragraph{Setup vs.\ query tradeoff}
\ApproachA incurs a substantially higher one-time setup cost because it uploads and initializes the full diagonal-rotation key set, while \ApproachB initializes only the reduced BSGS key subset plus offline pre-rotation keys.
The difference in per-query cost arises because \ApproachA must compute $511$ baby-step rotations per query and then sequentially accumulate $512$ diagonal products per matrix, whereas \ApproachB needs only $22$ baby-step rotations and evaluates each matrix via $23$ giant-step groups---of which $22$ require an on-demand GPU rotation.
The BSGS structure maps efficiently to GPU parallelism, replacing $512$ sequential products with ${\sim}23$ coarser iterations of $n_1$ products each, keeping \ApproachB's per-matrix cost well below \ApproachA's independent of~$G$.
\ApproachBprime shares the same online pipeline as \ApproachB and therefore has the same per-query rotation and multiplication costs; its advantage lies in a faster offline setup (no homomorphic pre-rotation) and fewer rotation keys (${\sim}53$ vs.\ ${\sim}75$), at the cost of a small NTT-conversion overhead during the online phase (see Section~\ref{sec:gpu-prerot-enroller}).
Concrete setup and query timing comparisons are given in Section~\ref{sec:gpu}.
\section{Experimental results}\label{sec:experiments}
We run experiments with both CPU and GPU support. For CPU, we use an Intel Xeon Platinum 8260 CPU at 2.4 GHz and 250\,GB of RAM, running Ubuntu 20.04.6 LTS. We use multi-threading for improved performance in all variants, and parameters satisfying a 128-bit security level. We rely on OpenFHE v1.2.3~\cite{OpenFHE} to implement our algorithms (building upon the framework from~\cite{martin2025hydia}). 

Our GPU experiments use two GPUs: an NVIDIA Quadro RTX 8000 with 48\,GB VRAM, and an NVIDIA H200 with 141\,GB HBM3e VRAM. We rely on FIDESLib v1.0.0~\cite{agullo2025fideslib} for GPU integration. FIDESLib internally builds a patched fork of OpenFHE v1.2.3, compiled with \texttt{-O3 -march=native} and \texttt{WITH\_NATIVEOPT=ON} for optimized GPU performance.

\subsection{Parameter selection}

We assume $\ell = 512$, following standard face feature extractors~\cite{martin2025hydia,deng2019arcface}.
For HyDia, we adopt the original parameters from~\cite{martin2025hydia}: ring dimension $2^{15}$ (batch size $2^{14}$), multiplicative depth 11, comparative depth $\kappa = 10$, and scaling factor 45. The remaining parameters are auto-selected for 128-bit security.
Our GPU experiments also test $\kappa = 8$ (total depth 9, see Figure~\ref{fig:gpu} and Tables~\ref{tab:benchmark-results-cpu-gpu-depth8-part1}--\ref{tab:benchmark-results-cpu-gpu-depth8}).
BSGS-Diagonal uses the same CKKS parameters, with depths $\kappa=11$ or $\kappa=9$ on CPU and $\kappa=8+2+1=11$ on GPU. The baby-step parameter is $n_1 = 23$ (manually tuned), giving $n_2 = \lceil 512/23 \rceil = 23$.

\begin{remark}\label{remark}
As mentioned above, we report results for comparison depths
$\kappa = 8$ and $\kappa = 10$. For identification, per-slot error is
independent of database size at both depths (see Section~\ref{sec:noise}). Consequently, all $\kappa = 8$ results, including timing, remain reliable across all tested scales.
For membership, accumulating per-comparison errors makes decisions sensitive to the mean bias $\mathbb{E}[E_c]$, which is larger at $\kappa = 8$ than $\kappa = 10$ and causes performance to degrade at smaller database sizes. 
However, because this bias is a fixed constant of the comparison polynomial and score distribution, it can be removed by subtracting $\mathbb{E}[E_c]$ from each comparison output before aggregation. This adjustment requires a single plaintext–ciphertext addition and no extra multiplicative depth.
Recentering the error to zero mean leaves only the $\sqrt{K}$ variance term of Section~\ref{sec:noise}, which remains negligible at all practical scales and, thus, eliminates this $\kappa = 8$ drawback.
\end{remark}

\paragraph{Datasets}
Experiments use synthetic datasets and the FRGC 2.0 RGB dataset~\cite{phillips2005overview}.
Synthetic datasets are parameterized by database size $K$, number of matches $K_m$, and a random seed. Each vector has dimension $\ell = 512$, with query components drawn uniformly from $[-99, 99]$, matching vectors perturbed by noise uniform in $[-2, 2]$, and non-matching vectors drawn independently. Database sizes range from $2^{10}$ to $2^{20}$.
For accuracy evaluation, we follow the HyDia setup~\cite{martin2025hydia}: a database of $44{,}228$ ArcFace embeddings~\cite{deng2019arcface} is enrolled server-side, and 50 probe queries are issued sequentially within the same session, yielding $50 \times 44{,}228 = 2{,}211{,}400$ pairwise comparisons in total.

\subsection{Benchmarking methodology}\label{sec:methodology}

Each configuration (approach $\times$ database size $\times$ comparison depth) is executed for $t = 11$ independent trials using multi-threaded OpenMP execution with 40 threads (empirically optimal on our 24-core / 48-thread Xeon Platinum 8260).
All trials share the same cryptographic key pair and encrypted database. The key pair and ciphertexts are generated once in trial~1 and serialized to disk, with trials $2$--$11$ deserializing directly.
For each trial, we collect wall-clock time, enrollment time, membership and identification query times (server-side only, excluding client encryption/decryption and network transfer), peak RAM, peak disk usage, and peak GPU VRAM (where applicable), sampled at $0.5$\,s intervals.
Every trial is automatically validated for correctness \footnote{A trial is marked \textsc{pass} only if both correctness conditions hold and the process exits successfully: (i)~the membership result matches the ground truth, and (ii)~the identification result returns exactly the set of planted matching indices. Failed trials are excluded from timing aggregation.} and failing trials are excluded from aggregation.

Since trial~1 includes one-time setup costs (key generation, database encryption, CUDA initialization), query-time metrics are reported as mean $\pm$ std over the last $N = 10$ trials (trials $2$--$11$), denoted \emph{last-$N$}. Enrollment time and peak disk usage are reported from trial~1 only. The wall-clock ``overall'' time is $\text{enroll}(\text{trial\,1}) + \text{mean}(\text{wall time, last-}N)$.

\begin{remark}
Membership and server timings exclude network transmission. Following Table~6 of~\cite{martin2025hydia}, network overhead is minimal at bandwidths of 1\,Gbps and above.
\end{remark}

\subsection{Setup Profiling}
To separate one-time and amortized costs from per-query latency, we report the full setup breakdown in Table~\ref{tab:setup-costs}. This includes key generation, rotation-key generation, database encryption and diagonal pre-rotation, GPU key upload, and GPU database caching (the last two being online setup metrics). In particular, BSGS-RTX-TBS performs diagonal pre-rotation on the server/GPU side during setup, whereas BSGS-RTX-TBE shifts this task to the Enroller before encryption. Hence, this architectural improvement results in a lower online setup time for BSGS-RTX-TBE across all evaluated database sizes, and consequently reduces the overall setup time compared to BSGS-RTX-TBS.

\begin{table*}[htbp]
\centering
\caption{Setup costs profiling for BSGS-RTX-TBS and BSGS-RTX-TBE strategies (Single-GPU).
\textbf{Key Gen}, \textbf{Rot. Key Gen.}, \textbf{DB Encrypt} and \textbf{Diag. PreRot} costs are incurred once and can be amortized across executions, whereas \textbf{GPU Key Upload} and \textbf{GPU DB Cache} setup costs are incurred in each trial. Values with $\pm$ denote the standard deviation over 10 trials.}
\label{tab:setup-costs}
\scriptsize
\setlength{\tabcolsep}{3.5pt}
\resizebox{\textwidth}{!}{%
\begin{tabular}{rrlrrrrrrr}
\toprule
$K$ & $K_m$ & Approach & Key Gen (s) & Rot. Key Gen. (s) & DB Encrypt (s) & Diag. PreRot (s) & GPU Key Upload (s) & GPU DB Cache (s) & Total Setup (s)\\
\midrule
\multirow{2}{*}{$2^{14}$} & \multirow{2}{*}{64} & BSGS-RTX-TBS & 0.4007 & 0.4479 & 1.0564 & - & 0.8489 $\pm$ 0.0247 & 2.9640 $\pm$ 0.0794 & 3.8129 $\pm$ 0.1025 \\
  &   & BSGS-RTX-TBE & 0.4010 & 0.3514 & 1.0304 & 0.0015 & 0.9750 $\pm$ 0.2735 & 2.3281 $\pm$ 0.1521 & 3.3032 $\pm$ 0.2429 \\
\midrule
\multirow{2}{*}{$2^{15}$} & \multirow{2}{*}{64} & BSGS-RTX-TBS & 0.3434 & 0.4418 & 1.9601 & - & 1.0723 $\pm$ 0.2601 & 5.2132 $\pm$ 0.2894 & 6.2855 $\pm$ 0.2996 \\
  &   & BSGS-RTX-TBE & 0.4155 & 0.4814 & 1.9953 & 0.0035 & 0.9532 $\pm$ 0.1755 & 4.8394 $\pm$ 0.1391 & 5.7926 $\pm$ 0.0697 \\
\midrule
\multirow{2}{*}{$2^{16}$} & \multirow{2}{*}{128} & BSGS-RTX-TBS & 0.3416 & 0.3997 & 3.9577 & - & 0.9812 $\pm$ 0.1965 & 10.0185 $\pm$ 0.6310 & 10.9997 $\pm$ 0.6619 \\
 &  & BSGS-RTX-TBE & 0.4204 & 0.3621 & 3.9644 & 0.0083 & 1.1416 $\pm$ 0.2340 & 9.3490 $\pm$ 0.5780 & 10.4906 $\pm$ 0.4779 \\
\bottomrule
\end{tabular}
}
\end{table*}

\subsection{Performance and memory tradeoffs with BSGS (CPU)}\label{sec:performance}
\paragraph{Client-side: rotation keys required for HyDia and BSGS-Diagonal}
For $\ell=512$ and $\texttt{numSlots} =16384$, HyDia requires $\ell-1=511$ rotation keys.
This number depends on the ring and embedding dimensions, but not on the database's size.


In BSGS-Diagonal, the number of rotation keys also depends on the BSGS parameter $n_1$, but remains independent of database size. Again, let $\ell=512$, $\texttt{numSlots}=16384$, and $n_1=23$.
The rotation indices consist of 
$S_{\mathrm{baby}}=\{1,\ldots,22\}$ and
$S_{\mathrm{giant}}=\{23,46,\ldots,506\}$ 
which results in 
$22 + 22 = 44$ rotation keys.

BSGS-Diagonal requires, thus, $91\%$ fewer rotation keys for any database size. 
Since a rotation key occupies about 30 MB in RAM according to our experiments, the memory savings at the client can reach 14 GB, significantly facilitating edge deployment.
\paragraph{Cost of the key reduction}
These client-side savings come with additional costs at the server. BSGS-Diagonal holds the full table of $\ell = 512$ rotated query ciphertexts during evaluation, meaning $\approx 4$\,GB for $8$\,MB ciphertexts, against $n_1 = 23$
($\approx 184$\,MB) for single-hoisted grouped BSGS. BSGS-Diagonal also
performs $n_1 n_2 - 1 = 511$ query rotations irrespective of $G$, above
the $(n_1-1) + G(n_2-1)$ rotations of the grouped variant for $G < 32$.
Hence, the grouped variant remains preferable for database sizes up to $2^{18}$
(see Table~\ref{tab:bsgs-diagonal-comparison}). 

\begin{figure}[t]
  \centering
  \includegraphics[width=\columnwidth]{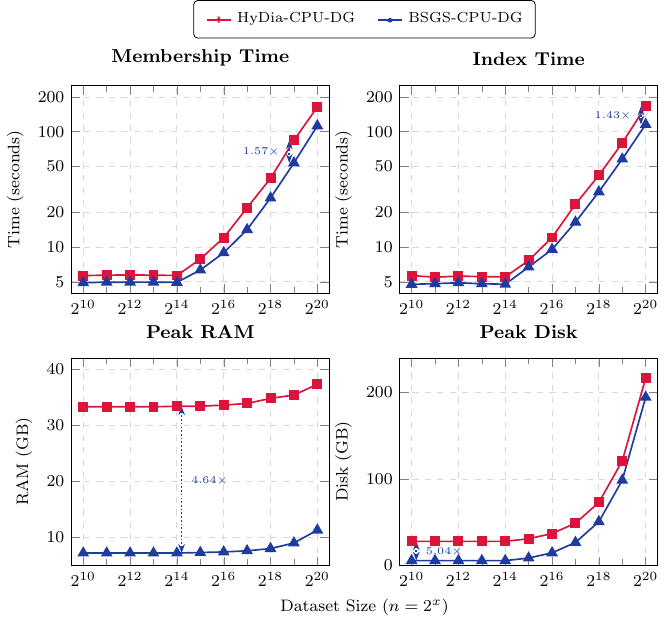}
  \caption{The original HyDia vs. our BSGS-Diagonal optimization on CPU for membership and index scenarios: execution time, peak RAM, and Disk usage for databases with $2^{10}$-$2^{20}$ entries and total multiplicative depth 11.}
\label{fig:cpu}
\end{figure}

\paragraph{Overall performance and peak RAM}
To evaluate BSGS-Diagonal against the diagonal method used by HyDia, we run both algorithms for membership and index scenarios and compare their execution time (in seconds), peak RAM (in GB), and peak disk usage (in GB) for datasets of size $2^{10}$ to $2^{20}$.
Figure~\ref{fig:cpu} illustrates our results. In terms of performance, BSGS-Diagonal slightly improves over HyDia, delivering up to 1.57$\times$ and 1.43$\times$ speed-ups for the database of size $2^{19}$ in the membership scenario and of size $2^{20}$ for the identification scenario, respectively. 
The biggest advantage of our method is the reduced memory consumption: we observe up to $4.6\times$ peak RAM reduction and $5\times$ lower disk usage (for databases of size $2^{10}$), and similar savings for all our databases up to size $2^{18}$.
Notably, with BSGS-Diagonal, the peak RAM never exceeds 8 GB for databases up to $2^{18}$ (as opposed to 33 GB for HyDia), making our algorithm much more suited for constrained edge devices.
Table~\ref{tab:benchmark-results-cpu-only-extended} in~\ref{appendix:benchmarks} lists our exact experimental results.


\subsection{GPU acceleration}\label{sec:gpu}
\paragraph{Rotation keys and GPU memory} HyDia-GPU uses $511$ rotation keys (identical to HyDia on CPU)
at ${\sim}37$\,MB each (${\sim}18.7$\,GB for keys alone).
With cached diagonals and intermediate buffers, \ApproachA on GPU requires ${\sim}24$\,GB of GPU memory for a database of $8{,}192$ vectors -- achievable on high-end GPUs (e.g., NVIDIA RTX~8000 with $48$\,GB or H200 with $141$\,GB). 

\paragraph{Transfer overhead}
\ApproachA uploads a single query ciphertext (${\sim}5$\,MB) per query and computes all $\ell{-}1$ baby-step rotations directly on GPU.
\ApproachB and \ApproachBprime instead upload $n_1 = 23$ baby-step ciphertexts (${\sim}115$\,MB) per query, pre-computed on CPU.
In both cases, this is far less than the $\ell = 512$ ciphertexts (${\sim}2.5$\,GB) required if all query rotations were computed on the CPU and uploaded to the GPU.

\paragraph{Setup vs.\ query cost}
\ApproachB is faster per query than \ApproachA: at $K{=}1{,}024$, \ApproachB achieves $0.25$\,s membership latency versus $0.63$\,s for \ApproachA ($2.5\times$ faster); at $K{=}65{,}536$, \ApproachA exceeds GPU memory entirely, while \ApproachB completes in ${\sim}3.0$\,s.
\ApproachA also incurs a substantially higher one-time setup cost: approximately $39$\,s 
versus ${\sim}9.5$\,s for \ApproachB and ${\sim}8.5$\,s for \ApproachBprime at $K{=}1{,}024$, due to the overhead of uploading and initializing all $516$ rotation keys.

\begin{figure}[t]
  \centering
  \includegraphics[width=\columnwidth]{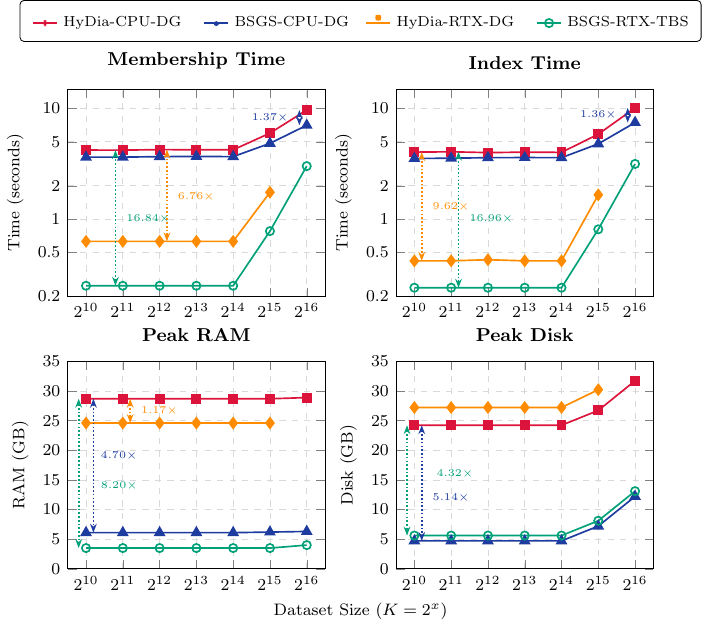}
  \caption{Original HyDia vs. our BSGS-Diagonal optimization on CPU and GPU, for membership and index scenarios: execution time, peak RAM, and Disk usage for $2^{10}$--$2^{16}$ databases (averages over 11 repetitions) and comparison depth $\compdepth=8$. GPU experiments use an NVIDIA Quadro RTX~8000.}
\label{fig:gpu}
\end{figure}

\paragraph{Performance improvement}
Figure~\ref{fig:gpu} illustrates the improvements obtained by adapting HyDia's original code and our new algorithm from CPU to GPU.
We improved \ApproachA by up to $6.8\times$ and $9.7\times$ for the membership and index scenarios, respectively, and \ApproachB by up to $17\times$ in both scenarios. 
The improvement for \ApproachB is more significant than for \ApproachA in approximately $2\times$ because BSGS requires far fewer rotation keys ($44$ vs.\ $511$), reducing the amount of key material residing in GPU memory and freeing more resources for parallel computation.
However, these peak speedups are observed at smaller databases ($K \leq 2^{14}$): at $K{=}2^{15}$, \ApproachA's membership speedup drops to $3.4\times$ and \ApproachB's to $7.7\times$; at $K{=}2^{16}$ \ApproachB's speedup decreases further to $3.2\times$, as larger databases increase GPU memory pressure and reduce the available parallelism.

This GPU-based improvement enables sub-second performance for both \ApproachA and \ApproachB on databases up to $2^{14}$, and sub-second performance for \ApproachB alone up to $2^{15}$.
Regarding peak RAM usage, \ApproachB reduces memory consumption by up to $8.2\times$ compared to the HyDia-CPU baseline (\ApproachBprime further reduces it to $11.0\times$), whereas \ApproachA yields a modest $1.17\times$ reduction. For peak disk usage, \ApproachBprime achieves up to $5.14\times$ less disk compared to HyDia-CPU (and \ApproachB up to $4.32\times$), whereas \ApproachA actually increases disk usage by ${\sim}12\%$ due to additional serialized key material from the GPU library.
Using GPU, we were only able to run \ApproachA and \ApproachB for databases of sizes up to $2^{15}$ and $2^{16}$, respectively, before running into out-of-GPU-memory errors caused by the large number of rotation keys and ciphertext intermediates, exceeding the available $48$\,GB of VRAM on the RTX~8000.
The full benchmark results for these experiments 
are presented in~\ref{appendix:cpu_gpu_depth8_table}, Tables~\ref{tab:benchmark-results-cpu-gpu-depth8-part1} and~\ref{tab:benchmark-results-cpu-gpu-depth8}.

\paragraph{Cost of the key reduction}
The GPU memory savings of \ApproachB{} and \ApproachBprime{}
carry three costs. First, the per-query transfer rises from ${\sim}5$\,MB to
${\sim}115$\,MB, as the baby-step rotations are computed on CPU to avoid
uploading the corresponding keys. Since both strategies have comparable
end-to-end latency, this leads to bandwidth rather than time costs. Second,
\ApproachB{} performs $n_2 - 1$ on-demand giant-step rotations per matrix
group, whereas \ApproachA{} performs none. Third, \ApproachBprime{} shifts the diagonal pre-rotation to the Enroller, incurring ${\sim}8\%$ enrollment
overhead and a one-time NTT conversion penalty at first use, estimated at
$20$--$25$\,ms per matrix group. Setup and enrollment costs for all variants
are reported in Table~\ref{tab:setup-costs}, complementing Table~\ref{tab:gpu-bsgs-complexity}'s per-query cost summaries.

\subsection{Accuracy evaluation on the synthetic datasets.}\label{sec:acc_synth}

\begin{figure}[tb]
  \centering
  \includegraphics[width=\columnwidth]{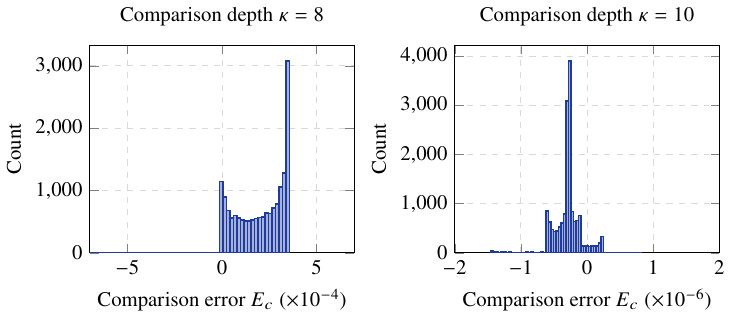}
  \caption{Error $E_c$ observed when using depths $\kappa = 8$ and $10$ for the homomorphic threshold comparison.}
  \label{fig:comparison_error_histograms}
\end{figure}

Following the analysis from Section~\ref{sec:noise}, we used our 128-bit security parameters with comparison depth $\kappa = 8$ and a sample of size $2^{14}$ to empirically evaluate the homomorphic comparison error $E_c$ (see Section~\ref{sec:noise}), finding a mean of $-2^{-21.71}$ and variance $2^{-44.48}$ (see Figure~\ref{fig:comparison_error_histograms} for the distribution of scores after the homomorphic threshold comparison, for a query with no matches).
Then, under the CLT approximation, we estimate that the probability of observing both false positive and false negative results in membership queries is negligible until $K = 3.4$ million entries.
After $K = 3.5$ million, the accumulated mean drift exceeds the decision reliability margin,
resulting in 100\% FN rate when $m = 1$.
However, as mentioned in Remark~\ref{remark}, this limitation can be avoided by a simple recentering operation.
At $K = 44{,}228$ this drift amounts to $\eta \cdot |\mathbb{E}[E_c]| \approx 0.014$,
far below the margin, consistent with the per-comparison parity reported in the following Section~\ref{sec:accuracy_frgc2}.

\subsection{Accuracy evaluation on FRGC 2.0}\label{sec:accuracy_frgc2}
To validate that our algorithmic restructuring and computational depth reduction preserve recognition quality, we evaluate the per-comparison classification accuracy of both HyDia-CPU and \cpubsgs{} on the FRGC~2.0 dataset, sending $50$ probe queries toward the full enrolled database of $44{,}228$ embeddings. Each of the $50\times 44,228=2,211,400$ query–entry pairs is classified independently, so the results characterize the per-entry (identification) decision.
For each approach, we test both comparison depths $\compdepth = 10$ and $\compdepth = 8$, and compare the encrypted evaluation against the corresponding plaintext baseline.
Table~\ref{tab:frgc2-accuracy} summarizes the results.

\begin{table*}[t]
\centering
\caption{Per-comparison accuracy on FRGC~2.0 ($50$ queries against $44{,}228$
enrolled entries, leading to $2{,}211{,}400$ pairwise comparisons). Counts are computed per query--entry pair and characterize the per-entry (identification) decision. All four approach--depth configurations produce
identical classification counts. ``Time/q'' is the amortized wall-clock time per query.}
\label{tab:frgc2-accuracy}
\renewcommand{\arraystretch}{1.15}
\setlength{\tabcolsep}{4pt}
\begin{tabular}{llc rrrr cccc c}
\toprule
{Approach} & {Mode} & $\boldsymbol{\compdepth}$ & {TP} & {FP} & {TN} & {FN} & {Prec.} & {Recall} & {F1} & {Acc.} & {Time/q} \\
\midrule
\multirow{2}{*}{HyDia-CPU}
  & Encrypted & 10 & 6,921 & 3 & 2,204,434 & 42 & 0.9996 & 0.9940 & 0.9968 & 0.999980 & \multirow{2}{*}{10.34\,s} \\
  & Plaintext & 10 & 6,921 & 2 & 2,204,435 & 42 & 0.9997 & 0.9940 & 0.9968 & 0.999980 & \\
\midrule
\multirow{2}{*}{HyDia-CPU}
  & Encrypted & 8  & 6,921 & 3 & 2,204,434 & 42 & 0.9996 & 0.9940 & 0.9968 & 0.999980 & \multirow{2}{*}{9.03\,s} \\
  & Plaintext & 8  & 6,921 & 2 & 2,204,435 & 42 & 0.9997 & 0.9940 & 0.9968 & 0.999980 & \\
\midrule
\multirow{2}{*}{\cpubsgs}
  & Encrypted & 10 & 6,921 & 3 & 2,204,434 & 42 & 0.9996 & 0.9940 & 0.9968 & 0.999980 & \multirow{2}{*}{8.39\,s} \\
  & Plaintext & 10 & 6,921 & 2 & 2,204,435 & 42 & 0.9997 & 0.9940 & 0.9968 & 0.999980 & \\
\midrule
\multirow{2}{*}{\cpubsgs}
  & Encrypted & 8  & 6,921 & 3 & 2,204,434 & 42 & 0.9996 & 0.9940 & 0.9968 & 0.999980 & \multirow{2}{*}{7.13\,s} \\
  & Plaintext & 8  & 6,921 & 2 & 2,204,435 & 42 & 0.9997 & 0.9940 & 0.9968 & 0.999980 & \\
\bottomrule
\end{tabular}
\end{table*}

Several observations follow from these results.
First, all four approach--depth configurations yield \emph{identical} classification counts ($\text{TP} = 6{,}921$, $\text{FN} = 42$, $\text{TN} = 2{,}204{,}434$, $\text{FP} = 3$ for encrypted evaluation), confirming that the BSGS restructuring introduces no accuracy degradation compared to the original HyDia diagonal method.
Second, reducing the Chebyshev comparison depth from $\compdepth = 10$ to $\compdepth = 8$ has no measurable effect on per-comparison classification accuracy, 
but yields a noticeable speedup (e.g., from $10.34$\,s to $9.03$\,s per query for HyDia-CPU, and from $8.39$\,s to $7.13$\,s for \cpubsgs).
This supports our choice of $\compdepth = 8$ in the GPU experiments.
Third, the encrypted evaluation closely follows the plaintext baseline: the only discrepancy is one additional false positive ($\text{FP} = 3$ vs.\ $2$), compensated by one fewer true negative, meaning at most one flipped decision out of $2{,}211{,}400$ comparisons. This discrepancy does not come from FHE noise, though, but from the polynomial approximation of the threshold comparison near the decision boundary: although the $\compdepth = 8$ and $\compdepth = 10$ configurations differ in noise magnitude, they flip the same single comparison.
Because this is a per-comparison effect, it is independent of database size.
Finally, both methods achieve per-comparison precision $> 99.95\%$, recall~$\approx 99.40\%$, $\text{F1} > 99.67\%$, and accuracy $> 99.998\%$, demonstrating that the encrypted pipeline preserves the recognition quality of the plaintext system on a standard biometric benchmark.

 \subsection{H200 Results}\label{sec:h200}

\begin{figure}[t]
  \centering
  \includegraphics[width=\columnwidth]{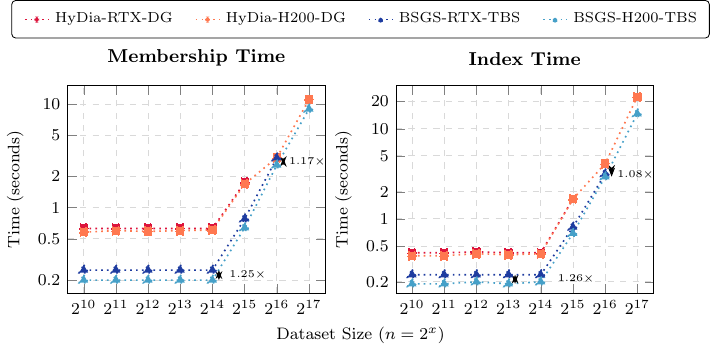}
  \caption{RTX~8000 vs.\ H200 comparison for membership time (left) and index time (right), $\compdepth{=}8$. H200 extends GPU coverage to $K{=}2^{17}$ and provides modest speedups ($1.08$--$1.25\times$) over the RTX~8000 at overlapping database sizes.}
\label{fig:rtxvsh200}
\end{figure}

The NVIDIA H200 (141\,GB HBM3e), with roughly $3\times$ more memory than the RTX~8000 (48\,GB), enables additional GPU experiments for larger databases.
Figure~\ref{fig:rtxvsh200} compares the RTX~8000 and H200 for HyDia-RTX-DG vs.\ HyDia-H200-DG and BSGS-RTX-TBS vs.\ BSGS-H200-TBS.

\paragraph{Extended database coverage}
H200 supports GPU experiments from $K{=}2^{15}$ to $K{=}2^{17}$ for HyDia and from $K{=}2^{16}$ to $K{=}2^{17}$ for BSGS, before out-of-memory errors are encountered.
At $K{=}2^{17}$, BSGS-H200-TBS achieves $8.93$\,s membership and $14.66$\,s index times, while HyDia-H200-DG requires $11.05$\,s and $22.66$\,s respectively.
This represents a $1.24\times$ and $1.54\times$ advantage for BSGS over HyDia.

\paragraph{RTX vs.\ H200 speedup}
For databases up to $K{=}2^{14}$, H200 offers a modest $1.25\times$ speedup over the RTX~8000 for BSGS membership (Figure~\ref{fig:rtxvsh200}, left), as computation at these sizes is not memory-bound.
At $K{=}2^{16}$ (the largest overlapping point for BSGS), H200 reduces membership time from $3.02$\,s to $2.58$\,s ($1.17\times$) and index time from $3.16$\,s to $2.93$\,s ($1.08\times$).
The speedup narrows at larger $K$ because the computation becomes increasingly dominated by arithmetic operations rather than memory transfers.

\paragraph{TBS vs.\ TBE on H200}
Tables~\ref{tab:benchmark-results-cpu-gpu-depth8-part1} and~\ref{tab:benchmark-results-cpu-gpu-depth8} include both BSGS-H200-TBS (server-side pre-rotation) and BSGS-H200-TBE (enroller-side pre-rotation).
For $K \leq 2^{16}$, TBE offers a lower overall RTT than TBS (e.g., $33.2$\,s vs.\ $35.9$\,s at $K{=}2^{16}$) and lower RAM ($3.5$\,GB vs.\ $3.9$\,GB), because the pre-rotation cost is shifted to enrollment.
At $K{=}2^{17}$, however, TBS achieves faster membership ($8.93$\,s vs.\ $9.73$\,s for TBE) as the server-side pre-rotation strategy benefits from H200's higher memory bandwidth at larger working sets.
Both variants maintain the same RAM and disk advantages over HyDia-H200-DG.

\subsection{Scalability}
\begin{figure}[t]
  \centering
  \includegraphics[width=\columnwidth]{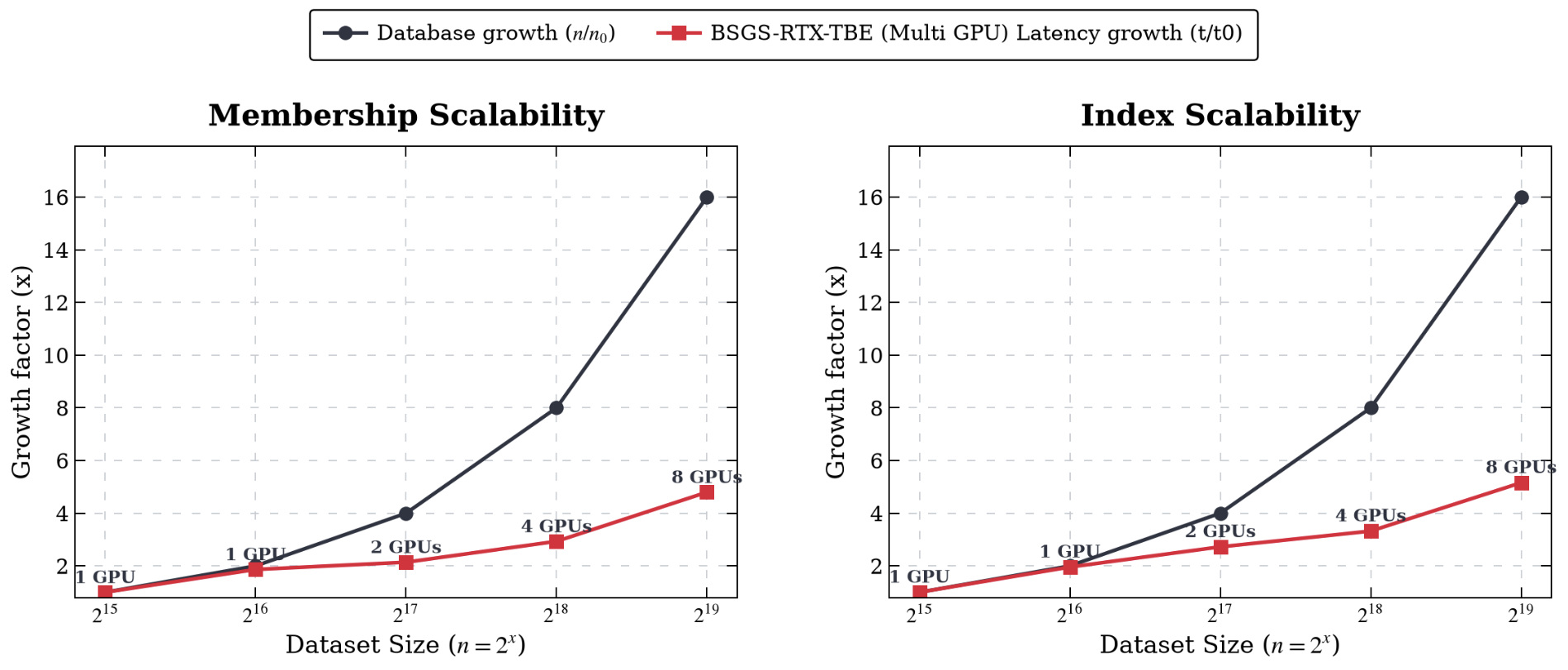}
\caption{\textbf{Scalability of BSGS-RTX-TBE (Multi-GPU).}
For each dataset size $n$, the database growth factor is computed as $G_n = \frac{n}{n_0}$, where $n_0$ is the smallest evaluated dataset.
Similarly, the latency growth factor is computed as $G_t = \frac{t(n)}{t(n_0)}$,
for both membership and index operations.
The method is considered scalable when time growth remains controlled with respect to data growth, i.e., when $G_t < G_n$ (or approximately sublinear), indicating that latency increases more slowly than the dataset size.}\label{fig:multigpu}
\end{figure}

To evaluate scalability, we implemented a conventional sharding strategy for our GPU strategies, HyDia-GPU, BSGS-RTX-TBS and BSGS-RTX-TBE, partitioning the database across multiple GPUs and processing the corresponding shards in parallel. The complete implementation is publicly available to facilitate independent validation and reproduction\footnote{Specifically, these optimizations were introduced in release v2.2.0, available at \url{https://github.com/FastHE-Search/FastHE-Search/releases/tag/v2.2.0}}. We used an AWS g6e.48xlarge instance equipped with eight GPUs, scaling the database size from $2^{16}$ to $2^{19}$ entries,  varying the number of GPUs from 1 to 2, 4, and 8, respectively. As the database size increases, both indexing and membership-query times exhibit sublinear growth, indicating that doubling the database size incurs only a small increment in execution time. For example, index search on a database of $2^{19}$ entries can be done in $0.2231 \pm 0.0116$~s for the BSGS-RTX-TBE approach, with $k_{\mathrm{match}}=256$. Compared to the CPU baselines for the same database size (Table~\ref{tab:benchmark-results-cpu-gpu-depth8}), this corresponds to up to $287\times$ and $211\times$ speedups over HyDia-CPU-DG and BSGS-CPU-DG, respectively, in the index scenario. Note that AWS GPU clusters have up to 8 GPUs per host node. To tackle databases of $2^{20}$ entries and above, a multi-node GPU architecture solution is necessary and we leave this task for future work. 
The scalability behavior of BSGS-RTX-TBE is illustrated in Figure~\ref{fig:multigpu}.

\section{Conclusion}
\label{sec:conclusion}
We improve the practicality of FHE-based encrypted facial matching in realistic client-server settings where both the database and queries are encrypted. Building upon HyDia, our BSGS-Diagonal algorithm substantially reduces rotation-key requirements and peak memory while slightly improving runtime, making the protocol better suited for resource-constrained clients. From a systems perspective, we design a GPU-resident pipeline that integrates OpenFHE with FIDESlib, adds GPU-native kernels for encrypted similarity evaluation and Chebyshev comparison, and minimizes host-device transfers. The combination yields up to $9\times$ speedup for HyDia and $21\times$ for our GPU BSGS variant (BSGS-RTX-TBS/TBE), bringing encrypted face recognition into the sub-second regime for databases having up to $2^{15}$ entries. To our knowledge, the proposed FIDESlib-based design is the first fully open-source GPU implementation supporting end-to-end encrypted similarity computation.

\appendix
\newpage
\section{Operation count for the diagonal method vs textbook BSGS}\label{appendix:comparison_tables}
\begin{table}[H]
\centering
\scriptsize
\begin{tabular}{@{}lcc@{}}
\toprule
Dimension & Diagonal & BSGS \\
\midrule
Multiplications & $\ell$ & $\ell$ \\
Additions & $\ell-1$ & $\ell-1$ \\
Rotations & $\ell-1$ & $(n_1-1)+(n_2-1)$ \\
Rotation keys & $\ell-1$  & $(n_1-1)+(n_2-1)$ \\
Relinearization keys & $\ell$ & $\ell$ \\
\hline
Asymptotic rotations &
$\Theta(\ell)$ &
$\Theta(n_1+n_2)=\Theta(\sqrt{\ell})$\\
Overall asymptotic cost & $\Theta(\ell)$ & $\Theta(\ell)$ \\Memory for keys (rot + relin) & $\Theta(\ell)$ & $\Theta(n_1+n_2)=\Theta(\sqrt{\ell})$ \\
\bottomrule
\end{tabular}
\caption{Asymptotic operation count and memory usage for diagonal encoding vs BSGS. We assume $n_2=\lceil \ell/n_1\rceil$.}
\label{tab:diag_vs_bsgs}
\end{table}

\begin{table}[h]
\centering
\scriptsize
\setlength{\tabcolsep}{3pt} 
\begin{tabular}{|c|c|cc|cc|}
\hline
\multirow{2}{*}{\textbf{DB size}} & \multirow{2}{*}{$G$} & \multicolumn{2}{c|}{\textbf{Total \# of Rotations}} & \multicolumn{2}{c|}{\textbf{Memory (\# of Rotated CTs)}} \\ \cline{3-6} 
 & & \multicolumn{1}{c|}{\textbf{\shortstack{Textbook\\BSGS}}} & \textbf{BSGS-Diagonal} & \multicolumn{1}{c|}{\textbf{\shortstack{Textbook\\BSGS}}} & \textbf{BSGS-Diagonal} \\ \hline
 $2^{14}$ & 1  & \multicolumn{1}{c|}{44}   & 511 & \multicolumn{1}{c|}{$23$} & $512$ \\ \hline
$2^{16}$ & 4  & \multicolumn{1}{c|}{110}  & 511 & \multicolumn{1}{c|}{$23$} & $512$ \\ \hline
$2^{18}$ & 16 & \multicolumn{1}{c|}{374}  & 511 & \multicolumn{1}{c|}{$23$} & $512$ \\ \hline
$2^{19}$ & 32 & \multicolumn{1}{c|}{726}  & 511 & \multicolumn{1}{c|}{$23$} & $512$ \\ \hline
$2^{20}$ & 64 & \multicolumn{1}{c|}{1430} & 511 & \multicolumn{1}{c|}{$23$} & $512$ \\ \hline
\end{tabular}
\caption{Comparison of rotations performed and memory usage (number of rotated query ciphertexts simultaneously stored) between single-hoisted textbook BSGS and BSGS-Diagonal for varying database sizes and $n_1 = n_2 = 23$.}
\label{tab:bsgs-diagonal-comparison}
\end{table}

\section{HyDia~\cite{martin2025hydia} pseudo-codes}\label{appendix:pseudo_code}
\begin{algorithm}[H]
\caption{Enroller (HyDia original):}
\label{alg:enroller-hydia}
\begin{algorithmic}[1]
\scriptsize
\Require Database $\mathcal{D}=\{x_i\in\mathbb{R}^{\ell}\}_{i=1}^{K}$, public key \texttt{pk}, crypto context \texttt{cc}
\Statex

\State \textbf{Normalize database (plaintext):}
\For{$i=1$ to $\texttt{numVectors}$}
  \State $x_i \gets \mathrm{Normalize}(x_i,\ \ell)$
\EndFor
\Statex

\State \textbf{Build $\ell\times \ell$ squares and extract diagonals (plaintext):}
\State $\{S^{(g)}\}\gets \mathrm{SplitIntoSquareMatrices}(\mathcal{D},\,\ell)$ \Comment $S^{(g)}\in\mathbb{R}^{\ell\times \ell}$
\For{each group $g$}
  \State $\mathrm{Diags}^{(g)} \gets \mathrm{Diagonals}(S^{(g)})$ \Comment list of $\ell$ diagonals, each length $\ell$
\EndFor
\Statex

\State \textbf{Concatenate diagonals into SIMD-sized rows:}
\State $\{\text{row}_i\}\gets \mathrm{ConcatenateRows}\!\left(\{\mathrm{Diags}^{(g)}\}\right)$
\Comment packs diagonals linearly into plaintext rows
\Statex

\State \textbf{Parallel encryption and async I/O:}
\For{$i = 0$ to $|\{\text{row}_i\}|-1$} \Comment loop is parallelized in implementation
  \State $ct \gets \mathrm{Encrypt}(\texttt{pk},\ \text{row}_i)$
  \State $\texttt{path} \gets \text{``serial/db\_diagonal/index''} + i + \text{``.bin''}$
  \State $\mathrm{WriteBinaryAsync}(ct,\ \texttt{path})$
\EndFor
\end{algorithmic}
\end{algorithm}

\begin{algorithm}[htbp]
\caption{\textsc{ComputeSimilarity}}\label{alg:compute-similarity}
\begin{algorithmic}[1]
\scriptsize
\Require Encrypted query $\vec{q}$,
        vector dimension $\ell$,
        number of database vectors $K$,
        batch size $\numslots$, cyclotomic order $\Omega$.

\State $G \gets \left\lceil \dfrac{K}{\numslots} \right\rceil$
      \Comment number of packed matrices / groups
\State $\mathsf{similarityCipher} \gets [\,]$

\State \textbf{Precompute for fast hoisted rotations of the query ciphertext}
\State $\mathsf{rotatedQueryCipher} \gets [\,]$
\State $\mathsf{rotatedQueryCipher}[0] \gets \vec{q}[0]$
\State $\mathsf{queryPrecomp} \gets \textsc{EvalFastRotationPrecompute}(\vec{q}[0])$
\Statex

\For {$1 \le i \le \ell-1$} \Comment can be parallelized
    \State \hspace{1em} $\mathsf{rotatedQueryCipher}[i] \gets
    \textsc{EvalFastRotation}(\vec{q}[0],\, i,\, \Omega,\, \mathsf{queryPrecomp})$
    \Comment rotate packed query by $i$ slots
\EndFor
\Statex

\For {$0 \le m \le G-1$}
    \State \hspace{1em} $\mathsf{similarityCipher}[m] \gets \textsc{ComputeSimilarityMatrix}(\mathsf{rotatedQueryCipher},\, m)$
    \Comment compute similarity scores for DB block $m$ (packed in $\numslots$ slots)
\EndFor
\Statex

\State \Return $\mathsf{similarityCipher}$
\end{algorithmic}
\end{algorithm}

\begin{algorithm}[htbp]
\caption{Membership Scenario}\label{alg:membership}
\begin{algorithmic}[1]
\scriptsize
\Require Encrypted query $\vec{q}$, threshold $\tau$,
         comparator depth $\kappa$, batch size $\numslots$.

\State $\mathsf{scoreCipher} \gets \textsc{ComputeSimilarity}(\vec{q})$
      \Comment vector of group ciphertexts, each packing scores in $\numslots$ slots

\For {$0 \leq g \leq |\mathsf{scoreCipher}|{-}1$}
    \State \hspace{1em} $\mathsf{scoreCipher}[g] \gets
    \textsc{ChebyshevCompare}(\mathsf{scoreCipher}[g],\, \tau,\, \kappa)$
    \Comment $\approx 1$ if score $\ge \tau$, else $\approx 0$ 
\EndFor
\State $\mathsf{membershipCipher} \gets \textsc{EvalAddManyInPlace}(\mathsf{scoreCipher})$
      \Comment sum all group ciphertexts into a single ciphertext

\State $\mathsf{membershipCipher} \gets \textsc{EvalSum}(\mathsf{membershipCipher},\, \numslots)$
      \Comment sum all $\numslots$ slots into slot 0

\State \Return $\mathsf{membershipCipher}$
      \Comment slot 0 encodes the total count of matches across all groups/slots
\end{algorithmic}
\end{algorithm}

\begin{algorithm}[htbp]
\caption{Index Scenario}\label{alg:index}
\begin{algorithmic}[1]
\scriptsize
\Require Encrypted query $\vec{q}$, threshold $\tau$,
         comparator depth $\kappa$.

\State $\mathsf{scoreCipher} \gets \textsc{ComputeSimilarity}(\vec{q})$
      \Comment vector of group ciphertexts, each packing scores in $\numslots$ blocks

\For {$0 \leq g \leq |\mathsf{scoreCipher}|{-}1\}$} 
    \State \hspace{1em} $\mathsf{scoreCipher}[g] \gets \textsc{ChebyshevCompare}(\mathsf{scoreCipher}[g],\, \tau,\, \kappa)$
    \Comment $\approx 1$ if score $\ge \tau$, else $\approx 0$
\EndFor
\Statex

\State \Return $\mathsf{scoreCipher}$
\end{algorithmic}
\end{algorithm}

\section{Full Benchmark Results}\label{appendix:benchmarks}
Tables~\ref{tab:benchmark-results-cpu-only-extended}, \ref{tab:benchmark-results-cpu-gpu-depth8-part1}, and~\ref{tab:benchmark-results-cpu-gpu-depth8} report the following per-configuration metrics.

Variant names follow the convention \emph{Method-Platform-Variant}: DG denotes the diagonal method, TBS denotes BSGS with server-side diagonal pre-rotation, and TBE denotes BSGS with enroller-side pre-rotation. Platforms are CPU, RTX (NVIDIA Quadro RTX 8000), or H200 (NVIDIA H200).

\emph{Metrics.}\quad $K$ is the database size, $K_m$ the number of planted matches. \textbf{RTT\,(s)} is overall wall-clock time, computed as $\text{enroll}(\text{trial\,1}) + \text{mean}(\text{wall time},\,\text{last-}N)$ over $N{=}10$ steady-state trials, covering key deserialization, query encryption, server-side computation, and decryption. \textbf{Membership\,(s)} and \textbf{Index\,(s)} are server-side homomorphic computation times for the two scenarios, reported as mean $\pm$ std over last-$N$. \textbf{RAM\,(GB)} is peak RSS of the process, mean $\pm$ std over last-$N$. \textbf{Disk\,(GB)} is the total size of the serialized cryptographic material (keys, database diagonals, intermediate ciphertexts) and is deterministic across trials ($\sigma = 0$). All timings exclude network transfer and client-side encryption/decryption.
\subsection{CPU-only benchmarks with BSGS-Diagonal}
Table~\ref{tab:benchmark-results-cpu-only-extended} reports our CPU-only benchmark results with $\compdepth=10$, \emph{i.e.,} with total multiplicative depth equal to 11 (as in the original HyDia work). We compare BSGS-CPU-DG against HyDia-CPU-DG, HERS~\cite{engelsma2022hers}, GROTE~\cite{ibarrondo2023grote}, Blind-Match~\cite{choi2024blind}), where BSGS-CPU-DG corresponds to BSGS-Diagonal.

\begin{table*}[htbp]
\centering
\caption{
CPU-only benchmark results with $\compdepth=10$. We compare HyDia-CPU-DG against BSGS-CPU-DG, where BSGS-CPU-DG corresponds to \cpubsgs. Values are reported as mean $\pm$ std over 11 trials. The parenthetical value in \textcolor{gray}{gray} is the ratio against HyDia-CPU-DG as the baseline.}

\label{tab:benchmark-results-cpu-only-extended}
\scriptsize
\setlength{\tabcolsep}{3.5pt}
\resizebox{\textwidth}{!}{%
\begin{tabular}{rrlrrrrr}
\toprule
$K$ & $K_m$ & Approach & RTT (s) & Membership (s) & Index (s) & RAM (GB) & Disk (GB) \\
\midrule
\multirow{5}{*}{$2^{10}$} & \multirow{5}{*}{16} & HyDia-CPU-DG & 30.60 {\scriptsize\textcolor{gray}{(1.00$\times$)}} & 5.67 $\pm$ 0.08 {\scriptsize\textcolor{gray}{(1.00$\times$)}} & 5.67 $\pm$ 0.24 {\scriptsize\textcolor{gray}{(1.00$\times$)}} & 33.3 {\scriptsize\textcolor{gray}{(1.00$\times$)}} & 27.84 {\scriptsize\textcolor{gray}{(1.00$\times$)}} \\
 & & BSGS-CPU-DG & 13.10 {\scriptsize\textcolor{gray}{(2.34$\times$)}} & 4.92 $\pm$ 0.03 {\scriptsize\textcolor{gray}{(1.15$\times$)}} & 4.76 $\pm$ 0.09 {\scriptsize\textcolor{gray}{(1.19$\times$)}} & 7.2 {\scriptsize\textcolor{gray}{(4.62$\times$)}} & 5.52 {\scriptsize\textcolor{gray}{(5.04$\times$)}} \\
 & & GROTE & 137.10 {\scriptsize\textcolor{gray}{(0.22$\times$)}} & 25.59 $\pm$ 0.20 {\scriptsize\textcolor{gray}{(0.22$\times$)}} & 30.01 $\pm$ 0.70 {\scriptsize\textcolor{gray}{(0.19$\times$)}} & 80.57 {\scriptsize\textcolor{gray}{(0.41$\times$)}} & 78.19 {\scriptsize\textcolor{gray}{(0.36$\times$)}} \\
 & & Blind-Match & 69.12 {\scriptsize\textcolor{gray}{(0.44$\times$)}} & 9.21 $\pm$ 0.03 {\scriptsize\textcolor{gray}{(0.62$\times$)}} & 9.10 $\pm$ 0.14 {\scriptsize\textcolor{gray}{(0.62$\times$)}} & 53.35 {\scriptsize\textcolor{gray}{(0.62$\times$)}} & 53.17 {\scriptsize\textcolor{gray}{(0.52$\times$)}} \\
 & & HERS & 40.03 {\scriptsize\textcolor{gray}{(0.76$\times$)}} & 7.23 $\pm$ 0.24 {\scriptsize\textcolor{gray}{(0.78$\times$)}} & 7.29 $\pm$ 0.41 {\scriptsize\textcolor{gray}{(0.78$\times$)}} & 30.97 {\scriptsize\textcolor{gray}{(1.08$\times$)}} & 27.84 {\scriptsize\textcolor{gray}{(1.00$\times$)}} \\
\midrule
\multirow{5}{*}{$2^{11}$} & \multirow{5}{*}{16} & HyDia-CPU-DG & 30.60 {\scriptsize\textcolor{gray}{(1.00$\times$)}} & 5.72 $\pm$ 0.07 {\scriptsize\textcolor{gray}{(1.00$\times$)}} & 5.49 $\pm$ 0.07 {\scriptsize\textcolor{gray}{(1.00$\times$)}} & 33.3 {\scriptsize\textcolor{gray}{(1.00$\times$)}} & 27.84 {\scriptsize\textcolor{gray}{(1.00$\times$)}} \\
 & & BSGS-CPU-DG & 13.30 {\scriptsize\textcolor{gray}{(2.30$\times$)}} & 4.97 $\pm$ 0.07 {\scriptsize\textcolor{gray}{(1.15$\times$)}} & 4.83 $\pm$ 0.06 {\scriptsize\textcolor{gray}{(1.14$\times$)}} & 7.2 {\scriptsize\textcolor{gray}{(4.62$\times$)}} & 5.52 {\scriptsize\textcolor{gray}{(5.04$\times$)}} \\
 & & GROTE & 178.33 {\scriptsize\textcolor{gray}{(0.17$\times$)}} & 43.22 $\pm$ 0.97 {\scriptsize\textcolor{gray}{(0.13$\times$)}} & 48.83 $\pm$ 1.00 {\scriptsize\textcolor{gray}{(0.11$\times$)}} & 81.96 {\scriptsize\textcolor{gray}{(0.41$\times$)}} & 78.49 {\scriptsize\textcolor{gray}{(0.35$\times$)}} \\
 & & Blind-Match & 81.94 {\scriptsize\textcolor{gray}{(0.37$\times$)}} & 14.77 $\pm$ 0.07 {\scriptsize\textcolor{gray}{(0.39$\times$)}} & 14.40 $\pm$ 0.24 {\scriptsize\textcolor{gray}{(0.38$\times$)}} & 53.35 {\scriptsize\textcolor{gray}{(0.62$\times$)}} & 53.38 {\scriptsize\textcolor{gray}{(0.52$\times$)}} \\
 & & HERS & 40.26 {\scriptsize\textcolor{gray}{(0.76$\times$)}} & 7.40 $\pm$ 0.04 {\scriptsize\textcolor{gray}{(0.77$\times$)}} & 7.60 $\pm$ 0.31 {\scriptsize\textcolor{gray}{(0.72$\times$)}} & 30.93 {\scriptsize\textcolor{gray}{(1.08$\times$)}} & 27.84 {\scriptsize\textcolor{gray}{(1.00$\times$)}} \\
\midrule
\multirow{5}{*}{$2^{12}$} & \multirow{5}{*}{32} & HyDia-CPU-DG & 30.70 {\scriptsize\textcolor{gray}{(1.00$\times$)}} & 5.74 $\pm$ 0.12 {\scriptsize\textcolor{gray}{(1.00$\times$)}} & 5.61 $\pm$ 0.21 {\scriptsize\textcolor{gray}{(1.00$\times$)}} & 33.3 {\scriptsize\textcolor{gray}{(1.00$\times$)}} & 27.84 {\scriptsize\textcolor{gray}{(1.00$\times$)}} \\
 & & BSGS-CPU-DG & 13.30 {\scriptsize\textcolor{gray}{(2.31$\times$)}} & 4.97 $\pm$ 0.06 {\scriptsize\textcolor{gray}{(1.15$\times$)}} & 4.91 $\pm$ 0.31 {\scriptsize\textcolor{gray}{(1.14$\times$)}} & 7.2 {\scriptsize\textcolor{gray}{(4.62$\times$)}} & 5.52 {\scriptsize\textcolor{gray}{(5.04$\times$)}} \\
 & & GROTE & 231.62 {\scriptsize\textcolor{gray}{(0.13$\times$)}} & 72.36 $\pm$ 0.45 {\scriptsize\textcolor{gray}{(0.08$\times$)}} & 77.55 $\pm$ 0.48 {\scriptsize\textcolor{gray}{(0.07$\times$)}} & 82.14 {\scriptsize\textcolor{gray}{(0.41$\times$)}} & 79.08 {\scriptsize\textcolor{gray}{(0.35$\times$)}} \\
 & & Blind-Match & 102.75 {\scriptsize\textcolor{gray}{(0.30$\times$)}} & 26.21 $\pm$ 0.85 {\scriptsize\textcolor{gray}{(0.22$\times$)}} & 26.28 $\pm$ 0.99 {\scriptsize\textcolor{gray}{(0.21$\times$)}} & 53.41 {\scriptsize\textcolor{gray}{(0.62$\times$)}} & 53.78 {\scriptsize\textcolor{gray}{(0.52$\times$)}} \\
 & & HERS & 39.99 {\scriptsize\textcolor{gray}{(0.77$\times$)}} & 7.27 $\pm$ 0.02 {\scriptsize\textcolor{gray}{(0.79$\times$)}} & 7.64 $\pm$ 0.46 {\scriptsize\textcolor{gray}{(0.73$\times$)}} & 30.95 {\scriptsize\textcolor{gray}{(1.08$\times$)}} & 27.84 {\scriptsize\textcolor{gray}{(1.00$\times$)}} \\
\midrule
\multirow{5}{*}{$2^{13}$} & \multirow{5}{*}{32} & HyDia-CPU-DG & 30.70 {\scriptsize\textcolor{gray}{(1.00$\times$)}} & 5.72 $\pm$ 0.09 {\scriptsize\textcolor{gray}{(1.00$\times$)}} & 5.56 $\pm$ 0.12 {\scriptsize\textcolor{gray}{(1.00$\times$)}} & 33.3 {\scriptsize\textcolor{gray}{(1.00$\times$)}} & 27.84 {\scriptsize\textcolor{gray}{(1.00$\times$)}} \\
 & & BSGS-CPU-DG & 13.30 {\scriptsize\textcolor{gray}{(2.31$\times$)}} & 4.98 $\pm$ 0.03 {\scriptsize\textcolor{gray}{(1.15$\times$)}} & 4.82 $\pm$ 0.06 {\scriptsize\textcolor{gray}{(1.15$\times$)}} & 7.2 {\scriptsize\textcolor{gray}{(4.62$\times$)}} & 5.52 {\scriptsize\textcolor{gray}{(5.04$\times$)}} \\
 & & GROTE & 346.43 {\scriptsize\textcolor{gray}{(0.09$\times$)}} & 133.21 $\pm$ 4.05 {\scriptsize\textcolor{gray}{(0.04$\times$)}} & 136.18 $\pm$ 1.15 {\scriptsize\textcolor{gray}{(0.04$\times$)}} & 83.22 {\scriptsize\textcolor{gray}{(0.40$\times$)}} & 80.27 {\scriptsize\textcolor{gray}{(0.35$\times$)}} \\
 & & Blind-Match & 150.22 {\scriptsize\textcolor{gray}{(0.20$\times$)}} & 50.19 $\pm$ 0.44 {\scriptsize\textcolor{gray}{(0.11$\times$)}} & 49.45 $\pm$ 0.07 {\scriptsize\textcolor{gray}{(0.11$\times$)}} & 53.56 {\scriptsize\textcolor{gray}{(0.62$\times$)}} & 54.59 {\scriptsize\textcolor{gray}{(0.51$\times$)}} \\
 & & HERS & 41.09 {\scriptsize\textcolor{gray}{(0.75$\times$)}} & 7.15 $\pm$ 0.02 {\scriptsize\textcolor{gray}{(0.80$\times$)}} & 7.24 $\pm$ 0.00 {\scriptsize\textcolor{gray}{(0.77$\times$)}} & 30.93 {\scriptsize\textcolor{gray}{(1.08$\times$)}} & 27.84 {\scriptsize\textcolor{gray}{(1.00$\times$)}} \\
\midrule
\multirow{5}{*}{$2^{14}$} & \multirow{5}{*}{64} & HyDia-CPU-DG & 30.90 {\scriptsize\textcolor{gray}{(1.00$\times$)}} & 5.68 $\pm$ 0.08 {\scriptsize\textcolor{gray}{(1.00$\times$)}} & 5.52 $\pm$ 0.04 {\scriptsize\textcolor{gray}{(1.00$\times$)}} & 33.4 {\scriptsize\textcolor{gray}{(1.00$\times$)}} & 27.84 {\scriptsize\textcolor{gray}{(1.00$\times$)}} \\
 & & BSGS-CPU-DG & 13.30 {\scriptsize\textcolor{gray}{(2.32$\times$)}} & 4.95 $\pm$ 0.03 {\scriptsize\textcolor{gray}{(1.15$\times$)}} & 4.77 $\pm$ 0.04 {\scriptsize\textcolor{gray}{(1.16$\times$)}} & 7.2 {\scriptsize\textcolor{gray}{(4.64$\times$)}} & 5.52 {\scriptsize\textcolor{gray}{(5.04$\times$)}} \\
 & & GROTE & 611.64 {\scriptsize\textcolor{gray}{(0.05$\times$)}} & 262.94 $\pm$ 8.85 {\scriptsize\textcolor{gray}{(0.02$\times$)}} & 267.32 $\pm$ 3.89 {\scriptsize\textcolor{gray}{(0.02$\times$)}} & 85.26 {\scriptsize\textcolor{gray}{(0.39$\times$)}} & 82.65 {\scriptsize\textcolor{gray}{(0.34$\times$)}} \\
 & & Blind-Match & 249.88 {\scriptsize\textcolor{gray}{(0.12$\times$)}} & 97.54 $\pm$ 1.45 {\scriptsize\textcolor{gray}{(0.06$\times$)}} & 98.03 $\pm$ 0.79 {\scriptsize\textcolor{gray}{(0.06$\times$)}} & 53.82 {\scriptsize\textcolor{gray}{(0.62$\times$)}} & 56.22 {\scriptsize\textcolor{gray}{(0.50$\times$)}} \\
 & & HERS & 40.01 {\scriptsize\textcolor{gray}{(0.77$\times$)}} & 7.31 $\pm$ 0.05 {\scriptsize\textcolor{gray}{(0.78$\times$)}} & 7.34 $\pm$ 0.05 {\scriptsize\textcolor{gray}{(0.75$\times$)}} & 30.93 {\scriptsize\textcolor{gray}{(1.08$\times$)}} & 27.84 {\scriptsize\textcolor{gray}{(1.00$\times$)}} \\
\midrule
\multirow{5}{*}{$2^{15}$} & \multirow{5}{*}{64} & HyDia-CPU-DG & 36.50 {\scriptsize\textcolor{gray}{(1.00$\times$)}} & 7.87 $\pm$ 0.09 {\scriptsize\textcolor{gray}{(1.00$\times$)}} & 7.67 $\pm$ 0.24 {\scriptsize\textcolor{gray}{(1.00$\times$)}} & 33.4 {\scriptsize\textcolor{gray}{(1.00$\times$)}} & 30.84 {\scriptsize\textcolor{gray}{(1.00$\times$)}} \\
 & & BSGS-CPU-DG & 18.70 {\scriptsize\textcolor{gray}{(1.95$\times$)}} & 6.34 $\pm$ 0.09 {\scriptsize\textcolor{gray}{(1.24$\times$)}} & 6.73 $\pm$ 0.68 {\scriptsize\textcolor{gray}{(1.14$\times$)}} & 7.3 {\scriptsize\textcolor{gray}{(4.58$\times$)}} & 8.52 {\scriptsize\textcolor{gray}{(3.62$\times$)}} \\
 & & GROTE & 1150.81 {\scriptsize\textcolor{gray}{(0.03$\times$)}} & 539.27 $\pm$ 19.73 {\scriptsize\textcolor{gray}{(0.01$\times$)}} & 527.91 $\pm$ 15.71 {\scriptsize\textcolor{gray}{(0.01$\times$)}} & 89.71 {\scriptsize\textcolor{gray}{(0.37$\times$)}} & 87.40 {\scriptsize\textcolor{gray}{(0.35$\times$)}} \\
 & & Blind-Match & 461.44 {\scriptsize\textcolor{gray}{(0.08$\times$)}} & 201.46 $\pm$ 1.26 {\scriptsize\textcolor{gray}{(0.04$\times$)}} & 200.33 $\pm$ 2.11 {\scriptsize\textcolor{gray}{(0.04$\times$)}} & 54.45 {\scriptsize\textcolor{gray}{(0.61$\times$)}} & 59.47 {\scriptsize\textcolor{gray}{(0.52$\times$)}} \\
 & & HERS & 53.45 {\scriptsize\textcolor{gray}{(0.68$\times$)}} & 12.72 $\pm$ 0.04 {\scriptsize\textcolor{gray}{(0.62$\times$)}} & 12.56 $\pm$ 0.36 {\scriptsize\textcolor{gray}{(0.61$\times$)}} & 30.93 {\scriptsize\textcolor{gray}{(1.08$\times$)}} & 30.84 {\scriptsize\textcolor{gray}{(1.00$\times$)}} \\
\midrule
\multirow{5}{*}{$2^{16}$} & \multirow{5}{*}{128} & HyDia-CPU-DG & 48.40 {\scriptsize\textcolor{gray}{(1.00$\times$)}} & 12.09 $\pm$ 0.29 {\scriptsize\textcolor{gray}{(1.00$\times$)}} & 12.12 $\pm$ 0.38 {\scriptsize\textcolor{gray}{(1.00$\times$)}} & 33.6 {\scriptsize\textcolor{gray}{(1.00$\times$)}} & 36.84 {\scriptsize\textcolor{gray}{(1.00$\times$)}} \\
 & & BSGS-CPU-DG & 27.40 {\scriptsize\textcolor{gray}{(1.77$\times$)}} & 8.94 $\pm$ 0.05 {\scriptsize\textcolor{gray}{(1.35$\times$)}} & 9.55 $\pm$ 0.69 {\scriptsize\textcolor{gray}{(1.27$\times$)}} & 7.4 {\scriptsize\textcolor{gray}{(4.54$\times$)}} & 14.52 {\scriptsize\textcolor{gray}{(2.54$\times$)}} \\
 & & GROTE & 2149.25 {\scriptsize\textcolor{gray}{(0.02$\times$)}} & 1023.93 $\pm$ 1.39 {\scriptsize\textcolor{gray}{(0.01$\times$)}} & 1036.86 $\pm$ 8.35 {\scriptsize\textcolor{gray}{(0.01$\times$)}} & 98.34 {\scriptsize\textcolor{gray}{(0.34$\times$)}} & 96.90 {\scriptsize\textcolor{gray}{(0.38$\times$)}} \\
 & & Blind-Match & 880.58 {\scriptsize\textcolor{gray}{(0.05$\times$)}} & 410.51 $\pm$ 13.26 {\scriptsize\textcolor{gray}{(0.03$\times$)}} & 391.32 $\pm$ 2.64 {\scriptsize\textcolor{gray}{(0.03$\times$)}} & 55.83 {\scriptsize\textcolor{gray}{(0.60$\times$)}} & 65.97 {\scriptsize\textcolor{gray}{(0.56$\times$)}} \\
 & & HERS & 75.87 {\scriptsize\textcolor{gray}{(0.64$\times$)}} & 22.42 $\pm$ 0.53 {\scriptsize\textcolor{gray}{(0.54$\times$)}} & 23.27 $\pm$ 0.35 {\scriptsize\textcolor{gray}{(0.52$\times$)}} & 31.00 {\scriptsize\textcolor{gray}{(1.08$\times$)}} & 36.84 {\scriptsize\textcolor{gray}{(1.00$\times$)}} \\
\midrule
\multirow{2}{*}{$2^{17}$} & \multirow{2}{*}{128} & HyDia-CPU-DG & 77.90 {\scriptsize\textcolor{gray}{(1.00$\times$)}} & 21.91 $\pm$ 0.75 {\scriptsize\textcolor{gray}{(1.00$\times$)}} & 23.54 $\pm$ 0.85 {\scriptsize\textcolor{gray}{(1.00$\times$)}} & 33.9 {\scriptsize\textcolor{gray}{(1.00$\times$)}} & 48.85 {\scriptsize\textcolor{gray}{(1.00$\times$)}} \\
 & & BSGS-CPU-DG & 45.60 {\scriptsize\textcolor{gray}{(1.71$\times$)}} & 14.18 $\pm$ 0.12 {\scriptsize\textcolor{gray}{(1.55$\times$)}} & 16.47 $\pm$ 0.77 {\scriptsize\textcolor{gray}{(1.43$\times$)}} & 7.6 {\scriptsize\textcolor{gray}{(4.46$\times$)}} & 26.53 {\scriptsize\textcolor{gray}{(1.84$\times$)}} \\
\midrule
\multirow{2}{*}{$2^{18}$} & \multirow{2}{*}{256} & HyDia-CPU-DG & 128.10 {\scriptsize\textcolor{gray}{(1.00$\times$)}} & 39.42 $\pm$ 1.22 {\scriptsize\textcolor{gray}{(1.00$\times$)}} & 41.87 $\pm$ 2.37 {\scriptsize\textcolor{gray}{(1.00$\times$)}} & 34.8 {\scriptsize\textcolor{gray}{(1.00$\times$)}} & 72.86 {\scriptsize\textcolor{gray}{(1.00$\times$)}} \\
 & & BSGS-CPU-DG & 114.10 {\scriptsize\textcolor{gray}{(1.12$\times$)}} & 26.66 $\pm$ 0.47 {\scriptsize\textcolor{gray}{(1.48$\times$)}} & 30.13 $\pm$ 0.30 {\scriptsize\textcolor{gray}{(1.39$\times$)}} & 8.0 {\scriptsize\textcolor{gray}{(4.35$\times$)}} & 50.54 {\scriptsize\textcolor{gray}{(1.44$\times$)}} \\
\midrule
\multirow{2}{*}{$2^{19}$} & \multirow{2}{*}{256} & HyDia-CPU-DG & 242.20 {\scriptsize\textcolor{gray}{(1.00$\times$)}} & 84.12 $\pm$ 3.38 {\scriptsize\textcolor{gray}{(1.00$\times$)}} & 79.79 $\pm$ 2.25 {\scriptsize\textcolor{gray}{(1.00$\times$)}} & 35.4 {\scriptsize\textcolor{gray}{(1.00$\times$)}} & 120.88 {\scriptsize\textcolor{gray}{(1.00$\times$)}} \\
 & & BSGS-CPU-DG & 173.00 {\scriptsize\textcolor{gray}{(1.40$\times$)}} & 53.64 $\pm$ 0.83 {\scriptsize\textcolor{gray}{(1.57$\times$)}} & 57.91 $\pm$ 0.73 {\scriptsize\textcolor{gray}{(1.38$\times$)}} & 9.0 {\scriptsize\textcolor{gray}{(3.93$\times$)}} & 98.57 {\scriptsize\textcolor{gray}{(1.23$\times$)}} \\
\midrule
\multirow{2}{*}{$2^{20}$} & \multirow{2}{*}{512} & HyDia-CPU-DG & 483.50 {\scriptsize\textcolor{gray}{(1.00$\times$)}} & 163.66 $\pm$ 3.53 {\scriptsize\textcolor{gray}{(1.00$\times$)}} & 165.61 $\pm$ 3.58 {\scriptsize\textcolor{gray}{(1.00$\times$)}} & 37.3 {\scriptsize\textcolor{gray}{(1.00$\times$)}} & 216.93 {\scriptsize\textcolor{gray}{(1.00$\times$)}} \\
 & & BSGS-CPU-DG & 356.00 {\scriptsize\textcolor{gray}{(1.36$\times$)}} & 112.33 $\pm$ 1.46 {\scriptsize\textcolor{gray}{(1.46$\times$)}} & 115.42 $\pm$ 1.83 {\scriptsize\textcolor{gray}{(1.43$\times$)}} & 11.3 {\scriptsize\textcolor{gray}{(3.30$\times$)}} & 194.61 {\scriptsize\textcolor{gray}{(1.11$\times$)}} \\
\bottomrule
\end{tabular}
}
\end{table*}

\subsection{Additional CPU/GPU benchmark table}\label{appendix:cpu_gpu_depth8_table}

Tables~\ref{tab:benchmark-results-cpu-gpu-depth8-part1} and~\ref{tab:benchmark-results-cpu-gpu-depth8} report our CPU and GPU benchmark results with $\compdepth=8$, \emph{i.e.,} the total multiplicative depth is equal to 9. For $K=2^{16}$, HyDia-RTX-DG observed an out-of-GPU-memory issue on the RTX 8000 (48\,GB); all RTX approaches are OOM at $K\geq 2^{17}$. The H200 (141\,GB) extends GPU coverage up to $K=2^{17}$.

\begin{table*}[htbp]
\centering
\caption{CPU/GPU benchmark results with $\compdepth=8$ (Part~I: $K=2^{10}$--$2^{14}$). Values are reported as mean $\pm$ std over 11 trials. The parenthetical value in \textcolor{gray}{gray} is the ratio against HyDia-CPU-DG as the baseline; the best ratio per metric and $K$ group is \textbf{bolded} (entire cell). RTX rows use an NVIDIA Quadro RTX~8000 (48\,GB VRAM); H200 rows use an NVIDIA H200 (141\,GB VRAM).}
\label{tab:benchmark-results-cpu-gpu-depth8-part1}
\scriptsize
\setlength{\tabcolsep}{3.5pt}
\resizebox{\textwidth}{!}{%
\begin{tabular}{rrlrrrrr}
\toprule
$K$ & $K_m$ & Approach & RTT (s) & Membership (s) & Index (s) & RAM (GB) & Disk (GB) \\
\midrule
\multirow{8}{*}{$2^{10}$} & \multirow{8}{*}{16} & HyDia-CPU-DG & 24.90 {\scriptsize\textcolor{gray}{(1.00$\times$)}} & 4.22 $\pm$ 0.02 {\scriptsize\textcolor{gray}{(1.00$\times$)}} & 4.04 $\pm$ 0.07 {\scriptsize\textcolor{gray}{(1.00$\times$)}} & 28.7 $\pm$ 0.0 {\scriptsize\textcolor{gray}{(1.00$\times$)}} & 24.23 $\pm$ 0.00 {\scriptsize\textcolor{gray}{(1.00$\times$)}} \\
 &  & HyDia-RTX-DG & 39.40 {\scriptsize\textcolor{gray}{(0.63$\times$)}} & 0.63 $\pm$ 0.01 {\scriptsize\textcolor{gray}{(6.70$\times$)}} & 0.42 $\pm$ 0.00 {\scriptsize\textcolor{gray}{(9.62$\times$)}} & 24.6 $\pm$ 0.0 {\scriptsize\textcolor{gray}{(1.17$\times$)}} & 27.23 $\pm$ 0.00 {\scriptsize\textcolor{gray}{(0.89$\times$)}} \\
 &  & HyDia-H200-DG & 44.10 {\scriptsize\textcolor{gray}{(0.56$\times$)}} & 0.58 $\pm$ 0.03 {\scriptsize\textcolor{gray}{(7.28$\times$)}} & 0.39 $\pm$ 0.01 {\scriptsize\textcolor{gray}{(10.36$\times$)}} & 24.6 $\pm$ 0.0 {\scriptsize\textcolor{gray}{(1.17$\times$)}} & 27.23 $\pm$ 0.00 {\scriptsize\textcolor{gray}{(0.89$\times$)}} \\
 &  & BSGS-CPU-DG & 10.10 {\scriptsize\textcolor{gray}{(2.47$\times$)}} & 3.64 $\pm$ 0.04 {\scriptsize\textcolor{gray}{(1.16$\times$)}} & 3.54 $\pm$ 0.07 {\scriptsize\textcolor{gray}{(1.14$\times$)}} & 6.1 $\pm$ 0.0 {\scriptsize\textcolor{gray}{(4.70$\times$)}} & \textbf{4.71 $\pm$ 0.00 {\scriptsize\textcolor{gray}{(5.14$\times$)}}} \\
 &  & BSGS-RTX-TBS & 9.50 {\scriptsize\textcolor{gray}{(2.62$\times$)}} & 0.25 $\pm$ 0.00 {\scriptsize\textcolor{gray}{(16.88$\times$)}} & 0.24 $\pm$ 0.00 {\scriptsize\textcolor{gray}{(16.83$\times$)}} & 3.5 $\pm$ 0.0 {\scriptsize\textcolor{gray}{(8.20$\times$)}} & 5.61 $\pm$ 0.00 {\scriptsize\textcolor{gray}{(4.32$\times$)}} \\
 &  & BSGS-RTX-TBE & 8.50 {\scriptsize\textcolor{gray}{(2.93$\times$)}} & 0.25 $\pm$ 0.00 {\scriptsize\textcolor{gray}{(16.88$\times$)}} & 0.24 $\pm$ 0.00 {\scriptsize\textcolor{gray}{(16.83$\times$)}} & 2.6 $\pm$ 0.0 {\scriptsize\textcolor{gray}{(11.04$\times$)}} & \textbf{4.71 $\pm$ 0.00 {\scriptsize\textcolor{gray}{(5.14$\times$)}}} \\
 &  & BSGS-H200-TBS & 9.30 {\scriptsize\textcolor{gray}{(2.68$\times$)}} & \textbf{0.20 $\pm$ 0.01 {\scriptsize\textcolor{gray}{(21.10$\times$)}}} & \textbf{0.19 $\pm$ 0.01 {\scriptsize\textcolor{gray}{(21.26$\times$)}}} & 3.4 $\pm$ 0.0 {\scriptsize\textcolor{gray}{(8.44$\times$)}} & 5.61 $\pm$ 0.00 {\scriptsize\textcolor{gray}{(4.32$\times$)}} \\
 &  & BSGS-H200-TBE & \textbf{7.90 {\scriptsize\textcolor{gray}{(3.15$\times$)}}} & \textbf{0.20 $\pm$ 0.01 {\scriptsize\textcolor{gray}{(21.10$\times$)}}} & \textbf{0.19 $\pm$ 0.01 {\scriptsize\textcolor{gray}{(21.26$\times$)}}} & \textbf{2.5 $\pm$ 0.0 {\scriptsize\textcolor{gray}{(11.48$\times$)}}} & \textbf{4.71 $\pm$ 0.00 {\scriptsize\textcolor{gray}{(5.14$\times$)}}} \\
\midrule
\multirow{8}{*}{$2^{11}$} & \multirow{8}{*}{16} & HyDia-CPU-DG & 25.00 {\scriptsize\textcolor{gray}{(1.00$\times$)}} & 4.21 $\pm$ 0.04 {\scriptsize\textcolor{gray}{(1.00$\times$)}} & 4.07 $\pm$ 0.08 {\scriptsize\textcolor{gray}{(1.00$\times$)}} & 28.7 $\pm$ 0.1 {\scriptsize\textcolor{gray}{(1.00$\times$)}} & 24.23 $\pm$ 0.00 {\scriptsize\textcolor{gray}{(1.00$\times$)}} \\
 &  & HyDia-RTX-DG & 39.30 {\scriptsize\textcolor{gray}{(0.64$\times$)}} & 0.63 $\pm$ 0.02 {\scriptsize\textcolor{gray}{(6.68$\times$)}} & 0.42 $\pm$ 0.00 {\scriptsize\textcolor{gray}{(9.69$\times$)}} & 24.6 $\pm$ 0.0 {\scriptsize\textcolor{gray}{(1.17$\times$)}} & 27.23 $\pm$ 0.00 {\scriptsize\textcolor{gray}{(0.89$\times$)}} \\
 &  & HyDia-H200-DG & 46.90 {\scriptsize\textcolor{gray}{(0.53$\times$)}} & 0.60 $\pm$ 0.03 {\scriptsize\textcolor{gray}{(7.02$\times$)}} & 0.39 $\pm$ 0.01 {\scriptsize\textcolor{gray}{(10.44$\times$)}} & 24.6 $\pm$ 0.0 {\scriptsize\textcolor{gray}{(1.17$\times$)}} & 27.23 $\pm$ 0.00 {\scriptsize\textcolor{gray}{(0.89$\times$)}} \\
 &  & BSGS-CPU-DG & 10.10 {\scriptsize\textcolor{gray}{(2.48$\times$)}} & 3.64 $\pm$ 0.03 {\scriptsize\textcolor{gray}{(1.16$\times$)}} & 3.56 $\pm$ 0.07 {\scriptsize\textcolor{gray}{(1.14$\times$)}} & 6.1 $\pm$ 0.0 {\scriptsize\textcolor{gray}{(4.70$\times$)}} & \textbf{4.71 $\pm$ 0.00 {\scriptsize\textcolor{gray}{(5.14$\times$)}}} \\
 &  & BSGS-RTX-TBS & 9.60 {\scriptsize\textcolor{gray}{(2.60$\times$)}} & 0.25 $\pm$ 0.00 {\scriptsize\textcolor{gray}{(16.84$\times$)}} & 0.24 $\pm$ 0.00 {\scriptsize\textcolor{gray}{(16.96$\times$)}} & 3.5 $\pm$ 0.0 {\scriptsize\textcolor{gray}{(8.20$\times$)}} & 5.61 $\pm$ 0.00 {\scriptsize\textcolor{gray}{(4.32$\times$)}} \\
 &  & BSGS-RTX-TBE & \textbf{8.50 {\scriptsize\textcolor{gray}{(2.94$\times$)}}} & 0.26 $\pm$ 0.00 {\scriptsize\textcolor{gray}{(16.19$\times$)}} & 0.24 $\pm$ 0.00 {\scriptsize\textcolor{gray}{(16.96$\times$)}} & 2.6 $\pm$ 0.0 {\scriptsize\textcolor{gray}{(11.04$\times$)}} & \textbf{4.71 $\pm$ 0.00 {\scriptsize\textcolor{gray}{(5.14$\times$)}}} \\
 &  & BSGS-H200-TBS & 8.90 {\scriptsize\textcolor{gray}{(2.81$\times$)}} & \textbf{0.20 $\pm$ 0.01 {\scriptsize\textcolor{gray}{(21.05$\times$)}}} & \textbf{0.19 $\pm$ 0.01 {\scriptsize\textcolor{gray}{(21.42$\times$)}}} & 3.4 $\pm$ 0.0 {\scriptsize\textcolor{gray}{(8.44$\times$)}} & 5.61 $\pm$ 0.00 {\scriptsize\textcolor{gray}{(4.32$\times$)}} \\
 &  & BSGS-H200-TBE & 8.60 {\scriptsize\textcolor{gray}{(2.91$\times$)}} & 0.22 $\pm$ 0.02 {\scriptsize\textcolor{gray}{(19.14$\times$)}} & 0.20 $\pm$ 0.01 {\scriptsize\textcolor{gray}{(20.35$\times$)}} & \textbf{2.5 $\pm$ 0.0 {\scriptsize\textcolor{gray}{(11.48$\times$)}}} & \textbf{4.71 $\pm$ 0.00 {\scriptsize\textcolor{gray}{(5.14$\times$)}}} \\
\midrule
\multirow{8}{*}{$2^{12}$} & \multirow{8}{*}{32} & HyDia-CPU-DG & 25.00 {\scriptsize\textcolor{gray}{(1.00$\times$)}} & 4.26 $\pm$ 0.08 {\scriptsize\textcolor{gray}{(1.00$\times$)}} & 4.00 $\pm$ 0.11 {\scriptsize\textcolor{gray}{(1.00$\times$)}} & 28.7 $\pm$ 0.0 {\scriptsize\textcolor{gray}{(1.00$\times$)}} & 24.23 $\pm$ 0.00 {\scriptsize\textcolor{gray}{(1.00$\times$)}} \\
 &  & HyDia-RTX-DG & 39.40 {\scriptsize\textcolor{gray}{(0.63$\times$)}} & 0.63 $\pm$ 0.01 {\scriptsize\textcolor{gray}{(6.76$\times$)}} & 0.43 $\pm$ 0.00 {\scriptsize\textcolor{gray}{(9.30$\times$)}} & 24.6 $\pm$ 0.0 {\scriptsize\textcolor{gray}{(1.17$\times$)}} & 27.23 $\pm$ 0.00 {\scriptsize\textcolor{gray}{(0.89$\times$)}} \\
 &  & HyDia-H200-DG & 44.80 {\scriptsize\textcolor{gray}{(0.56$\times$)}} & 0.59 $\pm$ 0.02 {\scriptsize\textcolor{gray}{(7.22$\times$)}} & 0.41 $\pm$ 0.01 {\scriptsize\textcolor{gray}{(9.76$\times$)}} & 24.6 $\pm$ 0.0 {\scriptsize\textcolor{gray}{(1.17$\times$)}} & 27.23 $\pm$ 0.00 {\scriptsize\textcolor{gray}{(0.89$\times$)}} \\
 &  & BSGS-CPU-DG & 10.50 {\scriptsize\textcolor{gray}{(2.38$\times$)}} & 3.68 $\pm$ 0.04 {\scriptsize\textcolor{gray}{(1.16$\times$)}} & 3.60 $\pm$ 0.05 {\scriptsize\textcolor{gray}{(1.11$\times$)}} & 6.1 $\pm$ 0.0 {\scriptsize\textcolor{gray}{(4.70$\times$)}} & \textbf{4.71 $\pm$ 0.00 {\scriptsize\textcolor{gray}{(5.14$\times$)}}} \\
 &  & BSGS-RTX-TBS & 9.60 {\scriptsize\textcolor{gray}{(2.60$\times$)}} & 0.25 $\pm$ 0.00 {\scriptsize\textcolor{gray}{(17.04$\times$)}} & 0.24 $\pm$ 0.00 {\scriptsize\textcolor{gray}{(16.67$\times$)}} & 3.5 $\pm$ 0.0 {\scriptsize\textcolor{gray}{(8.20$\times$)}} & 5.61 $\pm$ 0.00 {\scriptsize\textcolor{gray}{(4.32$\times$)}} \\
 &  & BSGS-RTX-TBE & 8.80 {\scriptsize\textcolor{gray}{(2.84$\times$)}} & 0.25 $\pm$ 0.00 {\scriptsize\textcolor{gray}{(17.04$\times$)}} & 0.24 $\pm$ 0.00 {\scriptsize\textcolor{gray}{(16.67$\times$)}} & 2.6 $\pm$ 0.0 {\scriptsize\textcolor{gray}{(11.04$\times$)}} & \textbf{4.71 $\pm$ 0.00 {\scriptsize\textcolor{gray}{(5.14$\times$)}}} \\
 &  & BSGS-H200-TBS & 9.20 {\scriptsize\textcolor{gray}{(2.72$\times$)}} & \textbf{0.20 $\pm$ 0.01 {\scriptsize\textcolor{gray}{(21.30$\times$)}}} & \textbf{0.20 $\pm$ 0.01 {\scriptsize\textcolor{gray}{(20.00$\times$)}}} & 3.4 $\pm$ 0.0 {\scriptsize\textcolor{gray}{(8.44$\times$)}} & 5.61 $\pm$ 0.00 {\scriptsize\textcolor{gray}{(4.32$\times$)}} \\
 &  & BSGS-H200-TBE & \textbf{8.40 {\scriptsize\textcolor{gray}{(2.98$\times$)}}} & 0.21 $\pm$ 0.01 {\scriptsize\textcolor{gray}{(20.29$\times$)}} & \textbf{0.20 $\pm$ 0.01 {\scriptsize\textcolor{gray}{(20.00$\times$)}}} & \textbf{2.5 $\pm$ 0.0 {\scriptsize\textcolor{gray}{(11.48$\times$)}}} & \textbf{4.71 $\pm$ 0.00 {\scriptsize\textcolor{gray}{(5.14$\times$)}}} \\
\midrule
\multirow{8}{*}{$2^{13}$} & \multirow{8}{*}{32} & HyDia-CPU-DG & 25.10 {\scriptsize\textcolor{gray}{(1.00$\times$)}} & 4.24 $\pm$ 0.07 {\scriptsize\textcolor{gray}{(1.00$\times$)}} & 4.04 $\pm$ 0.09 {\scriptsize\textcolor{gray}{(1.00$\times$)}} & 28.7 $\pm$ 0.0 {\scriptsize\textcolor{gray}{(1.00$\times$)}} & 24.23 $\pm$ 0.00 {\scriptsize\textcolor{gray}{(1.00$\times$)}} \\
 &  & HyDia-RTX-DG & 39.00 {\scriptsize\textcolor{gray}{(0.64$\times$)}} & 0.63 $\pm$ 0.01 {\scriptsize\textcolor{gray}{(6.73$\times$)}} & 0.42 $\pm$ 0.00 {\scriptsize\textcolor{gray}{(9.62$\times$)}} & 24.6 $\pm$ 0.0 {\scriptsize\textcolor{gray}{(1.17$\times$)}} & 27.23 $\pm$ 0.00 {\scriptsize\textcolor{gray}{(0.89$\times$)}} \\
 &  & HyDia-H200-DG & 44.80 {\scriptsize\textcolor{gray}{(0.56$\times$)}} & 0.60 $\pm$ 0.05 {\scriptsize\textcolor{gray}{(7.07$\times$)}} & 0.40 $\pm$ 0.01 {\scriptsize\textcolor{gray}{(10.10$\times$)}} & 24.6 $\pm$ 0.0 {\scriptsize\textcolor{gray}{(1.17$\times$)}} & 27.23 $\pm$ 0.00 {\scriptsize\textcolor{gray}{(0.89$\times$)}} \\
 &  & BSGS-CPU-DG & 10.30 {\scriptsize\textcolor{gray}{(2.44$\times$)}} & 3.69 $\pm$ 0.07 {\scriptsize\textcolor{gray}{(1.15$\times$)}} & 3.61 $\pm$ 0.06 {\scriptsize\textcolor{gray}{(1.12$\times$)}} & 6.1 $\pm$ 0.0 {\scriptsize\textcolor{gray}{(4.70$\times$)}} & \textbf{4.71 $\pm$ 0.00 {\scriptsize\textcolor{gray}{(5.14$\times$)}}} \\
 &  & BSGS-RTX-TBS & 9.70 {\scriptsize\textcolor{gray}{(2.59$\times$)}} & 0.25 $\pm$ 0.00 {\scriptsize\textcolor{gray}{(16.96$\times$)}} & 0.24 $\pm$ 0.00 {\scriptsize\textcolor{gray}{(16.83$\times$)}} & 3.5 $\pm$ 0.0 {\scriptsize\textcolor{gray}{(8.20$\times$)}} & 5.61 $\pm$ 0.00 {\scriptsize\textcolor{gray}{(4.32$\times$)}} \\
 &  & BSGS-RTX-TBE & 8.60 {\scriptsize\textcolor{gray}{(2.92$\times$)}} & 0.25 $\pm$ 0.00 {\scriptsize\textcolor{gray}{(16.96$\times$)}} & 0.24 $\pm$ 0.00 {\scriptsize\textcolor{gray}{(16.83$\times$)}} & 2.6 $\pm$ 0.0 {\scriptsize\textcolor{gray}{(11.04$\times$)}} & \textbf{4.71 $\pm$ 0.00 {\scriptsize\textcolor{gray}{(5.14$\times$)}}} \\
 &  & BSGS-H200-TBS & 9.30 {\scriptsize\textcolor{gray}{(2.70$\times$)}} & \textbf{0.20 $\pm$ 0.01 {\scriptsize\textcolor{gray}{(21.20$\times$)}}} & \textbf{0.19 $\pm$ 0.01 {\scriptsize\textcolor{gray}{(21.26$\times$)}}} & 3.4 $\pm$ 0.0 {\scriptsize\textcolor{gray}{(8.44$\times$)}} & 5.61 $\pm$ 0.00 {\scriptsize\textcolor{gray}{(4.32$\times$)}} \\
 &  & BSGS-H200-TBE & \textbf{8.30 {\scriptsize\textcolor{gray}{(3.02$\times$)}}} & 0.21 $\pm$ 0.01 {\scriptsize\textcolor{gray}{(20.19$\times$)}} & 0.20 $\pm$ 0.02 {\scriptsize\textcolor{gray}{(20.20$\times$)}} & \textbf{2.5 $\pm$ 0.0 {\scriptsize\textcolor{gray}{(11.48$\times$)}}} & \textbf{4.71 $\pm$ 0.00 {\scriptsize\textcolor{gray}{(5.14$\times$)}}} \\
\midrule
\multirow{8}{*}{$2^{14}$} & \multirow{8}{*}{64} & HyDia-CPU-DG & 25.10 {\scriptsize\textcolor{gray}{(1.00$\times$)}} & 4.24 $\pm$ 0.06 {\scriptsize\textcolor{gray}{(1.00$\times$)}} & 4.02 $\pm$ 0.10 {\scriptsize\textcolor{gray}{(1.00$\times$)}} & 28.7 $\pm$ 0.1 {\scriptsize\textcolor{gray}{(1.00$\times$)}} & 24.23 $\pm$ 0.00 {\scriptsize\textcolor{gray}{(1.00$\times$)}} \\
 &  & HyDia-RTX-DG & 41.70 {\scriptsize\textcolor{gray}{(0.60$\times$)}} & 0.63 $\pm$ 0.01 {\scriptsize\textcolor{gray}{(6.73$\times$)}} & 0.42 $\pm$ 0.00 {\scriptsize\textcolor{gray}{(9.57$\times$)}} & 24.6 $\pm$ 0.0 {\scriptsize\textcolor{gray}{(1.17$\times$)}} & 27.23 $\pm$ 0.00 {\scriptsize\textcolor{gray}{(0.89$\times$)}} \\
 &  & HyDia-H200-DG & 45.60 {\scriptsize\textcolor{gray}{(0.55$\times$)}} & 0.61 $\pm$ 0.05 {\scriptsize\textcolor{gray}{(6.95$\times$)}} & 0.41 $\pm$ 0.02 {\scriptsize\textcolor{gray}{(9.80$\times$)}} & 24.6 $\pm$ 0.0 {\scriptsize\textcolor{gray}{(1.17$\times$)}} & 27.23 $\pm$ 0.00 {\scriptsize\textcolor{gray}{(0.89$\times$)}} \\
 &  & BSGS-CPU-DG & 10.50 {\scriptsize\textcolor{gray}{(2.39$\times$)}} & 3.68 $\pm$ 0.08 {\scriptsize\textcolor{gray}{(1.15$\times$)}} & 3.60 $\pm$ 0.06 {\scriptsize\textcolor{gray}{(1.12$\times$)}} & 6.1 $\pm$ 0.0 {\scriptsize\textcolor{gray}{(4.70$\times$)}} & \textbf{4.71 $\pm$ 0.00 {\scriptsize\textcolor{gray}{(5.14$\times$)}}} \\
 &  & BSGS-RTX-TBS & 9.70 {\scriptsize\textcolor{gray}{(2.59$\times$)}} & 0.25 $\pm$ 0.00 {\scriptsize\textcolor{gray}{(16.96$\times$)}} & 0.24 $\pm$ 0.00 {\scriptsize\textcolor{gray}{(16.75$\times$)}} & 3.5 $\pm$ 0.0 {\scriptsize\textcolor{gray}{(8.20$\times$)}} & 5.61 $\pm$ 0.00 {\scriptsize\textcolor{gray}{(4.32$\times$)}} \\
 &  & BSGS-RTX-TBE & 8.70 {\scriptsize\textcolor{gray}{(2.89$\times$)}} & 0.26 $\pm$ 0.01 {\scriptsize\textcolor{gray}{(16.31$\times$)}} & 0.24 $\pm$ 0.00 {\scriptsize\textcolor{gray}{(16.75$\times$)}} & 2.6 $\pm$ 0.0 {\scriptsize\textcolor{gray}{(11.04$\times$)}} & \textbf{4.71 $\pm$ 0.00 {\scriptsize\textcolor{gray}{(5.14$\times$)}}} \\
 &  & BSGS-H200-TBS & 9.80 {\scriptsize\textcolor{gray}{(2.56$\times$)}} & \textbf{0.20 $\pm$ 0.01 {\scriptsize\textcolor{gray}{(21.20$\times$)}}} & 0.20 $\pm$ 0.02 {\scriptsize\textcolor{gray}{(20.10$\times$)}} & 3.4 $\pm$ 0.0 {\scriptsize\textcolor{gray}{(8.44$\times$)}} & 5.61 $\pm$ 0.00 {\scriptsize\textcolor{gray}{(4.32$\times$)}} \\
 &  & BSGS-H200-TBE & \textbf{8.60 {\scriptsize\textcolor{gray}{(2.92$\times$)}}} & \textbf{0.20 $\pm$ 0.01 {\scriptsize\textcolor{gray}{(21.20$\times$)}}} & \textbf{0.19 $\pm$ 0.01 {\scriptsize\textcolor{gray}{(21.16$\times$)}}} & \textbf{2.5 $\pm$ 0.0 {\scriptsize\textcolor{gray}{(11.48$\times$)}}} & \textbf{4.71 $\pm$ 0.00 {\scriptsize\textcolor{gray}{(5.14$\times$)}}} \\
\bottomrule
\end{tabular}
}
\end{table*}

\begin{table*}[htbp]
\centering
\caption{CPU/GPU benchmark results with $\compdepth=8$ (Part~II: $K=2^{15}$--$2^{20}$). For $K=2^{16}$, HyDia-RTX-DG ran out of GPU memory; for $K\ge 2^{17}$, all RTX approaches ran out of GPU memory. Best ratio per metric and $K$ group is \textbf{bolded} (entire cell).}
\label{tab:benchmark-results-cpu-gpu-depth8}
\scriptsize
\setlength{\tabcolsep}{3.5pt}
\resizebox{\textwidth}{!}{%
\begin{tabular}{rrlrrrrr}
\toprule
$K$ & $K_m$ & Approach & RTT (s) & Membership (s) & Index (s) & RAM (GB) & Disk (GB) \\
\midrule
\multirow{8}{*}{$2^{15}$} & \multirow{8}{*}{64} & HyDia-CPU-DG & 29.90 {\scriptsize\textcolor{gray}{(1.00$\times$)}} & 6.00 $\pm$ 0.12 {\scriptsize\textcolor{gray}{(1.00$\times$)}} & 5.87 $\pm$ 0.12 {\scriptsize\textcolor{gray}{(1.00$\times$)}} & 28.7 $\pm$ 0.0 {\scriptsize\textcolor{gray}{(1.00$\times$)}} & 26.73 $\pm$ 0.00 {\scriptsize\textcolor{gray}{(1.00$\times$)}} \\
 &  & HyDia-RTX-DG & 53.00 {\scriptsize\textcolor{gray}{(0.56$\times$)}} & 1.75 $\pm$ 0.02 {\scriptsize\textcolor{gray}{(3.43$\times$)}} & 1.66 $\pm$ 0.02 {\scriptsize\textcolor{gray}{(3.54$\times$)}} & 24.6 $\pm$ 0.0 {\scriptsize\textcolor{gray}{(1.17$\times$)}} & 30.23 $\pm$ 0.00 {\scriptsize\textcolor{gray}{(0.88$\times$)}} \\
 &  & HyDia-H200-DG & 64.30 {\scriptsize\textcolor{gray}{(0.47$\times$)}} & 1.69 $\pm$ 0.06 {\scriptsize\textcolor{gray}{(3.55$\times$)}} & 1.66 $\pm$ 0.07 {\scriptsize\textcolor{gray}{(3.54$\times$)}} & 24.6 $\pm$ 0.0 {\scriptsize\textcolor{gray}{(1.17$\times$)}} & 30.23 $\pm$ 0.00 {\scriptsize\textcolor{gray}{(0.88$\times$)}} \\
 &  & BSGS-CPU-DG & \textbf{14.60 {\scriptsize\textcolor{gray}{(2.05$\times$)}}} & 4.85 $\pm$ 0.06 {\scriptsize\textcolor{gray}{(1.24$\times$)}} & 4.80 $\pm$ 0.14 {\scriptsize\textcolor{gray}{(1.22$\times$)}} & 6.2 $\pm$ 0.0 {\scriptsize\textcolor{gray}{(4.63$\times$)}} & \textbf{7.21 $\pm$ 0.00 {\scriptsize\textcolor{gray}{(3.71$\times$)}}} \\
 &  & BSGS-RTX-TBS & 16.80 {\scriptsize\textcolor{gray}{(1.78$\times$)}} & 0.78 $\pm$ 0.01 {\scriptsize\textcolor{gray}{(7.69$\times$)}} & 0.81 $\pm$ 0.01 {\scriptsize\textcolor{gray}{(7.25$\times$)}} & 3.5 $\pm$ 0.0 {\scriptsize\textcolor{gray}{(8.20$\times$)}} & 8.11 $\pm$ 0.00 {\scriptsize\textcolor{gray}{(3.30$\times$)}} \\
 &  & BSGS-RTX-TBE & 15.30 {\scriptsize\textcolor{gray}{(1.95$\times$)}} & 0.81 $\pm$ 0.01 {\scriptsize\textcolor{gray}{(7.41$\times$)}} & \textbf{0.81 $\pm$ 0.01 {\scriptsize\textcolor{gray}{(7.25$\times$)}}} & 2.6 $\pm$ 0.0 {\scriptsize\textcolor{gray}{(11.04$\times$)}} & \textbf{7.21 $\pm$ 0.00 {\scriptsize\textcolor{gray}{(3.71$\times$)}}} \\
 &  & BSGS-H200-TBS & 16.70 {\scriptsize\textcolor{gray}{(1.79$\times$)}} & \textbf{0.64 $\pm$ 0.01 {\scriptsize\textcolor{gray}{(9.38$\times$)}}} & \textbf{0.69 $\pm$ 0.02 {\scriptsize\textcolor{gray}{(8.51$\times$)}}} & 3.4 $\pm$ 0.0 {\scriptsize\textcolor{gray}{(8.44$\times$)}} & 8.11 $\pm$ 0.00 {\scriptsize\textcolor{gray}{(3.30$\times$)}} \\
 &  & BSGS-H200-TBE & 15.40 {\scriptsize\textcolor{gray}{(1.94$\times$)}} & 0.68 $\pm$ 0.02 {\scriptsize\textcolor{gray}{(8.82$\times$)}} & \textbf{0.69 $\pm$ 0.02 {\scriptsize\textcolor{gray}{(8.51$\times$)}}} & \textbf{2.5 $\pm$ 0.0 {\scriptsize\textcolor{gray}{(11.48$\times$)}}} & \textbf{7.21 $\pm$ 0.00 {\scriptsize\textcolor{gray}{(3.71$\times$)}}} \\
\midrule
\multirow{7}{*}{$2^{16}$} & \multirow{7}{*}{128} & HyDia-CPU-DG & 41.30 {\scriptsize\textcolor{gray}{(1.00$\times$)}} & 9.71 $\pm$ 0.26 {\scriptsize\textcolor{gray}{(1.00$\times$)}} & 10.15 $\pm$ 0.61 {\scriptsize\textcolor{gray}{(1.00$\times$)}} & 28.9 $\pm$ 0.1 {\scriptsize\textcolor{gray}{(1.00$\times$)}} & 31.74 $\pm$ 0.00 {\scriptsize\textcolor{gray}{(1.00$\times$)}} \\
 &  & HyDia-H200-DG & 98.50 {\scriptsize\textcolor{gray}{(0.42$\times$)}} & 3.04 $\pm$ 0.05 {\scriptsize\textcolor{gray}{(3.19$\times$)}} & 4.10 $\pm$ 0.03 {\scriptsize\textcolor{gray}{(2.48$\times$)}} & 24.6 $\pm$ 0.0 {\scriptsize\textcolor{gray}{(1.17$\times$)}} & 36.23 $\pm$ 0.00 {\scriptsize\textcolor{gray}{(0.88$\times$)}} \\
 &  & BSGS-CPU-DG & \textbf{22.80 {\scriptsize\textcolor{gray}{(1.81$\times$)}}} & 7.07 $\pm$ 0.08 {\scriptsize\textcolor{gray}{(1.37$\times$)}} & 7.48 $\pm$ 0.18 {\scriptsize\textcolor{gray}{(1.36$\times$)}} & 6.3 $\pm$ 0.0 {\scriptsize\textcolor{gray}{(4.59$\times$)}} & \textbf{12.21 $\pm$ 0.00 {\scriptsize\textcolor{gray}{(2.60$\times$)}}} \\
 &  & BSGS-RTX-TBS & 38.50 {\scriptsize\textcolor{gray}{(1.07$\times$)}} & 3.02 $\pm$ 0.14 {\scriptsize\textcolor{gray}{(3.22$\times$)}} & 3.16 $\pm$ 0.05 {\scriptsize\textcolor{gray}{(3.21$\times$)}} & 4.0 $\pm$ 0.0 {\scriptsize\textcolor{gray}{(7.22$\times$)}} & 13.11 $\pm$ 0.00 {\scriptsize\textcolor{gray}{(2.42$\times$)}} \\
 &  & BSGS-RTX-TBE & 36.10 {\scriptsize\textcolor{gray}{(1.14$\times$)}} & 3.03 $\pm$ 0.03 {\scriptsize\textcolor{gray}{(3.20$\times$)}} & 3.13 $\pm$ 0.04 {\scriptsize\textcolor{gray}{(3.24$\times$)}} & 3.6 $\pm$ 0.0 {\scriptsize\textcolor{gray}{(8.03$\times$)}} & \textbf{12.21 $\pm$ 0.00 {\scriptsize\textcolor{gray}{(2.60$\times$)}}} \\
 &  & BSGS-H200-TBS & 35.90 {\scriptsize\textcolor{gray}{(1.15$\times$)}} & \textbf{2.58 $\pm$ 0.08 {\scriptsize\textcolor{gray}{(3.76$\times$)}}} & 2.93 $\pm$ 0.08 {\scriptsize\textcolor{gray}{(3.46$\times$)}} & 3.9 $\pm$ 0.0 {\scriptsize\textcolor{gray}{(7.41$\times$)}} & 13.11 $\pm$ 0.00 {\scriptsize\textcolor{gray}{(2.42$\times$)}} \\
 &  & BSGS-H200-TBE & 33.20 {\scriptsize\textcolor{gray}{(1.24$\times$)}} & 2.60 $\pm$ 0.05 {\scriptsize\textcolor{gray}{(3.73$\times$)}} & \textbf{2.91 $\pm$ 0.03 {\scriptsize\textcolor{gray}{(3.49$\times$)}}} & \textbf{3.5 $\pm$ 0.0 {\scriptsize\textcolor{gray}{(8.26$\times$)}}} & \textbf{12.21 $\pm$ 0.00 {\scriptsize\textcolor{gray}{(2.60$\times$)}}} \\
\midrule
\multirow{5}{*}{$2^{17}$} & \multirow{5}{*}{128} & HyDia-CPU-DG & 64.60 {\scriptsize\textcolor{gray}{(1.00$\times$)}} & 17.95 $\pm$ 0.58 {\scriptsize\textcolor{gray}{(1.00$\times$)}} & 17.56 $\pm$ 0.71 {\scriptsize\textcolor{gray}{(1.00$\times$)}} & 29.1 $\pm$ 0.1 {\scriptsize\textcolor{gray}{(1.00$\times$)}} & 41.74 $\pm$ 0.00 {\scriptsize\textcolor{gray}{(1.00$\times$)}} \\
 &  & HyDia-H200-DG & 245.40 {\scriptsize\textcolor{gray}{(0.26$\times$)}} & 11.05 $\pm$ 0.18 {\scriptsize\textcolor{gray}{(1.62$\times$)}} & 22.66 $\pm$ 0.26 {\scriptsize\textcolor{gray}{(0.77$\times$)}} & 24.6 $\pm$ 0.0 {\scriptsize\textcolor{gray}{(1.18$\times$)}} & 48.24 $\pm$ 0.00 {\scriptsize\textcolor{gray}{(0.87$\times$)}} \\
 &  & BSGS-CPU-DG & \textbf{43.30 {\scriptsize\textcolor{gray}{(1.49$\times$)}}} & 11.67 $\pm$ 0.25 {\scriptsize\textcolor{gray}{(1.54$\times$)}} & \textbf{13.53 $\pm$ 0.66 {\scriptsize\textcolor{gray}{(1.30$\times$)}}} & \textbf{6.4 $\pm$ 0.1 {\scriptsize\textcolor{gray}{(4.55$\times$)}}} & \textbf{22.21 $\pm$ 0.00 {\scriptsize\textcolor{gray}{(1.88$\times$)}}} \\
 &  & BSGS-H200-TBS & 97.70 {\scriptsize\textcolor{gray}{(0.66$\times$)}} & \textbf{8.93 $\pm$ 0.22 {\scriptsize\textcolor{gray}{(2.01$\times$)}}} & 14.66 $\pm$ 0.13 {\scriptsize\textcolor{gray}{(1.20$\times$)}} & 7.5 $\pm$ 0.0 {\scriptsize\textcolor{gray}{(3.88$\times$)}} & 23.12 $\pm$ 0.00 {\scriptsize\textcolor{gray}{(1.81$\times$)}} \\
 &  & BSGS-H200-TBE & 93.70 {\scriptsize\textcolor{gray}{(0.69$\times$)}} & 9.73 $\pm$ 0.28 {\scriptsize\textcolor{gray}{(1.84$\times$)}} & 15.03 $\pm$ 0.54 {\scriptsize\textcolor{gray}{(1.17$\times$)}} & 6.5 $\pm$ 0.0 {\scriptsize\textcolor{gray}{(4.48$\times$)}} & \textbf{22.21 $\pm$ 0.00 {\scriptsize\textcolor{gray}{(1.88$\times$)}}} \\
\midrule
\multirow{2}{*}{$2^{18}$} & \multirow{2}{*}{256} & HyDia-CPU-DG & 130.90 {\scriptsize\textcolor{gray}{(1.00$\times$)}} & 34.91 $\pm$ 1.31 {\scriptsize\textcolor{gray}{(1.00$\times$)}} & 35.71 $\pm$ 1.34 {\scriptsize\textcolor{gray}{(1.00$\times$)}} & 29.8 $\pm$ 0.2 {\scriptsize\textcolor{gray}{(1.00$\times$)}} & 61.75 $\pm$ 0.00 {\scriptsize\textcolor{gray}{(1.00$\times$)}} \\
 &  & BSGS-CPU-DG & \textbf{102.50 {\scriptsize\textcolor{gray}{(1.28$\times$)}}} & \textbf{22.01 $\pm$ 0.48 {\scriptsize\textcolor{gray}{(1.59$\times$)}}} & \textbf{24.64 $\pm$ 0.53 {\scriptsize\textcolor{gray}{(1.45$\times$)}}} & \textbf{6.7 $\pm$ 0.1 {\scriptsize\textcolor{gray}{(4.45$\times$)}}} & \textbf{42.22 $\pm$ 0.00 {\scriptsize\textcolor{gray}{(1.46$\times$)}}} \\
\midrule
\multirow{2}{*}{$2^{19}$} & \multirow{2}{*}{256} & HyDia-CPU-DG & 200.90 {\scriptsize\textcolor{gray}{(1.00$\times$)}} & 64.21 $\pm$ 2.92 {\scriptsize\textcolor{gray}{(1.00$\times$)}} & 63.96 $\pm$ 2.69 {\scriptsize\textcolor{gray}{(1.00$\times$)}} & 31.3 $\pm$ 0.3 {\scriptsize\textcolor{gray}{(1.00$\times$)}} & 101.77 $\pm$ 0.00 {\scriptsize\textcolor{gray}{(1.00$\times$)}} \\
 &  & BSGS-CPU-DG & \textbf{146.90 {\scriptsize\textcolor{gray}{(1.37$\times$)}}} & \textbf{44.49 $\pm$ 0.39 {\scriptsize\textcolor{gray}{(1.44$\times$)}}} & \textbf{47.09 $\pm$ 0.24 {\scriptsize\textcolor{gray}{(1.36$\times$)}}} & \textbf{7.9 $\pm$ 0.1 {\scriptsize\textcolor{gray}{(3.96$\times$)}}} & \textbf{82.24 $\pm$ 0.00 {\scriptsize\textcolor{gray}{(1.24$\times$)}}} \\
\midrule
\multirow{2}{*}{$2^{20}$} & \multirow{2}{*}{512} & HyDia-CPU-DG & 384.60 {\scriptsize\textcolor{gray}{(1.00$\times$)}} & 126.35 $\pm$ 3.74 {\scriptsize\textcolor{gray}{(1.00$\times$)}} & 127.81 $\pm$ 3.96 {\scriptsize\textcolor{gray}{(1.00$\times$)}} & 34.6 $\pm$ 0.5 {\scriptsize\textcolor{gray}{(1.00$\times$)}} & 181.81 $\pm$ 0.00 {\scriptsize\textcolor{gray}{(1.00$\times$)}} \\
 &  & BSGS-CPU-DG & \textbf{306.30 {\scriptsize\textcolor{gray}{(1.26$\times$)}}} & \textbf{88.99 $\pm$ 2.70 {\scriptsize\textcolor{gray}{(1.42$\times$)}}} & \textbf{95.85 $\pm$ 5.51 {\scriptsize\textcolor{gray}{(1.33$\times$)}}} & \textbf{10.1 $\pm$ 0.0 {\scriptsize\textcolor{gray}{(3.43$\times$)}}} & \textbf{162.29 $\pm$ 0.00 {\scriptsize\textcolor{gray}{(1.12$\times$)}}} \\
\bottomrule
\end{tabular}
}
\end{table*}


\section{Declaration of generative AI and AI-assisted technologies in the manuscript preparation process}

During the preparation of this work, the authors used AI-assisted tools, specifically OpenAI GPT and Anthropic Claude, to help implement a subset of the optimization ideas proposed in this work in C++ and CUDA. After using these tools, the authors reviewed and edited all generated code as needed and take full responsibility for the content of the published article.
\section{Acknowledgment}
Funding: Marcos A. Simplicio Jr was supported in part by the Brazilian National Council for Scientific and Technological Development (CNPq) under Grant 307732/2023-1, and by the Coordination for the Improvement of Higher Education Personnel (CAPES) under Finance Code 001.


\bibliographystyle{elsarticle-num-names} 
\bibliography{sample}






\end{document}